\documentclass[10pt,journal,compsoc,preprint]{IEEEtran}

\ifCLASSOPTIONcompsoc
  \usepackage[nocompress]{cite}
\else
  \usepackage{cite}
\fi

\usepackage{flushend}
\usepackage{microtype}                 
\PassOptionsToPackage{warn}{textcomp}  
\usepackage{textcomp}                  
\usepackage{mathptmx}                  
\usepackage{times}                     
         
\usepackage{cite}                      

\usepackage{times}                     
\usepackage{tabu}                      
\usepackage{booktabs}    
\usepackage{todonotes}

\usepackage{hyperref}
\usepackage{ dsfont }
\usepackage{xspace}
\usepackage{multirow}
\usepackage{subcaption}
\usepackage{etoolbox}
\usepackage{rotating}
\usepackage{enumitem}
\usepackage[bottom]{footmisc}
\usepackage{makecell}
\usepackage{amsmath}
\usepackage{graphbox} 
\usepackage{bm}
\usepackage{siunitx}

\definecolor{eval_1b9e77}{HTML}{1b9e77}
\definecolor{eval_d95f02}{HTML}{d95f02}
\definecolor{eval_7570b3}{HTML}{7570b3}
\definecolor{eval_e7298a}{HTML}{e7298a}
\definecolor{eval_66a61e}{HTML}{66a61e}
\definecolor{eval_e6ab02}{HTML}{e6ab02}
\definecolor{eval_a6761d}{HTML}{a6761d}

\newcommand{\bluegreen}{\textcolor{eval_1b9e77}{\textbf{\#1B9E77}}\xspace}
\newcommand{\orange}{\textcolor{eval_d95f02}{\textbf{\#D95F02}}\xspace}
\newcommand{\purple}{\textcolor{eval_7570b3}{\textbf{\#7570B3}}\xspace}
\newcommand{\pink}{\textcolor{eval_e7298a}{\textbf{\#E7298A}}\xspace}
\newcommand{\green}{\textcolor{eval_66a61e}{\textbf{\#66A61E}}\xspace}
\newcommand{\yellow}{\textcolor{eval_e6ab02}{\textbf{\#E6AB02}}\xspace}
\newcommand{\brown}{\textcolor{eval_a6761d}{\textbf{\#A6761D}}\xspace}

\renewcommand{\diamond}{Diamond~\includegraphics[width=0.25cm]{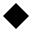}\xspace}
\newcommand{\square}{Square~\includegraphics[width=0.25cm]{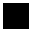}\xspace}
\renewcommand{\triangle}{Triangle~\includegraphics[width=0.25cm]{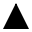}\xspace}
\newcommand{\clover}{Clover~\includegraphics[width=0.25cm]{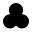}\xspace}
\renewcommand{\circle}{Circle~\includegraphics[width=0.25cm]{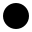}\xspace}

\newcommand{\resnet}{\textit{ResNet}\xspace}

\newcommand{\typecol}{\textit{color}\xspace}
\newcommand{\typeshape}{\textit{shape}\xspace}
\newcommand{\typered}{\textit{red.}\xspace}
\newcommand{\typeconj}{\textit{conj.}\xspace}

\newcommand{\nbCols}{\textit{\#colors}\xspace}
\newcommand{\nbShapes}{\textit{\#shapes}\xspace}
\newcommand{\nbDiffStimuli}{\textit{\#diffStimuli}\xspace}

\newcommand{\Hred}{\textit{H\textsubscript{red}}\xspace}
\newcommand{\Hconj}{\textit{H\textsubscript{conj}}\xspace}
\newcommand{\Htype}{\textit{H\textsubscript{type}}\xspace}
\newcommand{\Hcol}{\textit{H\textsubscript{color}}\xspace}
\newcommand{\Hshape}{\textit{H\textsubscript{shape}}\xspace}

\begin{document}
\title{Impacts of the Numbers of Colors and Shapes on Outlier Detection: from Automated to User Evaluation}

\author{Loann Giovannangeli, Romain Giot, David Auber and Romain Bourqui%
\thanks{L. Giovannangeli, R. Giot, D. Auber and R. Bourqui are with the LaBRI UMR CNRS 5800, University of Bordeaux, France.}%
\thanks{\{loann.giovannangeli; romain.giot; david.auber; romain.bourqui\}@labri.fr}%
}%

\markboth{preprint}{}
\IEEEtitleabstractindextext{%

\begin{abstract}
The design of efficient representations is well established as a fruitful way to explore and analyze complex or large data. In these representations, data are encoded with various visual attributes depending on the needs of the representation itself. To make coherent design choices about visual attributes, the visual search field proposes guidelines based on the human brain perception of features. However, information visualization representations frequently need to depict more data than the amount these guidelines have been validated on. Since, the information visualization community has extended these guidelines to a wider parameter space.

This paper contributes to this theme by extending visual search theories to an information visualization context. We consider a visual search task where subjects are asked to find an unknown outlier in a grid of randomly laid out distractor. Stimuli are defined by color and shape features for the purpose of visually encoding categorical data. The experimental protocol is made of a parameters space reduction step (\textit{i.e.}, sub-sampling) based on a machine learning model, and a user evaluation to measure capacity limits and validate hypotheses. The results show that the major difficulty factor is the number of visual attributes that are used to encode the outlier. When redundantly encoded, the display heterogeneity has no effect on the task. When encoded with one attribute, the difficulty depends on that attribute heterogeneity until its capacity limit (7 for color, 5 for shape) is reached. Finally, when encoded with two attributes simultaneously, performances drop drastically even with minor heterogeneity.

\end{abstract}

}

\maketitle

\IEEEdisplaynontitleabstractindextext

\begin{figure*}
\centering
  \begin{subfigure}[b]{.230\linewidth}
		\includegraphics[width=\linewidth]{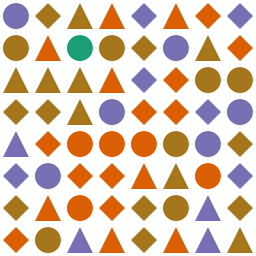}
		\caption{\typecol type; 3 shapes; 4 colors}
	\end{subfigure}
	\hspace{0.25cm}
	\begin{subfigure}[b]{.230\linewidth}
		\includegraphics[width=\linewidth]{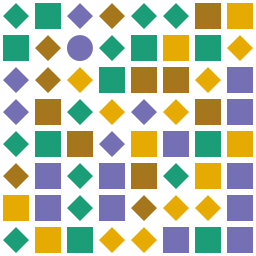}
		\caption{\typeshape type; 3 shapes; 4 colors}
	\end{subfigure}
	\hspace{0.25cm}
	\begin{subfigure}[b]{.230\linewidth}
		\includegraphics[width=\linewidth]{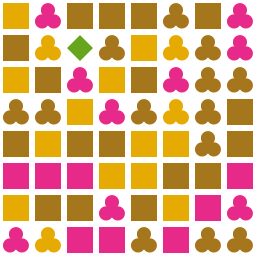}
		\caption{\typered type; 3 shapes; 4 colors}
	\end{subfigure}
	\hspace{0.25cm}
	\begin{subfigure}[b]{.230\linewidth}
		\includegraphics[width=\linewidth]{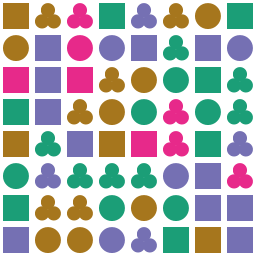}
		\caption{\typeconj type; 3 shapes; 4 colors}
	\end{subfigure}
	\caption{\label{fig:typesExamples} 
	Experimental object examples for the 4 possible \emph{types} of outliers. The outlier is always at position 10 (\emph{i.e.}, second row, third column from the top left corner) in these examples. 
	(a) In \typecol type images, the outlier color is unique; 
	(b) in \typeshape type images, the outlier shape is unique; 
	(c) in \textit{redundant} type (\typered) images, both the outlier color and shape are unique; and 
	(d) in \textit{conjunction} type (\typeconj) images, the outlier combination of color and shape is unique.}
  \label{fig:teaser}
\end{figure*}

\section{Introduction}
\label{sec:introduction}
The perception and visual search communities have conducted many studies to theorize how our brain processes scenes (\textit{i.e.}, images) in such a way that we immediately identify recognizable objects but have to focus on more specific (complex) objects. These studies have led to different theories \cite{treisman1980theory,duncan1989search}, among which the Treisman and Gelade \textit{Feature-integration theory of attention}~\cite{treisman1980theory} has served as a major reference for a long time. In this theory, representations are made of several visual attributes (\textit{e.g.}, color, direction, movement), which themselves are composed of features (\textit{e.g.}, color: red, blue; direction: left, right). The brain can process all features of a visual attribute in parallel (\emph{i.e. feature search}), whereas focal attention is needed to combine features of different attributes (\emph{i.e. conjunction search}). Hence, in terms of performance, this theory contrasts \textit{feature} with \textit{conjunction} search, as the first enables parallel search on all stimuli, while the latter involves serially processing the stimuli with focal attention.

In the information visualization research field, one of the main goals is to ease the search for data that are not trivially queryable. To efficiently represent these data, experts make visual choices driven by visual perception theories, as shown by Ware \cite{colinware2012book} or Healey and Enns \cite{healey2012recap}.
Various recommendations about visual attribute efficiency have been produced to help experts in their choices when highlighting data in their representations~\cite{haroz2012outlier,gramazio2014relation,huber2005motion,itoh2004grouping,mackinlay1986position
}.
However, visual search researchers are mostly interested in small-scale, fine-grained displays and measure subject differences in performance in milliseconds. Treisman and Gelade~\cite{treisman1980theory} claimed that \textit{``we cannot normally locate an item which differs from a field of distractors without also knowing at least on which dimension (color or shape) that difference exists"}, but their experiment was made of trials for which the time limit was set to 3 seconds. Such a time limit does not reflect the order of magnitude at which information visualization representations aim to be efficient. Since the two domains have different goals and do not measure efficiency at the same scale, we believe that such claims have limited applicability in information visualization. This paper aims to study the capacity limits of the \textit{color} and \textit{shape} visual attributes in a more information visualization--like context: when the number of distractors is relatively high and the number of data classes (\emph{i.e.}, visually different stimuli) increases. In this context, the \textit{capacity limit} of a visual attribute refers to the maximum number of different features of that attribute that can be in a representation without damaging its readability.

Color \cite{ware1988color,healey1996color} and shape \cite{chernoff1973shape,post1995shape} are two widely used visual attributes for encoding data in representations (\emph{e.g.}, scatter plots\cite{gleicher2013scatterplot}, geographic maps \cite{bertin1983semiology}, graphs \cite{altunbay2009colorGraph}, and parallel coordinates \cite{zhou2008parallelCoord}). However, it is often unclear how well these visual attributes remain efficient as visualizations become increasingly complex (\textit{e.g.}, the number of data items or classes to represent increases).

In this paper, we adapt some terminology from the literature to align ourselves with the information visualization community. We call \emph{parameters} what Treisman (\emph{e.g.}, \cite{treisman1977multidim, treisman1980theory}) calls \emph{dimensions}, and we use \emph{value} instead of \emph{feature}, \emph{outlier} instead of \emph{target} and \emph{distractor} instead of \emph{nontarget}. A \emph{parameter} therefore refers to either a visual attribute (\emph{e.g.}, color, shape) or any other variable of the representation (\emph{e.g.}, number of shapes), both of which have \emph{values} (\emph{e.g.}, color: red, shape: circle, number of shapes: $4$).

The experiment aims to observe the capacity limits of human attention when processing a visual search task on stimuli defined by color and shape. A naive capacity limit approximation exists when the number of features of a visual attribute in a display exceeds the short-term memory capacity. Subjects would have to \textit{remember} the stimuli while continuously performing the visual search, and this would make the task harder to solve. This limit is commonly assumed to be $7 \pm 2$ (\textit{i.e.} Miller~\cite{miller1956magical} magical number). We suspect that the capacity limits of the color and shape attributes are lower than expected and are likely to be exceeded in many information visualization representations. The experiment studies a visual search task in which the goal is to identify an outlier stimulus in a grid of randomly laid out distractor stimuli. There is always exactly one outlier per grid, whose color, shape and encoding are unknown to the subject. The originality of this work is that we exhaustively study the configurations of the considered parameter values to measure their effects on subject performance under the scope of information visualization constraints.

The main parameters that define the considered task are the number of shapes, the number of colors, the shape of the outlier, the color of the outlier, the position of the outlier and the type. The first two parameters refer to the total number of colors or shapes that are displayed in a grid. The next three are the outlier stimulus properties (shape, color, position). Finally, type encodes what makes the outlier unique in a grid; it has $4$ possible values (shown in \autoref{fig:teaser}), meaning that the outlier color, the outlier shape, both the outlier color and shape or their combination (color-shape) are unique in the grid.

In this paper, we study how the number of shapes, number of colors, and type affect subject performance when solving the task. As the exhaustive number of parameter configurations is high, it is first shortened based on a computed difficulty metric that relies on a deep neural network (DNN) model, which was trained to solve the task with all the configurations. The model performances provide insights into the effects of the parameters values on hypothetical performances obtained when solving the task. We call this process a \textit{difficulty} metric since it assesses the difficulty of solving a task with regard to an experimental object, and this is different from \textit{quality} metrics that are used to assess the readability of representations. Relying on this metric, the user evaluation design space and hypotheses are refined. Finally, the validity of the proposed hypotheses is verified, and the capacity limits of color and shape are studied through a  user evaluation.

The contributions of this paper are the experiment, along with the validation of the hypotheses, the measurements of the capacity limits of color and shape, the extension of some well-established visual search claims to an information visualization context and, to a lesser extent, the idea of a computable difficulty metric based on the performances of a DNN. The novelty of this work comes from the large parameter space it covers (\num{3290} configurations that are repeated \num{64} times, leading to \num{210560} experimental objects) and the computation of a dedicated difficulty metric that learns the task difficulty factors from the data to set up an end-user evaluation.

The remainder of this paper is structured as follows. \autoref{sec:related_work} presents related work from the visual search literature and its relation to information visualization. \autoref{sec:methodology} presents the design of the computed difficulty metric, and \autoref{sec:dataset} and \ref{sec:automated_pre_eval} describe its steps in detail. \autoref{sec:user_eval} presents the experimental evaluation setup and results. Finally, \autoref{sec:discussion} discusses these results, and \autoref{sec:conclusion} draws conclusions and presents future work leads.

\section{Related Work}
\label{sec:related_work}
The task in this experiment relates to the identification of a target stimulus in a background of nontarget stimuli, and it has various application domains. In perception, researchers have used this task to understand how the brain processes displays, while in the field of information visualization, studies are focused on optimizing the time required to solve the task. In this section, we present some literature on the two domains. We also present recent works about the use of deep neural networks (DNNs) for evaluating the readability of representations.

\subsection{Visual Search in Perception}
The most widely accepted theory in the visual search research field is the \emph{Feature-integration theory of attention} by Treisman and Gelade \cite{treisman1980theory}. It distinguishes \emph{feature search}, where the brain processes all stimuli features of a visual attribute in parallel, from \emph{conjunction search}, where the brain focuses serially on stimuli to combine the features of different visual attributes. They claimed that the search time in conjunction search can be up to three times longer if the similarities between the stimuli are too high. Their results were built on evaluations made using simplistic displays with a short time limit to solve the task and performance deltas measured in milliseconds. As mentioned in \autoref{sec:introduction}, we believe that such results have limited applicability in information visualization contexts.

On the other hand, Duncan and Humphreys \cite{duncan1989search} leveraged Treisman theory to build their own based on stimuli similarities and templates. They showed that as the target to nontarget (T-N, \textit{i.e.}, outlier to distractor) similarity increases, the task becomes more difficult. This situation is even worse if the nontarget to nontarget (N-N) similarity increases, except in cases where the T-N similarity remains small. Furthermore, the number of possible nontargets in the representation, which they called \textit{nontarget heterogeneity}, severely affects the task difficulty. Finally, they stated that if a target can be identified by a specific dimension (\textit{i.e.}, a relevant dimension), heterogeneity in other dimensions (\emph{i.e.}, irrelevant dimensions) should only have a minor impact on the search task; this corroborated the results of Treisman \cite{treisman1977multidim}. These are important results for our task design, as we consider a large parameter space. We expect to observe the harmful effects of \textit{nontarget heterogeneity}, which is attenuated when it occurs in irrelevant parameters.

Even though these theories diverge on their conceptions of the visual search process in the human brain, they also agree on some points. First, both agree that conjunction search is harder than feature search. They also agree that a shape is decomposed into simpler sub-features (\emph{e.g.}, vertical lines, tilted lines, curves) during the stimulus identification process \cite{treisman1988tilted}, which makes it a complex dimension. Finally, both claim that similarity between stimuli is a factor of upmost importance in visual search task difficulty.

Pashler \cite{pashler1988texture} has shown that, whether a target is known \emph{or not}, varying an irrelevant dimension in a representation only induces minimal interference in the results. As our task requires the subjects to identify \textit{unknown} outlier stimuli, heterogeneity in an irrelevant dimension should not significantly affect their performances. Their experiment only considered a few variations in the visual attributes they monitored (\textit{e.g.}, two shapes, two background colors). We expect our experiment to either extend or refute their claim in an information visualization context.

Regarding feature search, Quinlan and Humphreys \cite{quinlan1987combination} found that the visual search of a target defined by shape is slightly linearly related to the total number of stimuli, whereas the latter has no impact on a target defined by color. This statement is partially observed, as we find that an outlier defined by color takes significantly less time to find than an outlier defined by shape for a fixed number of stimuli (see \autoref{sec:capacity_limits}). This corroborates previous studies, as shape is a visual attribute that can be decomposed into lower level features \cite{treisman1988tilted}, making it more prone to confusion (\textit{i.e.}, illusory conjunctions) than other attributes. For conjunction search, they found that error rates increase with the number of stimuli and that response time is linearly related with it. Moreover, they showed that in conjunction search, T-N similarities have more impact on subject performances than in single feature search. Finally, they pointed out that the more visual attributes the target shares with the distractors, the more difficult the task is; this fact is essential to the \Htype hypothesis in this experiment (see \autoref{sec:autom_hypothesis}). \\

\subsection{Visual Search in Information Visualization}

The perception field of research is a cornerstone of the information visualization community, as it is mandatory to understand how the brain perceives and processes images to build efficient visualizations. To that extent, Healey and Enns \cite{healey2012recap} drew a landscape of the visual perception literature that was dedicated to computer graphics applications. These works were fine-grained designed: in most cases, subjects were confronted with tasks with at most 3 colors or letters (\emph{i.e.}, shapes) and time limits between 1 and 3 seconds; reaction time differences were measured in milliseconds. To create guidelines for visualization designers, the information visualization community has kept running its own measurements at a broader scale of human perception than in the above experiment.

Haroz and Whitney \cite{haroz2012outlier} studied how \emph{colored groups} and \emph{motion} influence the effectiveness of information visualizations. Their experiments involved thousands of trials where the task was to find a target stimulus in a representation. Some parameters, such as the number of colors or the layout of the \emph{color groups}, were studied. They found that grouping colors (\emph{i.e.}, classes) significantly eased the task when the target to find was unknown. Moreover, when colors were grouped, it was easier to access overall information, such as the total number of colors/classes.

Inspired by Haroz and Whitney \cite{haroz2012outlier}, Gramazio \emph{et al.} \cite{gramazio2014relation} studied how the same task was sensitive to representation size. The number of stimuli in a representation, their layout, their size and the number of colors varied across the experiment. They studied these variations in both a grid and a scatter plot. Their results corroborated previous works, as they claimed that grouped classes led to better performances than ungrouped classes. They also found a correlation between the number of displayed stimuli and the task difficulty, and this correlation was attenuated when stimulus classes were grouped.

Demiralp \textit{et al.}~\cite{demiralp2014kernels} introduced the notion of a \textit{perceptual kernel}, a distance matrix that represents the perceived distances between members of a set of stimuli composed of one or several visual attributes. In their experiment, they estimated the perceptual kernels for the color, shape and size visual attributes, as well as their pairwise combinations. They showed that color and shape have very different kernels. In the shape kernel, we observe several ``distant" clusters of ``close" shapes, whereas the distances between colors are more evenly distributed. On the other hand, all stimuli are close to many others in the color-shape kernel. Their experiment considered 4 colors and 4 shapes (\textit{i.e.}, 16 stimuli), but only 4 clusters could be distinguished in the kernel, meaning that all stimuli had high levels of similarity with others. We expect that varying the number of shapes or colors should not have the same effect on the performances in our experiment as their kernels are different; mixing both attributes should have a significant impact.

According to Mackinlay \cite{mackinlay1986position}, \emph{position} is the best parameter for visually encoding data in representations. For example, in Western culture, as one reads from left to right and top to bottom, one could assume that cells placed in the top left corner of the grid are processed first. In this experiment, the stimulus layout in the representations is fixed ($8 \times 8$ grid), and we do not study the impact of the \textit{outlier position} on the results since it is not an assumption our study aims to validate. To mitigate its consequences on the results, its values are uniformly distributed in the dataset during both the deep neural network model learning phase (see \autoref{sec:methodology}, \autoref{fig:dataDistribOutPos}) and the user evaluation (see \autoref{sec:user_eval}, \autoref{fig:eval_outpos_distrib}).

\subsection{Neural Networks for Visualization Evaluation}
\label{sec:related_work_dnn}

Behrisch \textit{et al.}~\cite{behrisch2018quality} conducted a recent survey on quality metrics for information visualization and claimed that deep neural networks (DNNs) were a promising direction for evaluating the quality of a representation. On the same line of research, Haehn \emph{et al.}~\cite{haehn2018cnn} reproduced the Cleveland and McGill~\cite{cleveland1984study} study with different convolutional neural networks (CNNs) to evaluate how these networks performed compared to humans on various elementary graphical perception tasks (\emph{e.g.}, position relative to a scale, angle, or area). They found that CNNs and humans behave differently on these elementary graphical perception elements but were still enthusiastic about evaluating representations with DNNs. Later, Haleem \textit{et al.}~\cite{haleem2019evaluating} trained a CNN to predict various graph node-link representation quality metrics while feeding it with laid out graph images only (\textit{i.e.}, the CNN did not have access to the node coordinates, edges, etc.). Their model reached an accuracy above $85\%$ at a 95\% confidence level. These quality metrics were designed to encode some graph readability information for humans (although we already noted that they do not always accurately reflect human perception capabilities); their study proved that CNNs \textit{can} strongly approach them and thus significantly approach human perception capabilities. Finally, Giovannangeli \emph{et al.}~\cite{giovannangeli2020methodo} partially reproduced two evaluations comparing node-link to adjacency matrix graph representations~\cite{ghoniem2005readability,okoe2018node} with CNNs on counting and connectivity tasks. They proposed an automated method to compare visualization techniques and concluded that humans and machine-learning-based computer vision techniques can be correlated on the tasks they considered.

All these studies remained cautious about their results and raised several limitations. The task definition, data generation process, network architecture, hyperparameters, initial weights, etc. can lead to different network strategies and performances. As this research field was recently developed, it is still not well understood how CNNs and humans can be correlated, and we currently know more about their differences than correlations.

\section{DNN as a metric: Design Space Insights}
\label{sec:methodology}
As the evaluation parameter space of this experiment is large (\num{3290} configurations repeated \num{64} times), it would be complicated to perform this study directly with human subjects. Hence, the first step is to reduce the evaluation parameter space to make it feasible for human subjects to complete in a reasonable amount of time. To limit arbitrary design choices, the parameter space reduction process is based on a difficulty metric.
\subsection{DNN-based Difficulty Metric}
\label{sec:dnn_quality_metric}

Inspired by the Giovannangeli \emph{et al.}~\cite{giovannangeli2020methodo} method and following the recommendations of Haehn \emph{et al.}~\cite{haehn2018cnn} and Haleem \textit{et al.}~\cite{haleem2019evaluating}, we use an approach based on a 
convolutional neural network (CNN)
 to build a difficulty metric that assesses the efficiency of the parameter values in solving this task. The advantage of such an approach is that it does not require any a priori information about what configuration of parameters makes the task easier or harder. Its drawbacks will be discussed in \autoref{sec:resnet_limitations}. A CNN is selected and trained to solve the task. Its predictions are used as a basis to compute the metric. Relying on this metric, hypotheses are set up and refined prior to being verified during a standard user evaluation. The metric also provides insights into the configurations of parameters that lead to significantly different performances, and this helps to prune irrelevant configurations during the parameter space reduction process. We acknowledge that metrics do not perfectly reflect human behaviors on visual search tasks and that each of them has its limits. Nevertheless, they are automated and reproducible means for assessing the relations between experimental object parameters and task difficulty.

\subsection{Difficulty Metric Design Procedure}

The first step is to generate annotated data for training the DNN model to learn the task (see \autoref{sec:dataset}). Following the recommendations of~\cite{giovannangeli2020methodo}, hundreds of thousands of data samples are generated while trying to keep the parameter distribution uniform. The objective is to ensure (i) that the model truly learns to solve the task and does not learn the ground truth distribution and (ii) that the model does not perform better with a given parameter value because it has been seen more often in the training dataset.
Then, a generic DNN architecture from the literature (\resnet~\cite{he2016resnet}) is trained to optimize its accuracy on the task, and we keep its tuning to a minimum to avoid biasing the model with any a priori knowledge we could have about the task difficulty.
The trained model is then evaluated to ensure that it learned to solve the task so we can analyze its performances.
Its results are aggregated in different ways to statistically study the effect of each parameter on the task difficulty. The outcome of the statistical study is finally used as a difficulty metric (see \autoref{sec:automated_pre_eval}). Based on this metric, the hypotheses are refined, and the user evaluation parameter space is reduced to contain a limited number of arbitrary design choices. Finally, a user evaluation (see \autoref{sec:user_eval}) is conducted to verify these hypotheses.

\section{Task and Dataset}
\label{sec:dataset}
In this section, we detail the parameters that are considered during the experiment and their values.

\subsection{Task}
\label{sec:task}
The chosen task consists of identifying an outlier in an $8 \times 8$ grid of colored shapes drawn in an image of $256 \times 256$ pixels.
These properties enable (i) the consideration of a reasonable number of values for our key parameters (presented immediately after) and (ii) a good trade-off between image readability for a user and the possibility of feeding the image to a deep learning model on our runtime infrastructure (Titan X 12 GB GPU).
In such an image, a colored shape (\emph{i.e.}, stimulus) is considered an outlier if there is no other stimulus with the same \textit{color} and \textit{shape} visual attributes. The dimension(s) on which the outlier is made unique varies according to the \textit{type} parameter.

\textbf{Type} relates to the dimension(s) that make the outlier unique. It has 4 possible values: (i) \typecol, where the \emph{color} of the outlier is unique in the grid; (ii) \typeshape, where its \emph{shape} is unique in the grid; (iii) \textit{redundant}, where \emph{both} its color and shape are unique in the grid, (this refers to redundant encoding)~\cite{nothelfer2017redundant}; and (iv) \textit{conjunction}, where its color-shape \emph{combination} is unique in the grid.
In the following sections, we refer to these parameter values as the \typecol, \typeshape, \typered and \typeconj types. Examples of type values are provided in \autoref{fig:teaser}.

Each image contains exactly one outlier and $63$ distractors.
A colored shape is considered a distractor if it appears at least twice in the grid (otherwise, it is an outlier). There are at most $31$ different color-shape combinations for distractors in a grid.

\subsection{Data Space Definition}
\label{sec:data_space}

\begin{table}[!tb]
\centering
\resizebox{\columnwidth}{!}{
\begin{tabular}{|c|c|c|c|c|c|}
\hline
\multicolumn{3}{|c|}{\textbf{Visual attribute values}}& \multicolumn{3}{c|}{\textbf{Image}} \\
\hline
 \textbf{Shape} & \textbf{Color} & \textbf{Position}  & \multicolumn{1}{c|}{\textbf{Type}} & \textbf{\#colors} &  \textbf{\#shapes} \\
\hline
 \includegraphics[width=0.3cm]{figures/eval_colors_shapes/triangle.png}&  \bluegreen & 0 & color & 1 & 1  \\
 
 \includegraphics[width=0.3cm]{figures/eval_colors_shapes/circle.png}&   \orange & \smash{\vdots}  & shape & 2 & 2  \\
 
 \includegraphics[width=0.3cm]{figures/eval_colors_shapes/square.png}&  \purple & 63 & redundant (red.) & 3 & 3   \\
 
 \includegraphics[width=0.3cm]{figures/eval_colors_shapes/clover.png}&  \pink & & conjunction (conj.) & 4 & 4  \\
 
 \includegraphics[width=0.3cm]{figures/eval_colors_shapes/diamond.png}& \green &  & & 5 & 5    \\
 
 \includegraphics[width=0.3cm]{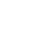} &   \yellow &  & & 6  &     \\
  \includegraphics[width=0.3cm]{figures/eval_colors_shapes/empty.png} &   \brown &  & & 7  &      \\
\hline
\end{tabular}
}
-\caption{\label{tab:parametersValues} All parameters values considered in this study. Color values are given as hexadecimal RGB codes. Shape and color values can be used by either the outlier, through the \emph{outlier color} and \emph{outlier shape} parameters, or by the distractor stimuli.}
\end{table}

The experimental objects of this study are images representing a grid. Each object is defined by six parameter values (see \autoref{tab:parametersValues}).

\textbf{Outlier shape} values are chosen among a set of shapes (see \autoref{tab:parametersValues} column 1). A shape can be defined by many sub-features (\emph{e.g.}, lines, orientation, size). In this experiment, every shape appears in a single orientation, and its size is set to the maximum value to fit in a $32 \times 32$ pixels cell using a 3 pixels padding. Five shapes are selected - \triangle, \circle, \square, \clover and \diamond - to mix the use of straight vertical/horizontal, diagonal and curved lines.

\textbf{Outlier color} values are chosen among a set of colors (see \autoref{tab:parametersValues} column 2). Some methods already exist for finding an efficient color set to represent targets (\emph{e.g.}, that of Bauer \emph{et al.} \cite{bauer1996color}). In this experiment, colors are chosen from the ColorBrewer\footnote[1]{\url{https://colorbrewer2.org/\#type=qualitative\&scheme=Dark2\&n=7}, consulted on January 2021} website~\cite{harrower2003colorbrewer}, a well-known color palette provider. More specifically, colors are taken from the \emph{7 qualitative classes} palette named \emph{Dark2}, as the colors should be as independent as possible (\emph{i.e.}, categorical) in this experiment. This palette was also one of the proposed sets with high saturation. It is important to note that, from the beginning, we planned to exclude colorblind subjects from the evaluation.

The \textbf{outlier position} relates to the position of the outlier in the experimental object. In this study, the position varies between $0$ and $63$ corresponding to the row-major order of the grid.

The \textbf{type} of an image relates to the outlier characteristics that make it unique in that experimental object (see \autoref{sec:task}).

The \textbf{number of colors} (\nbCols) relates to the total number of distinct colors used in an experimental object. In this experiment, the number of colors varies between $1$ and $7$. It is noteworthy that if an experimental object type is \typecol or \typered, the number of colors cannot be set to $1$ as a color must be reserved for the outlier.

The \textbf{number of shapes} (\nbShapes) relates to the total number of distinct shapes used in an experimental object. In this experiment, the number of shapes is between $1$ and $5$. For an experimental object of type \typeshape or \typered, the number of shapes cannot be set to $1$ as one is reserved for the outlier.

\subsection{Dataset Generation}
\label{sec:dataset_gen}

\begin{figure}[!bt]
	\centering
	\begin{subfigure}[b]{.49\linewidth}
		\includegraphics[width=\linewidth]{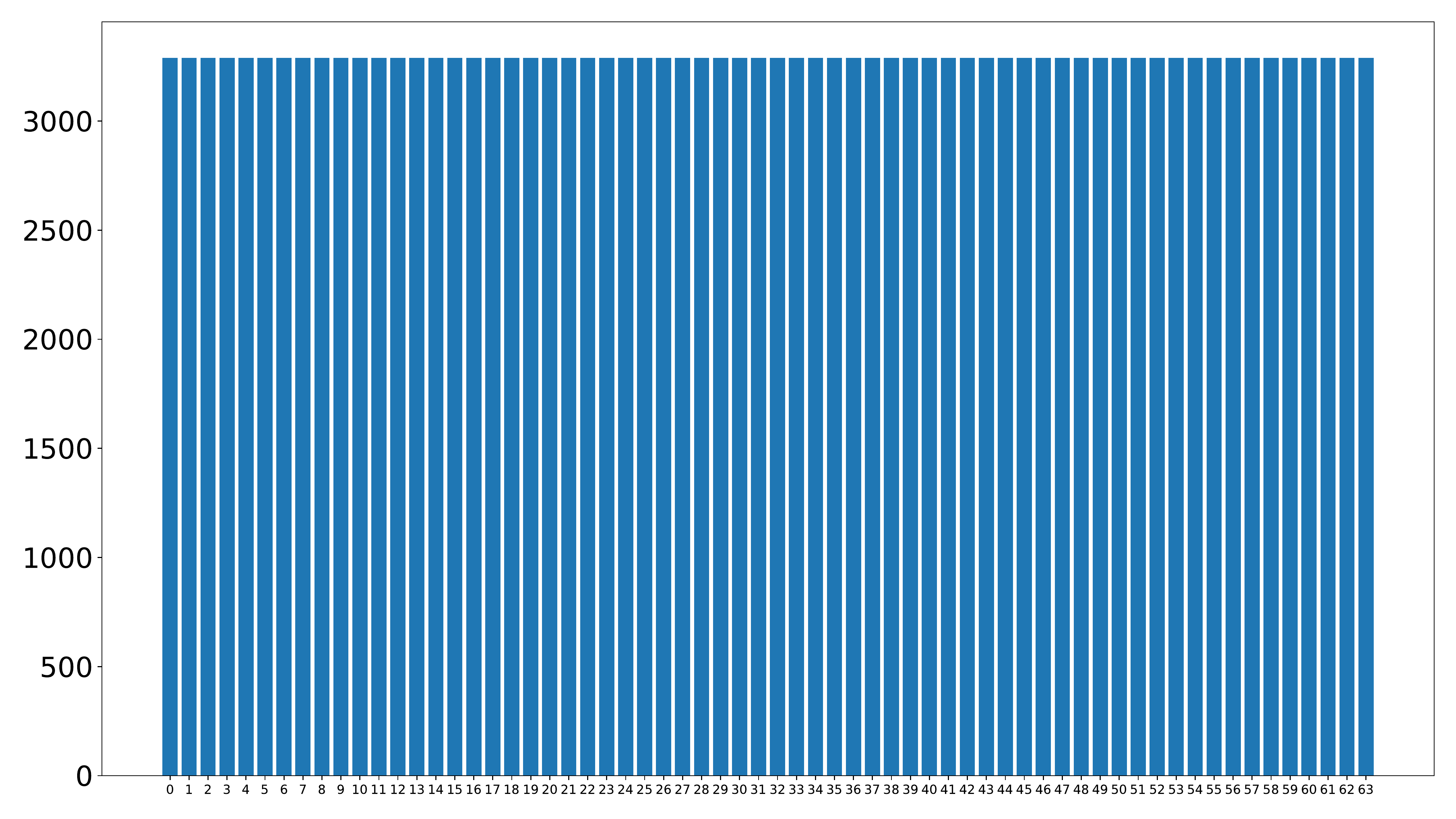}
		\caption{\label{fig:dataDistribOutPos}\textit{Outlier position} distribution}
	\end{subfigure}
	\begin{subfigure}[b]{.49\linewidth}
		\includegraphics[width=\linewidth]{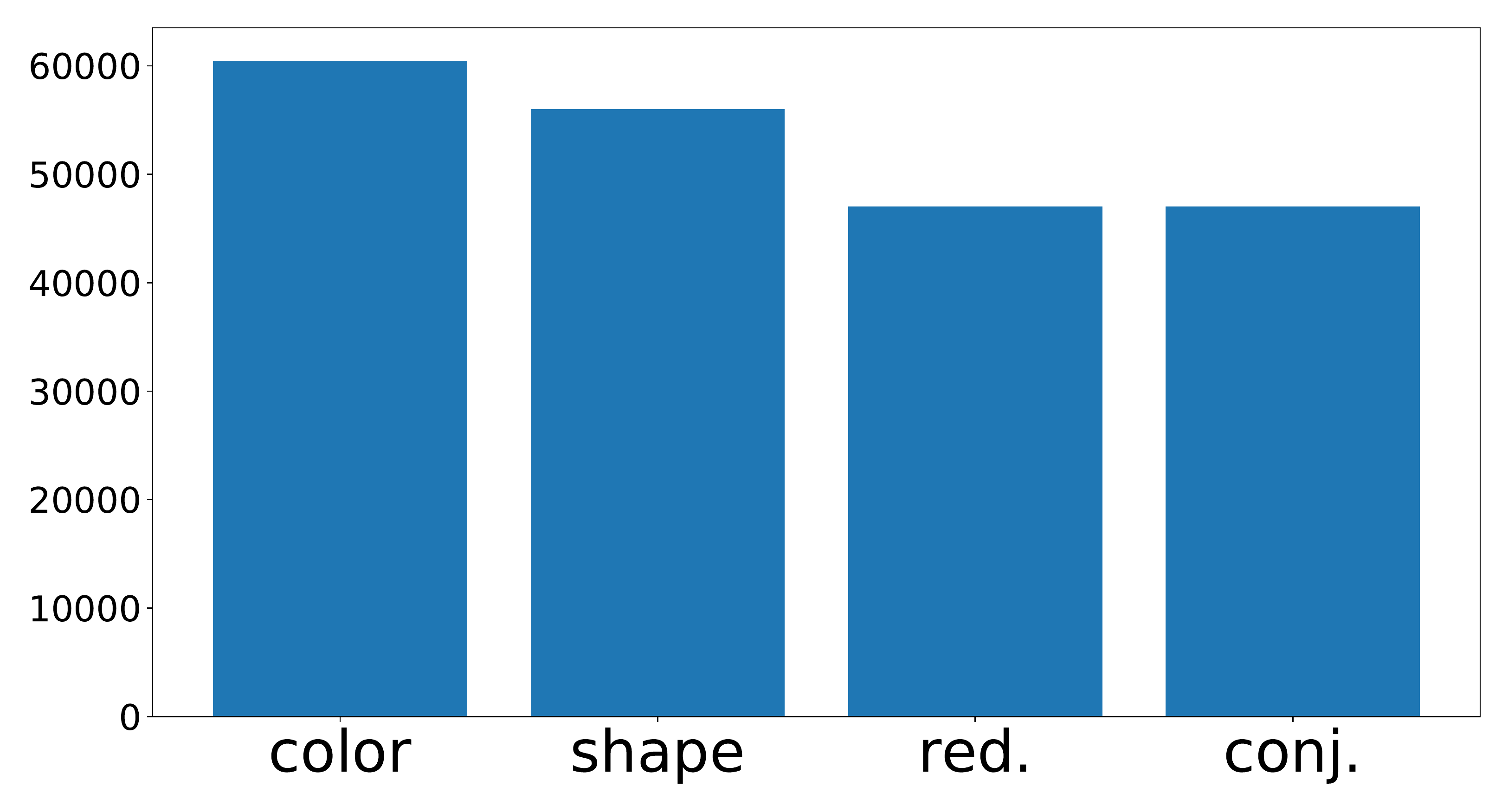}
		\caption{\label{fig:dataDistribType}\textit{Type} distribution}
	\end{subfigure}
	\begin{subfigure}[b]{.49\linewidth}
		\includegraphics[width=\linewidth]{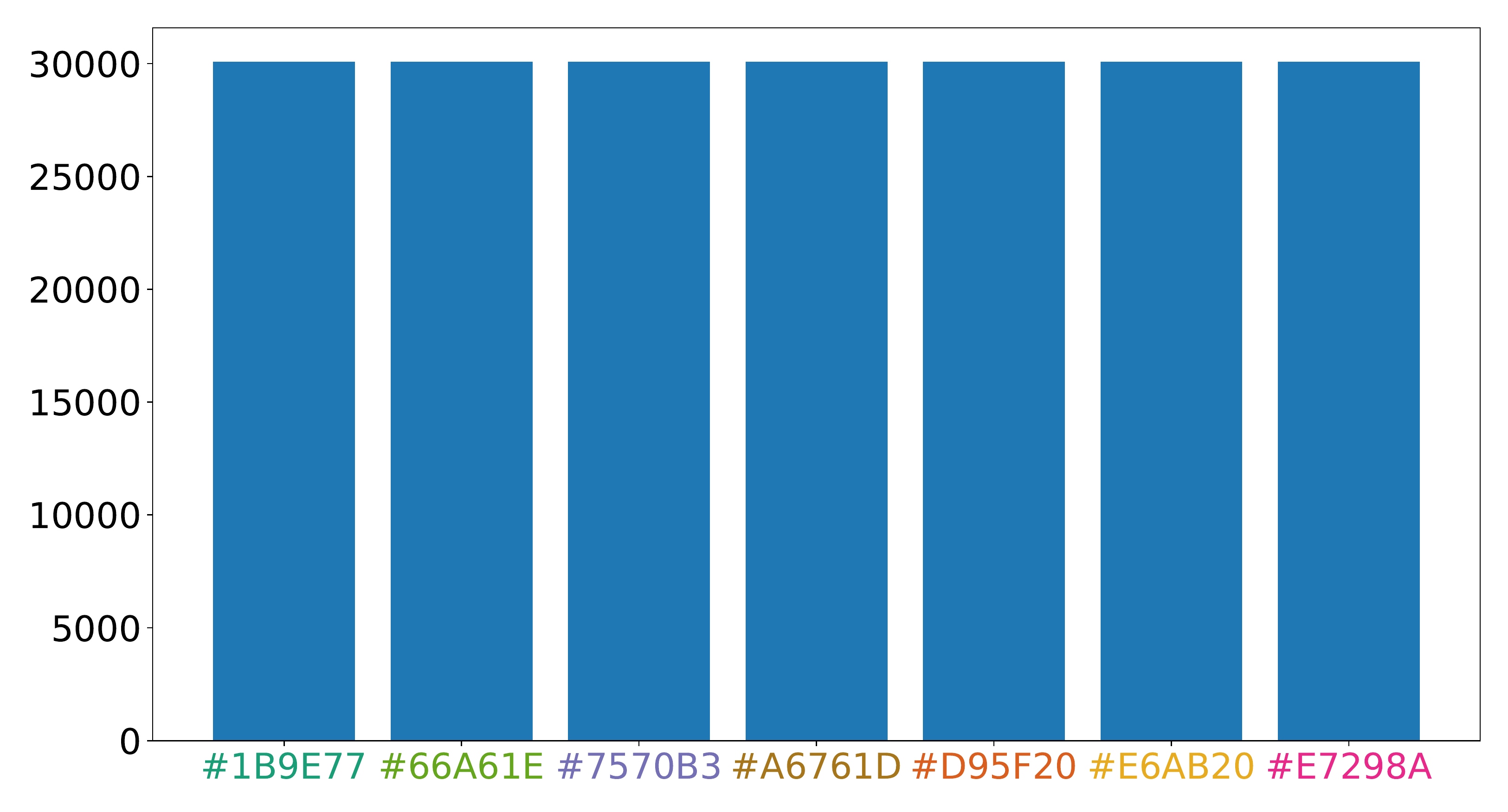}
		\caption{\label{fig:dataDistribOutCol}\textit{Outlier color} distribution}
	\end{subfigure}
	\begin{subfigure}[b]{.49\linewidth}
		\includegraphics[width=\linewidth]{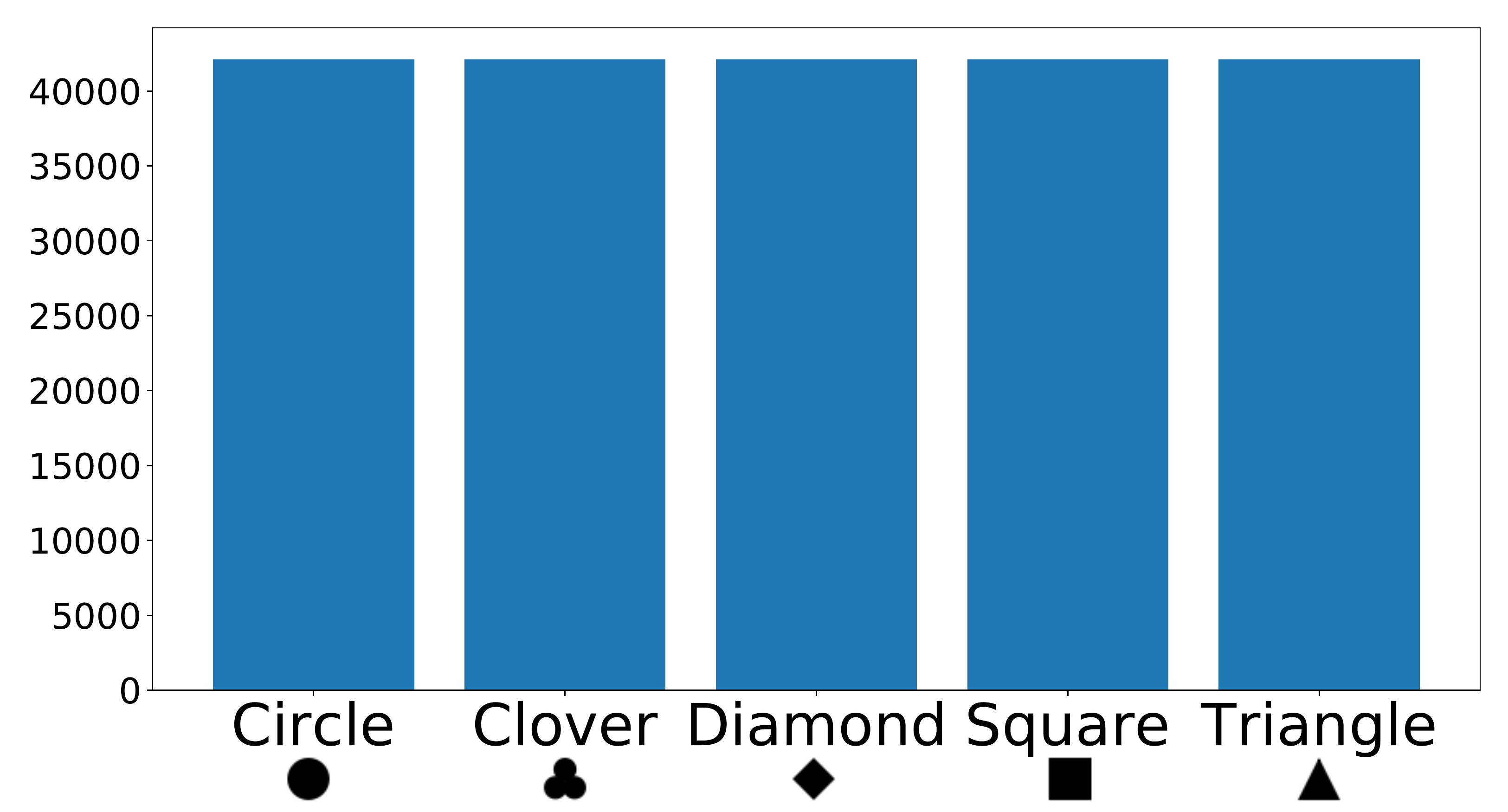}
		\caption{\label{fig:dataDistribOutShape}\textit{Outlier shape} distribution}
	\end{subfigure}
	\begin{subfigure}[b]{.49\linewidth}
		\includegraphics[width=\linewidth]{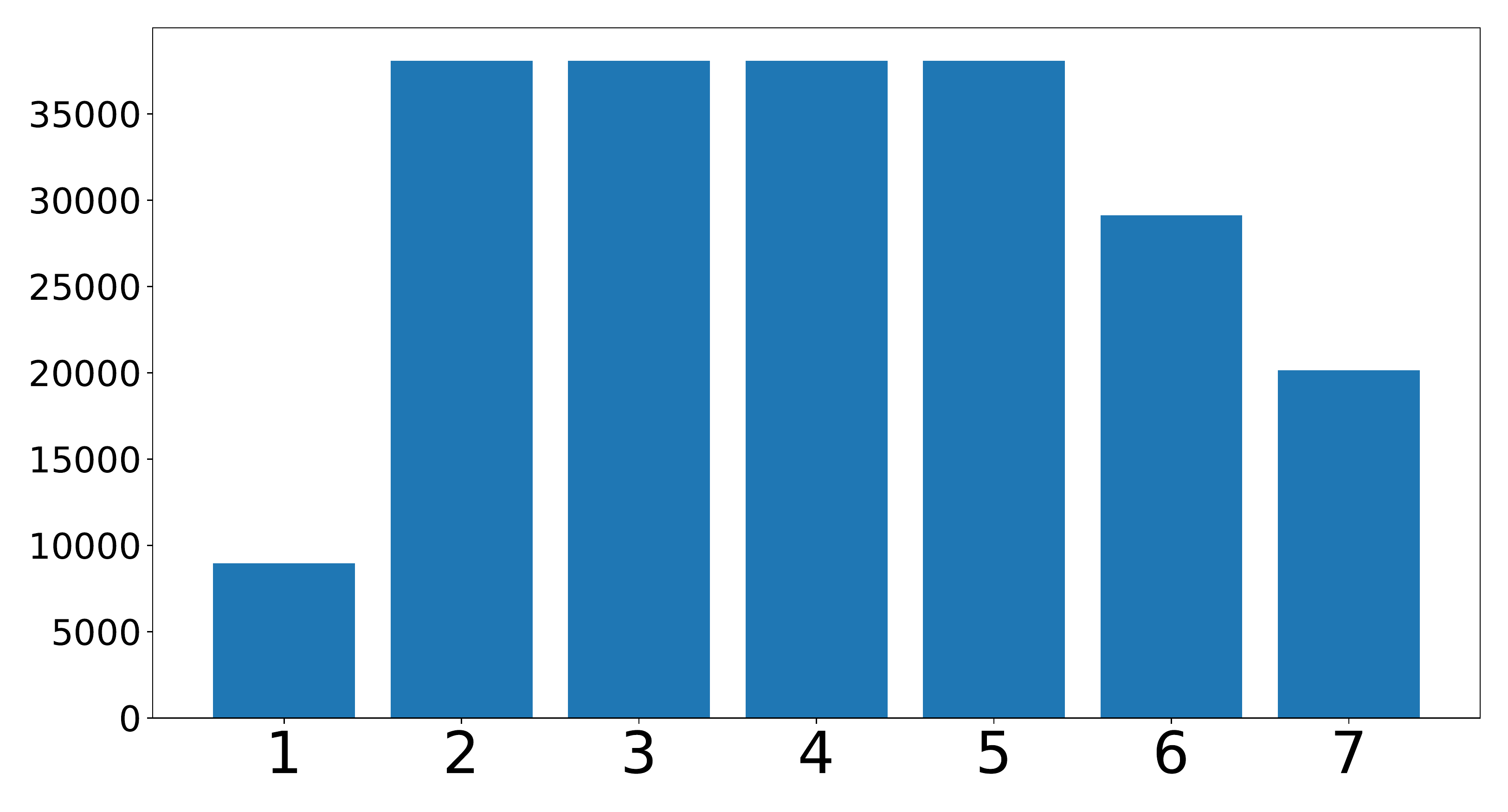}
		\caption{\label{fig:dataDistribNbCol}\nbCols distribution}
	\end{subfigure}
	\begin{subfigure}[b]{.49\linewidth}
		\includegraphics[width=\linewidth]{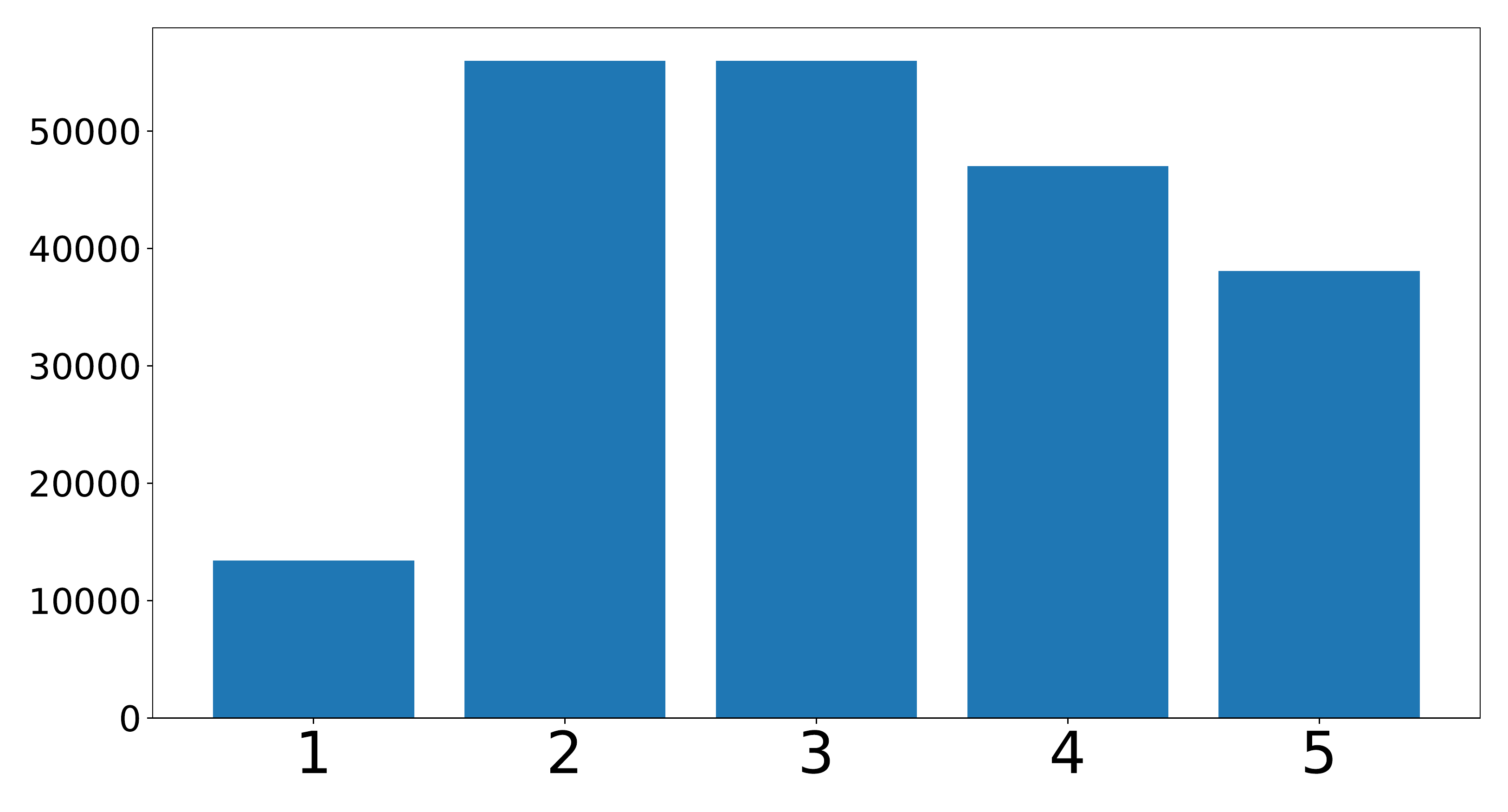}
		\caption{\label{fig:dataDistribNbShape}\nbShapes distribution}
	\end{subfigure}
	\caption{\label{fig:dataDistrib}Parameter value distributions in the \num{210560} images.}
\end{figure}

When generating the experimental dataset, the six parameter values were balanced to minimize distribution bias and train the model correctly.
The main concern was to balance outlier shape-color-position occurrences (see \autoref{fig:dataDistribOutPos}) to prevent the deep learning model from learning to find some stimuli or locations more easily than others because they were more common in the dataset.

The generation process also followed some constraints. Obviously, images with $1$ color and $1$ shape could not be generated, but less evident cases could not be generated either. Images of type \typered cannot be generated with either $1$ color and $2$ shapes or $2$ colors and $1$ shape. For images of type \typeconj, the combinations of parameter values using $1$ color or $1$ shape were not considered, as they would result in images of type \typeshape or \typecol, respectively. In addition, type \typeconj images could not be generated using $7$ colors and $5$ shapes in an $8 \times 8$ grid. One of the $7*5=35$ combinations should be reserved for the outlier, and $34$ should appear twice (for the distractors), so this would lead to at least $69$ stimuli.

The configurations using ($4$ shapes, $7$ colors) and ($5$ shapes, $6$ colors) were removed. From our experience, knowledge of the literature and pilot experiments, we strongly expect that the capacity limits we aim to study will be reached before (\emph{i.e.}, with lower parameter values) needing to use these high-valued configurations.

These constraints explain why the \emph{type}, \nbCols and \nbShapes values are not fully balanced, as shown in Figures~\ref{fig:dataDistribType},~\ref{fig:dataDistribNbCol} and~\ref{fig:dataDistribNbShape}.

By generating one image per combination of parameter values (see \autoref{tab:parametersValues}) while excluding those described above, we ended up with \num{210560} different images. As stated in \autoref{sec:introduction}, this study was not designed to investigate the effect of the \emph{outlier position} on the task. This parameter was only used to generate several samples with other parameter value combinations and is balanced uniformly to mitigate its consequences on the results. Therefore, the experiment studies $3290$ different parameter combinations, repeated $64$ times each.

Finally, the dataset was randomly split into 3 subsets for supervised learning purposes (\textit{hold-out} validation~\cite{arlot2010holdout}): \emph{train} (to learn the model), \emph{validation} (to prevent overfitting during training) and \emph{test} (to evaluate the model on unseen data).

\section{Difficulty Metric}
\label{sec:automated_pre_eval}
This section first presents how we selected our network architecture and trained it to solve the task. Then, it presents the trained model results as well as the statistical study that drove us to refine our hypotheses and reduce the user evaluation parameter space.

\subsection{Model Selection and Training}

As mentioned in~\cite{giovannangeli2020methodo}, a generic deep neural network (DNN) should be used rather than an architecture dedicated to the task(s) (or visualization technique(s)) to be studied.
To that extent and following the recommendation of~\cite{haehn2018cnn}, we tried several network architectures (\emph{e.g.}, \textit{LeNet} \cite{lecun1998lenet} and \textit{VGG-16/19} \cite{simonyan2014vgg}) and selected \resnet (from He \emph{et al.}~\cite{he2016resnet}) as it correctly learned to solve the task.
He \emph{et al.} \cite{he2019saliency} showed that such a model pretrained for image recognition already encodes some saliency information, which we expected to speed up the learning process with regard to spatial identification. The default weights of \resnet were set to their pretrained values on ImageNet~\cite{russakovsky2015imagenet}.

The \resnet architecture was tuned~\cite{he2016resnet}. Its input layer was set to fit the generated image resolution, and two successive dense layers were added after its output to fit the required number of classes for prediction.
We consider the identification of the outlier as a classification problem rather than a regression problem, where there would be a notion of distance between the predictions and their ground truths.
The size of the last dense layer was therefore set to predict the outlier position, \emph{i.e.}, to predict $64$ classes, and the size of the penultimate dense layer was set to 1024.

While the optimizer and default tuning of the learning phase were not modified, the batch size was set to $64$ (instead of $256$).
We used the \emph{early stopping} function of the Keras library~\cite{chollet2015keras} with a \textit{patience} of 15 epochs to end the training process.

\subsection{Results}

\begin{figure}[!tb]
\centering
\includegraphics[align=c,width=0.50\linewidth]{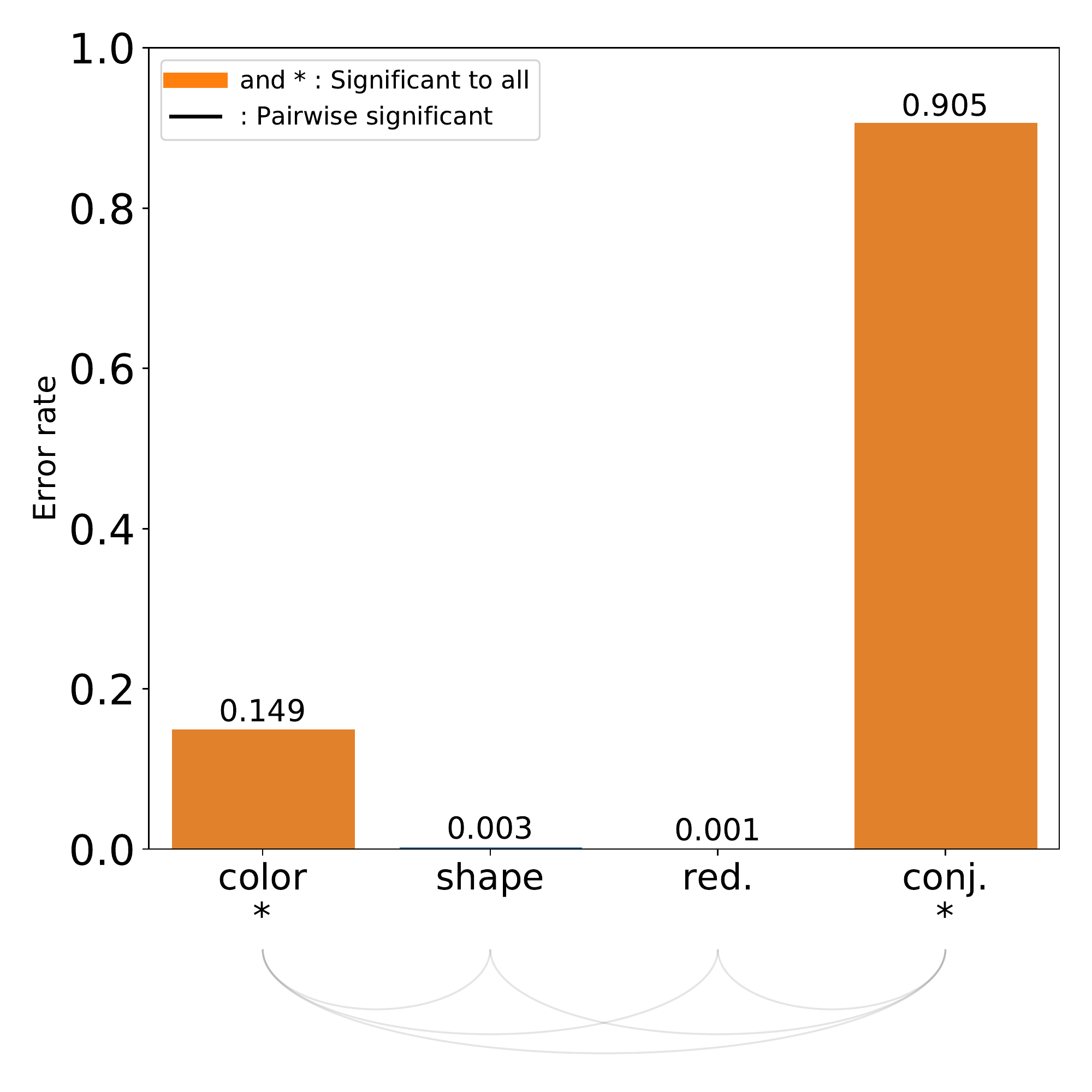}
\caption{Trained \resnet error rates (ER) on the \emph{test} set for the \emph{type} values. An arc between two labels means that the pairwise comparison between the ER of the two parameter values is significant ($p{\text -}value < 0.05$) according to a Wilcoxon rank-sum test. \textit{Reading example}: the \typecol type is significantly easier than the \typeconj type and significantly harder than the \typeshape and \typered types. There is no significant difference between the \typered and \typeshape types.}
\label{fig:perf_resnet_type}
\end{figure}

\begin{figure*}
 \centering

	\begin{tabular}{|c|c|c|c|c|}
	\hline
	& \nbCols & \nbShapes & \emph{Outlier Color} \\
		\hline
{\shortstack{\begin{sideways}\hspace{-0.5cm}Overall\end{sideways}}}
&
		{\includegraphics[align=c,width=.235\linewidth]{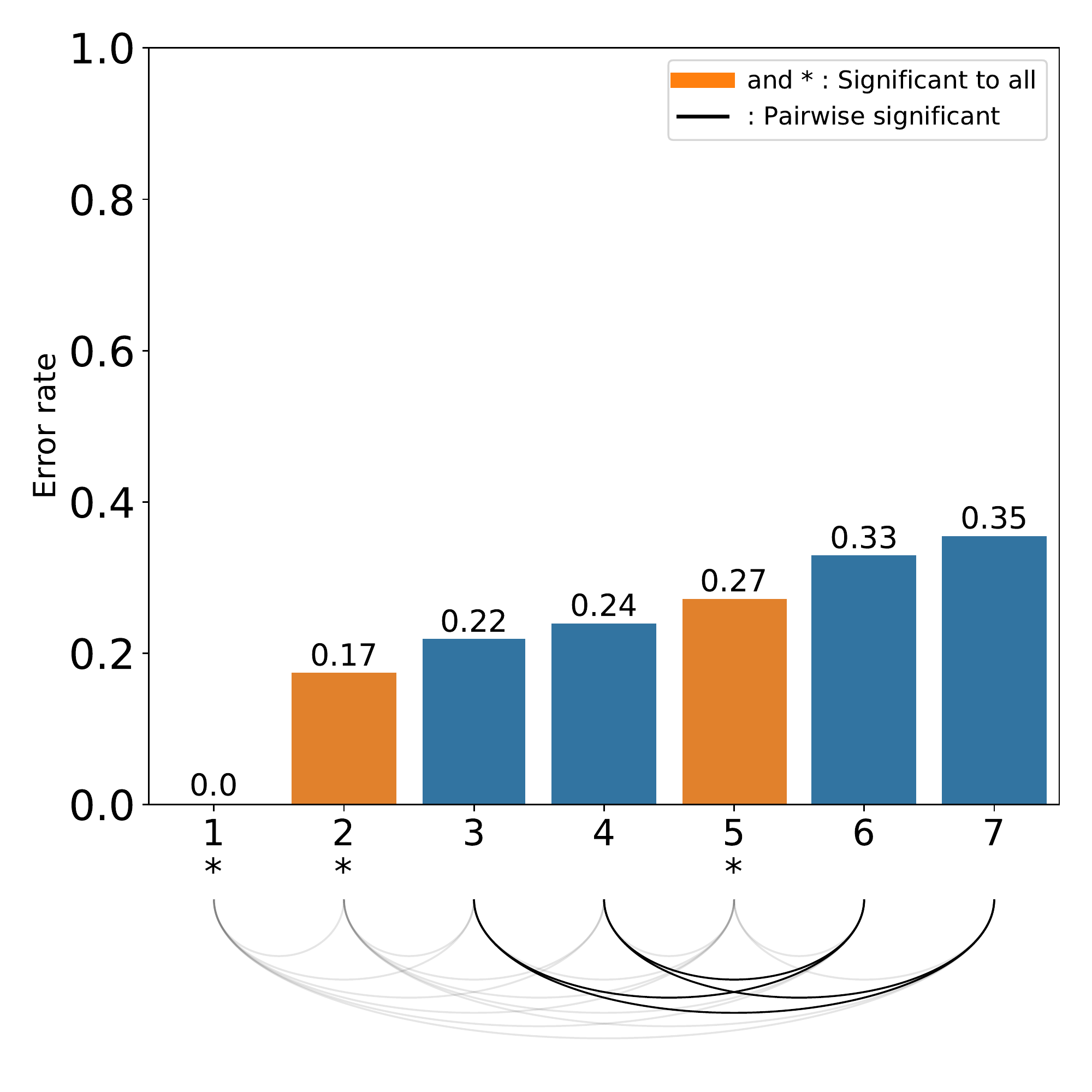}}&%
		{\includegraphics[align=c,width=.235\linewidth]{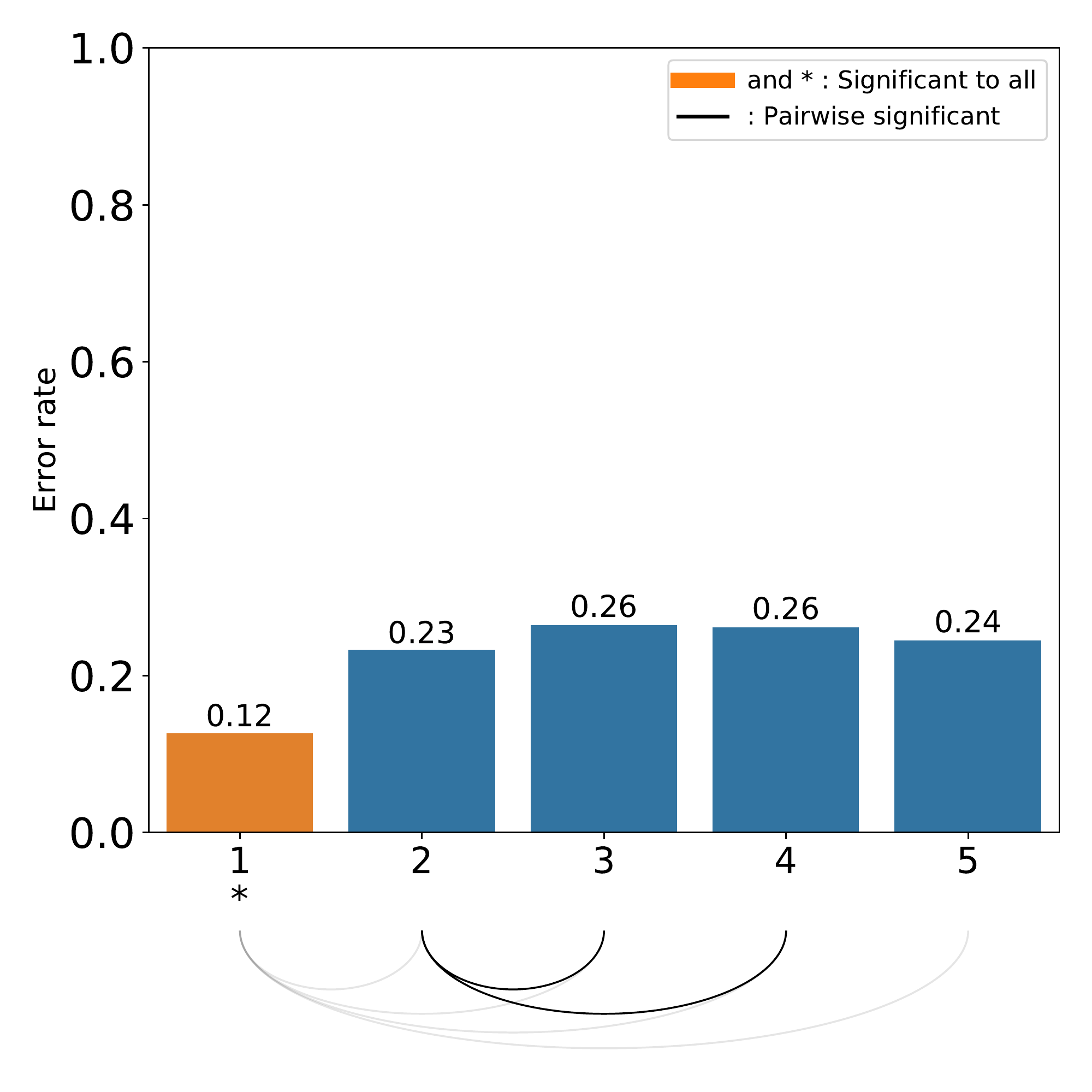}}&
		{\includegraphics[align=c,width=.235\linewidth]{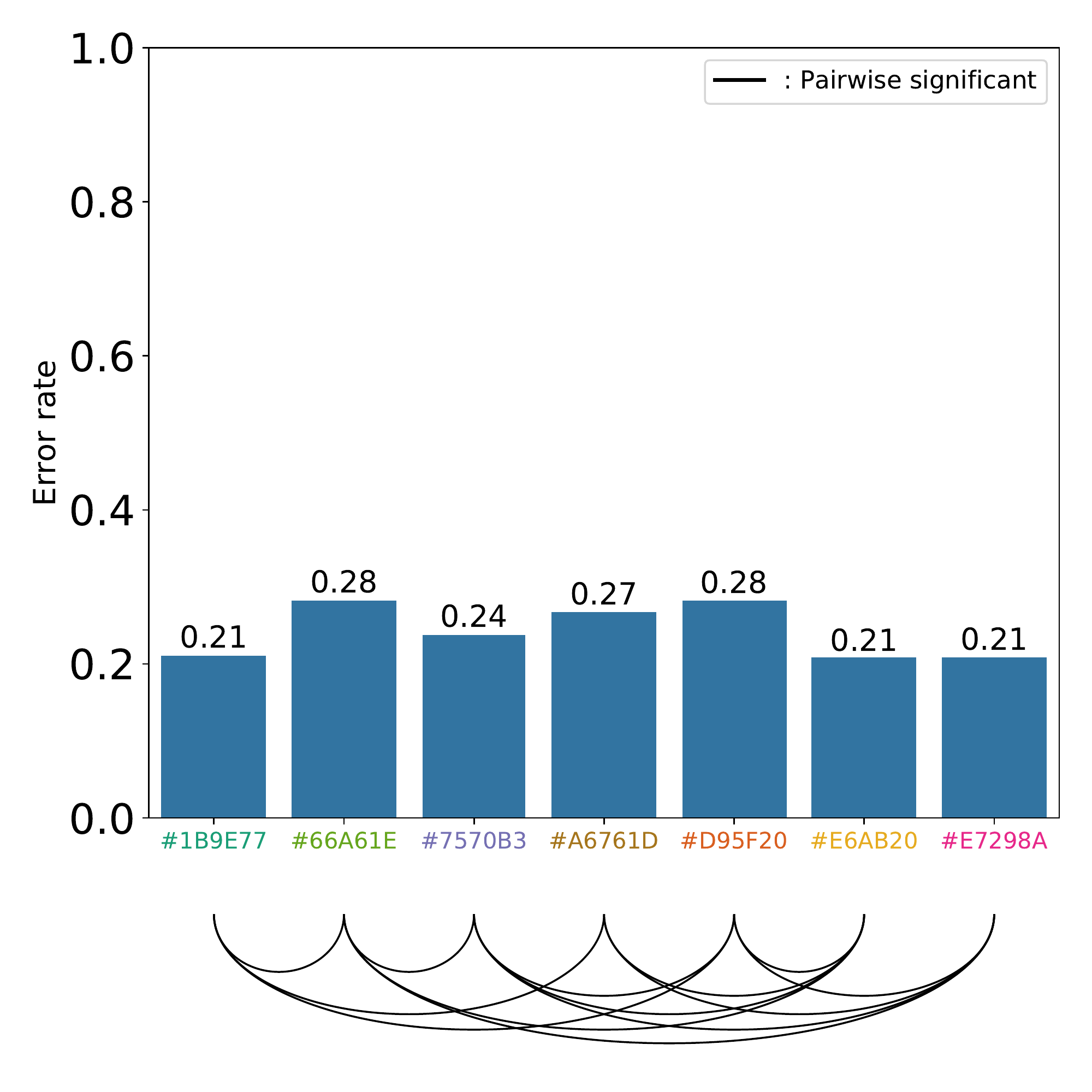}}\\	
	\hline
{\shortstack{\begin{sideways}\hspace{-0.5cm} \emph{Color} type\end{sideways}}}
&
		{\includegraphics[align=c,width=.235\linewidth]{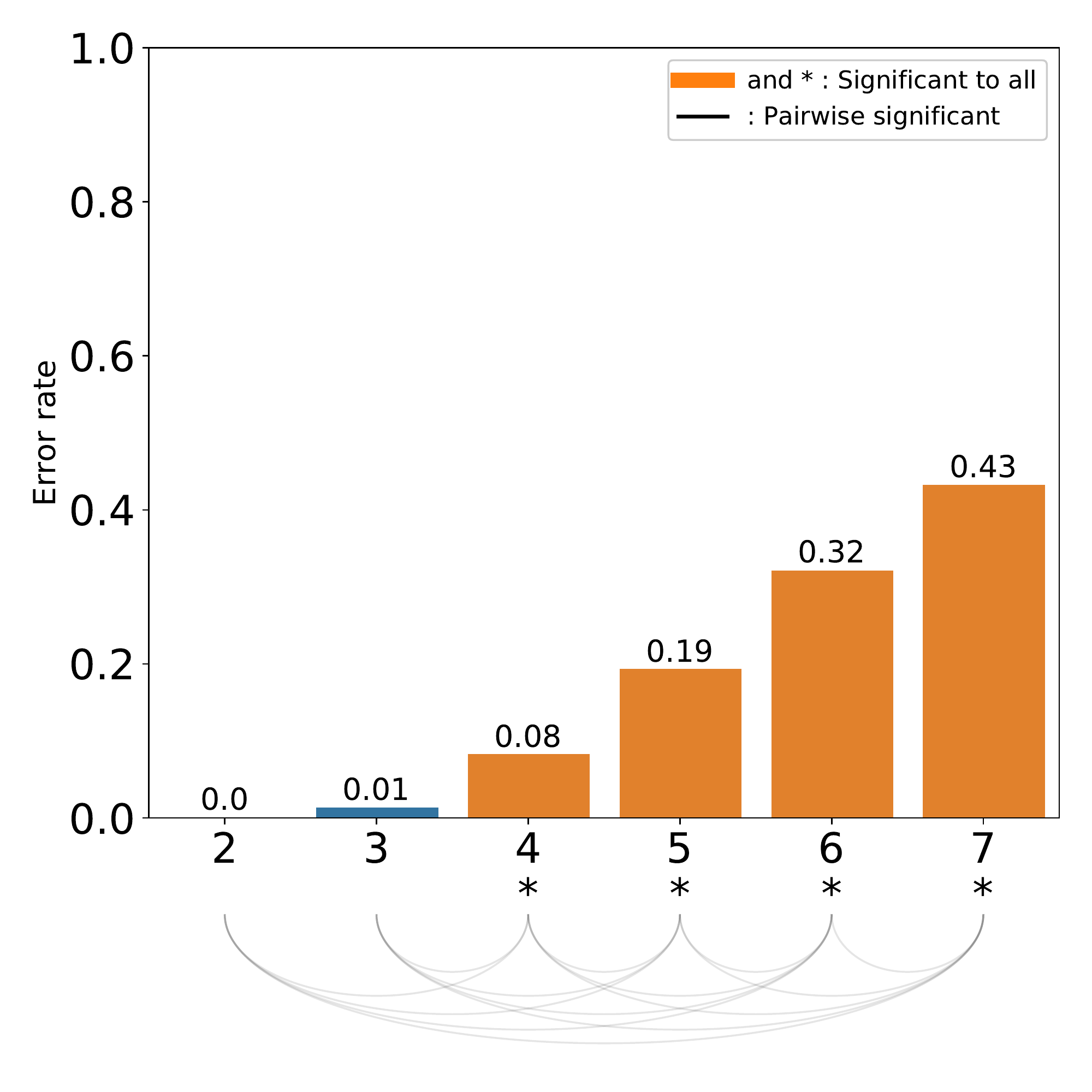}}&%
        {\includegraphics[align=c,width=.235\linewidth]{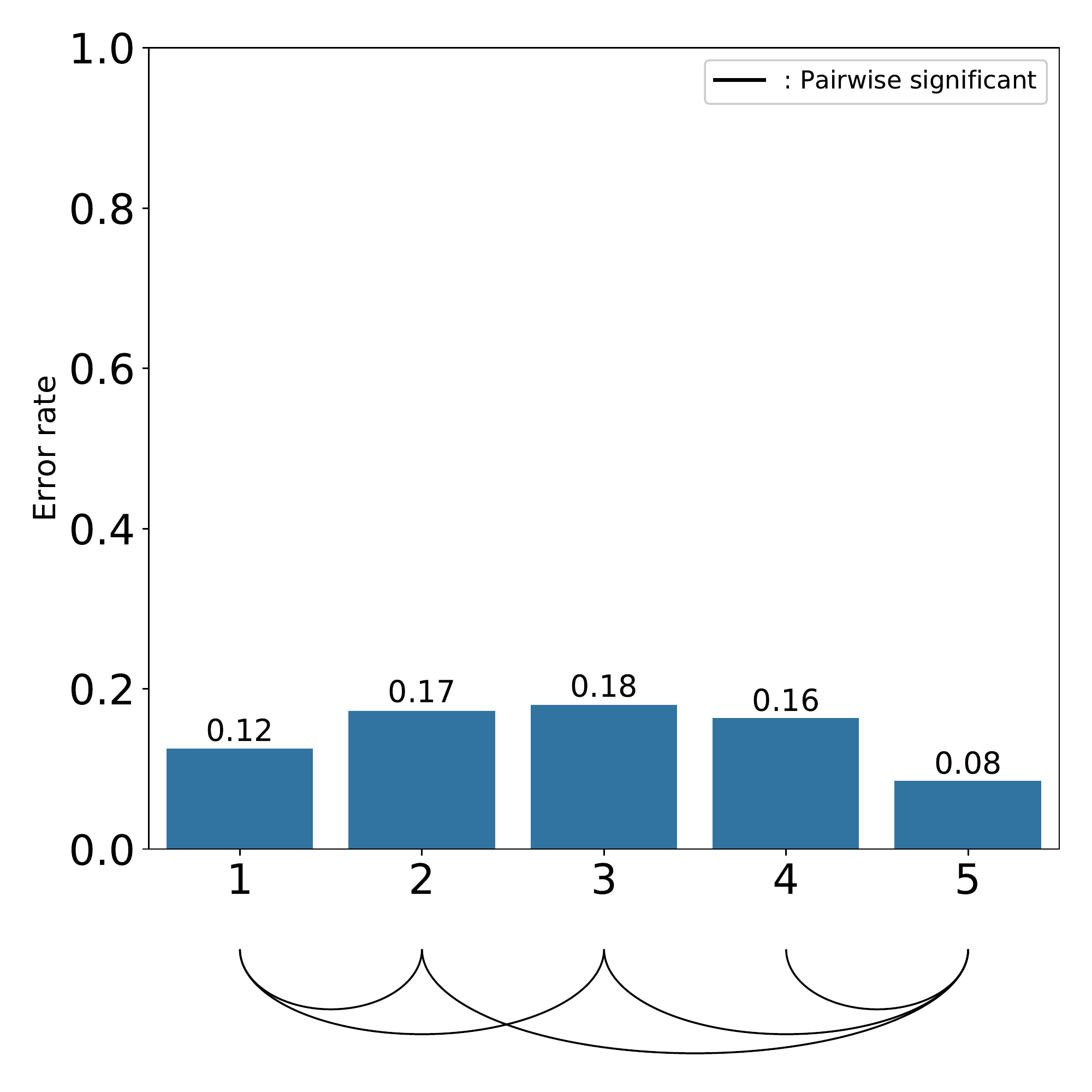}}&%
        {\includegraphics[align=c,width=.235\linewidth]{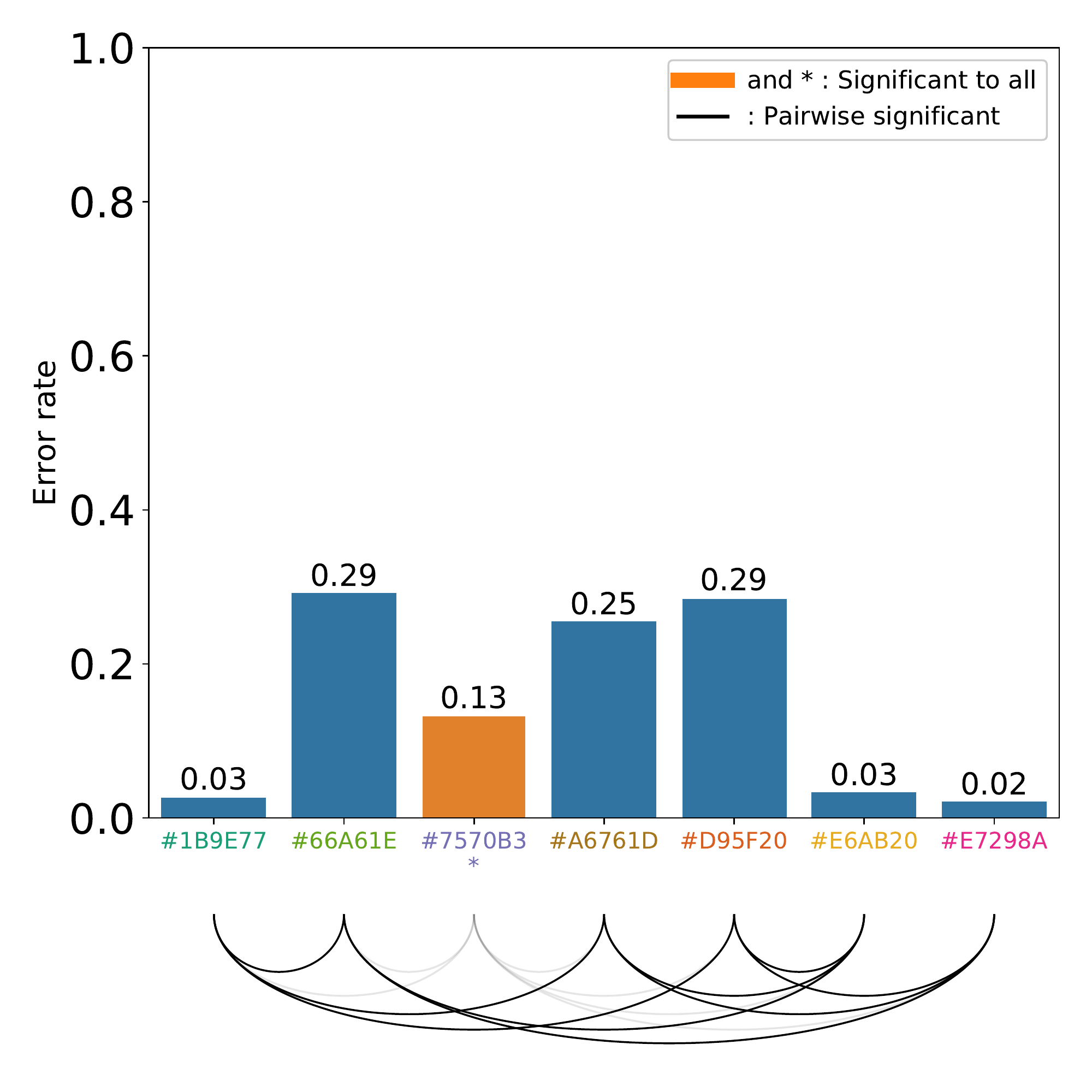}}\\
	\hline
{\shortstack{\begin{sideways}\hspace{-0.5cm} \emph{Shape} type\end{sideways}}}

&
		{\includegraphics[align=c,width=.235\linewidth]{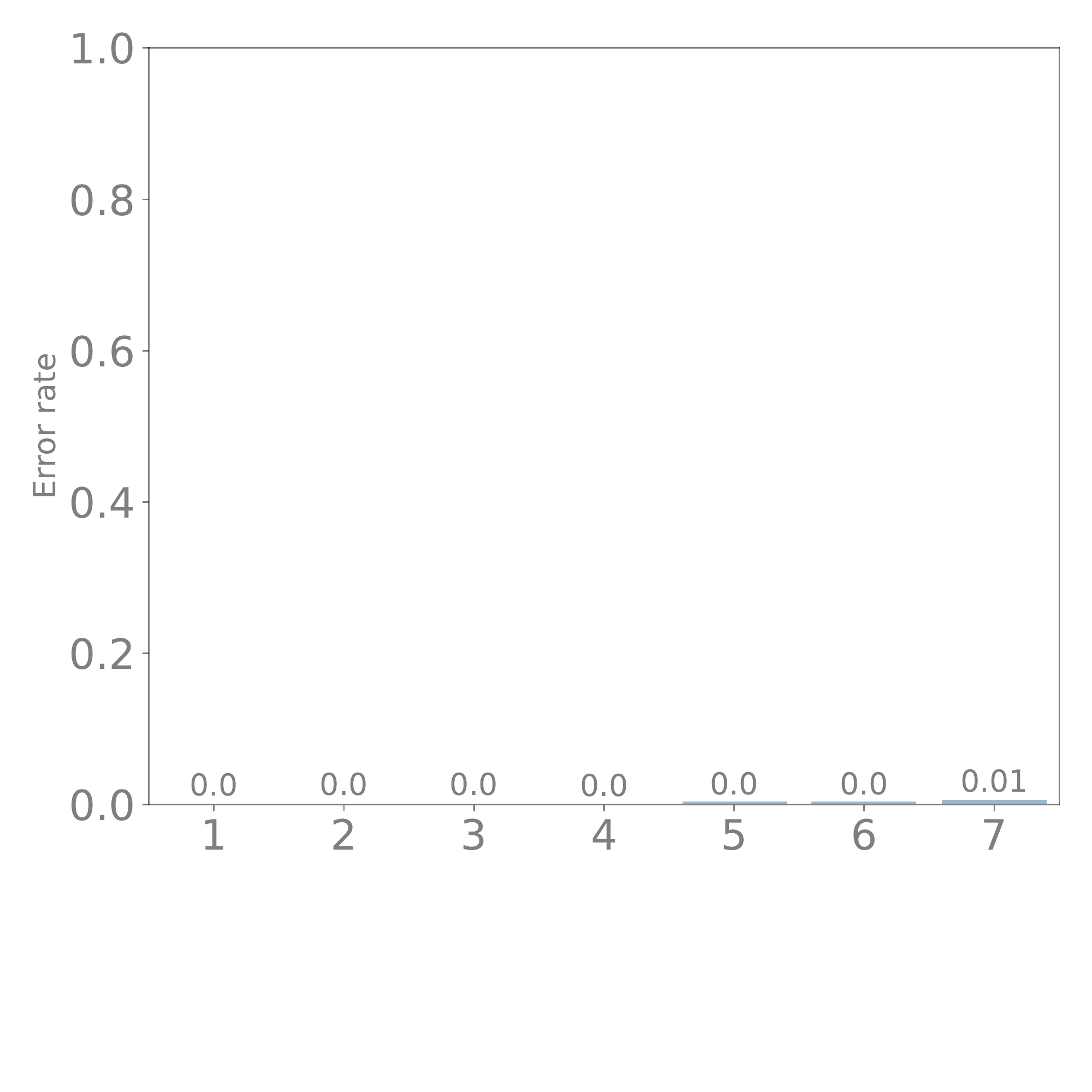}}&
        {\includegraphics[align=c,width=.235\linewidth]{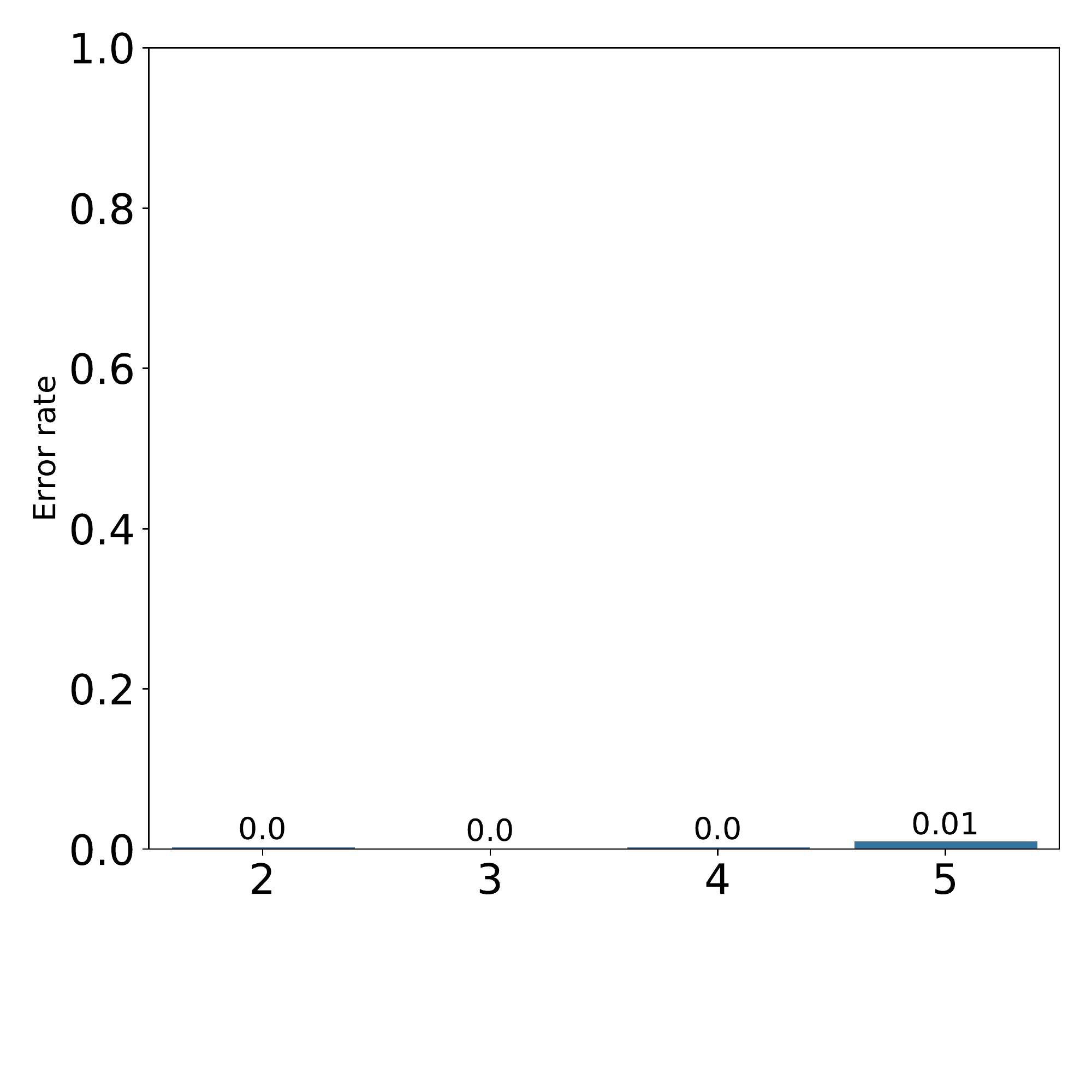}}&
        {\includegraphics[align=c,width=.235\linewidth]{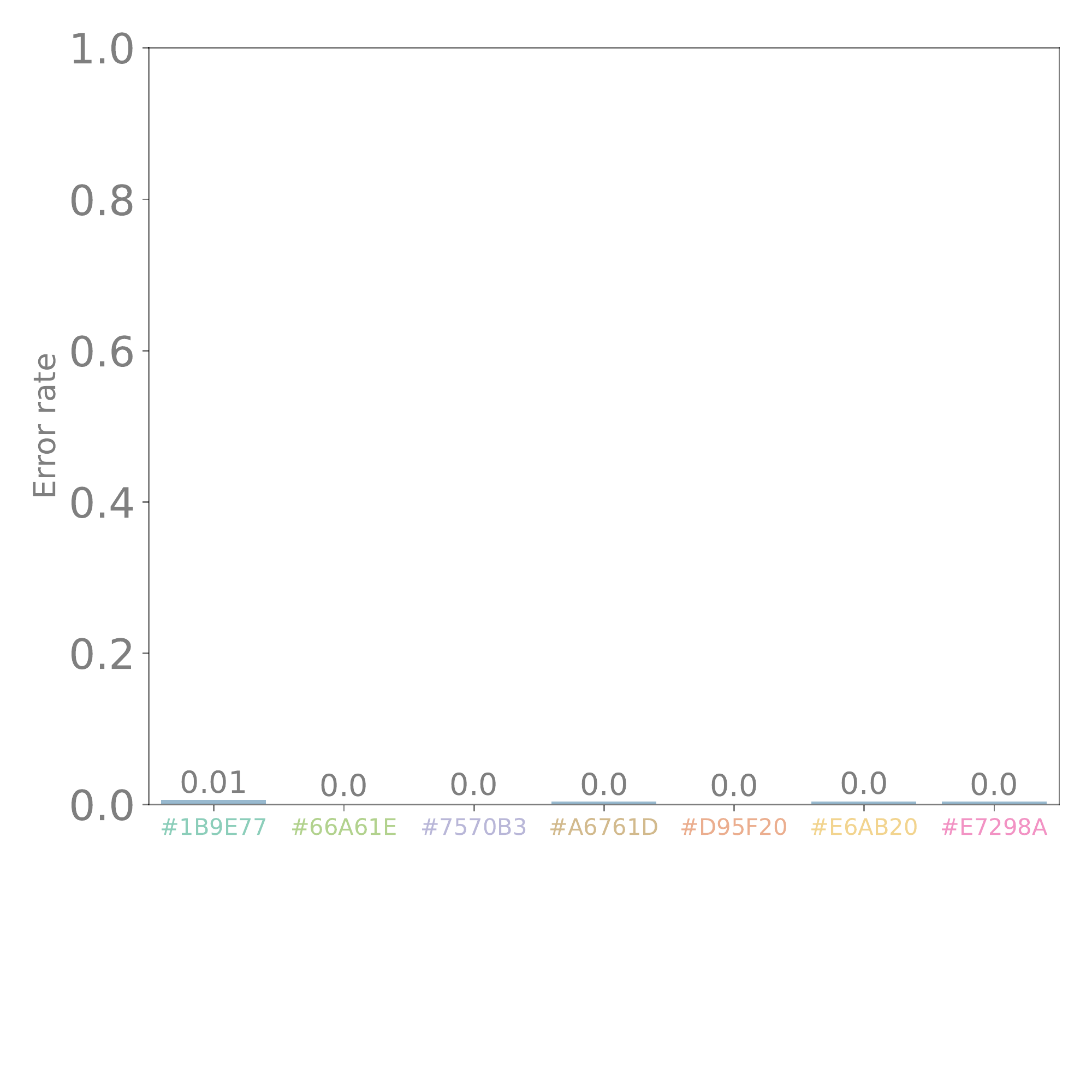}}\\
    \hline
{\shortstack{\begin{sideways}\hspace{-0.5cm} \emph{Red.} type\end{sideways}}}

&
		{\includegraphics[align=c,width=.235\linewidth]{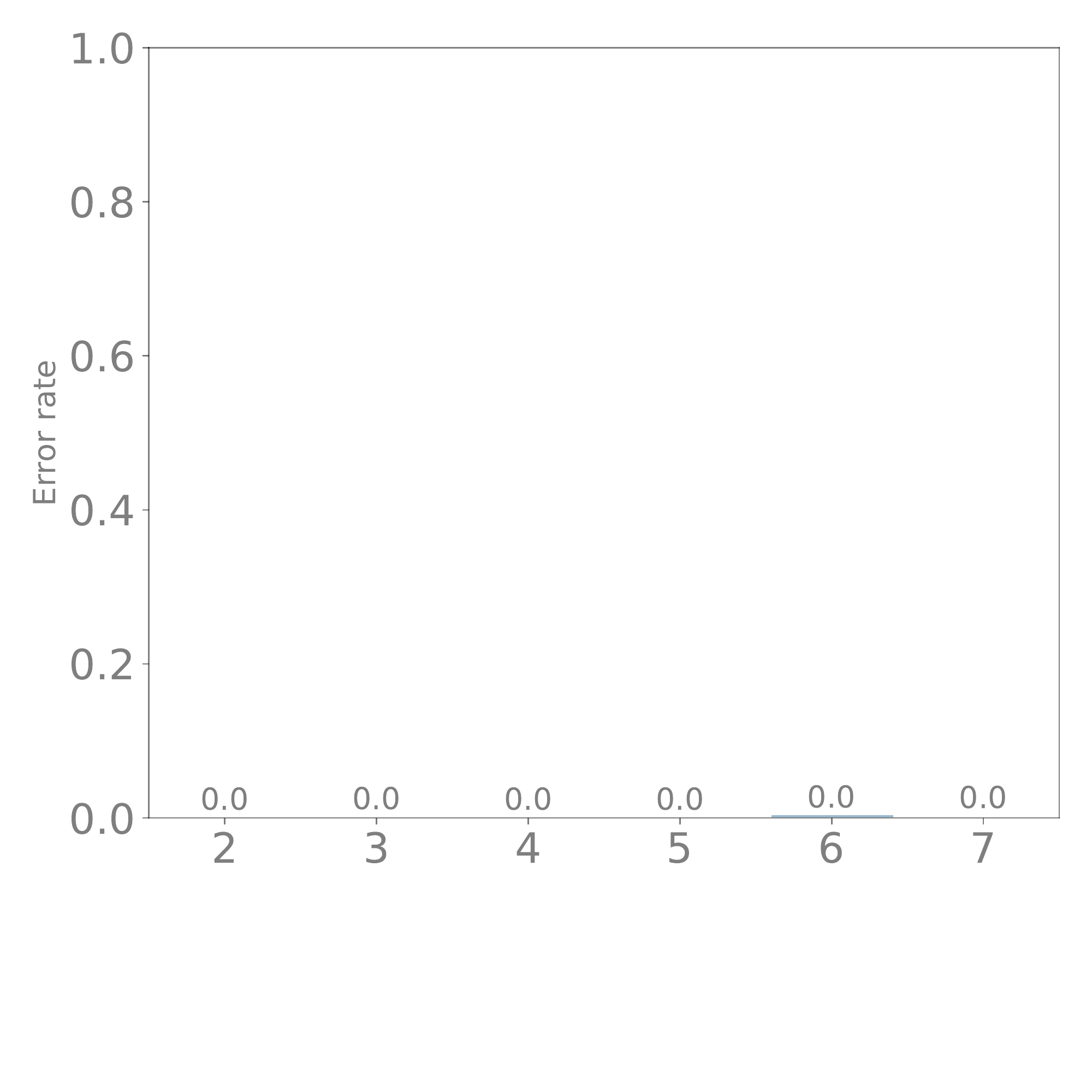}}&	
		{\includegraphics[align=c,width=.235\linewidth]{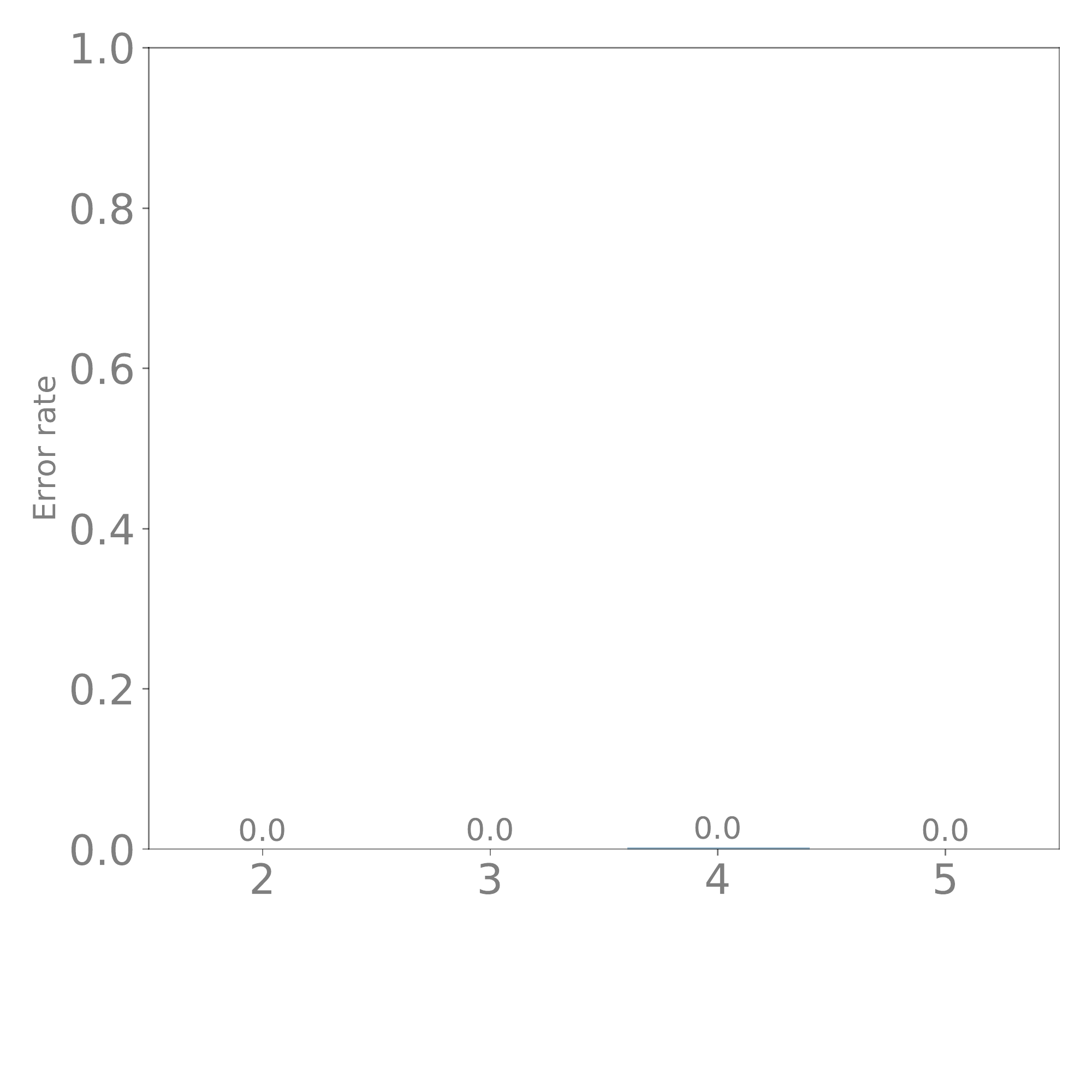}}&	
		{\includegraphics[align=c,width=.235\linewidth]{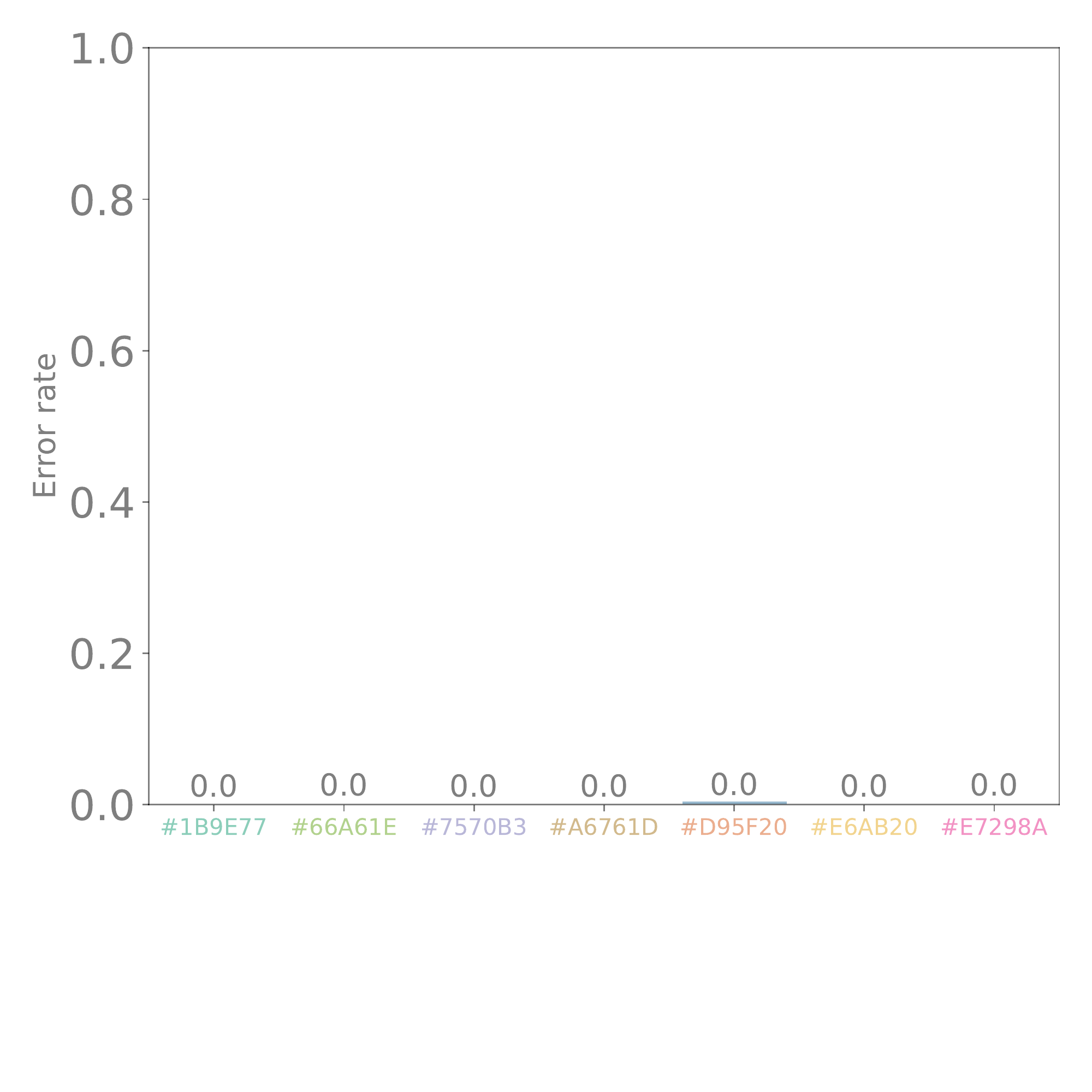}}\\
    \hline
{\shortstack{\begin{sideways}\hspace{-0.5cm} \emph{Conj.} type\end{sideways}}}

&
		{\includegraphics[align=c,width=.235\linewidth]{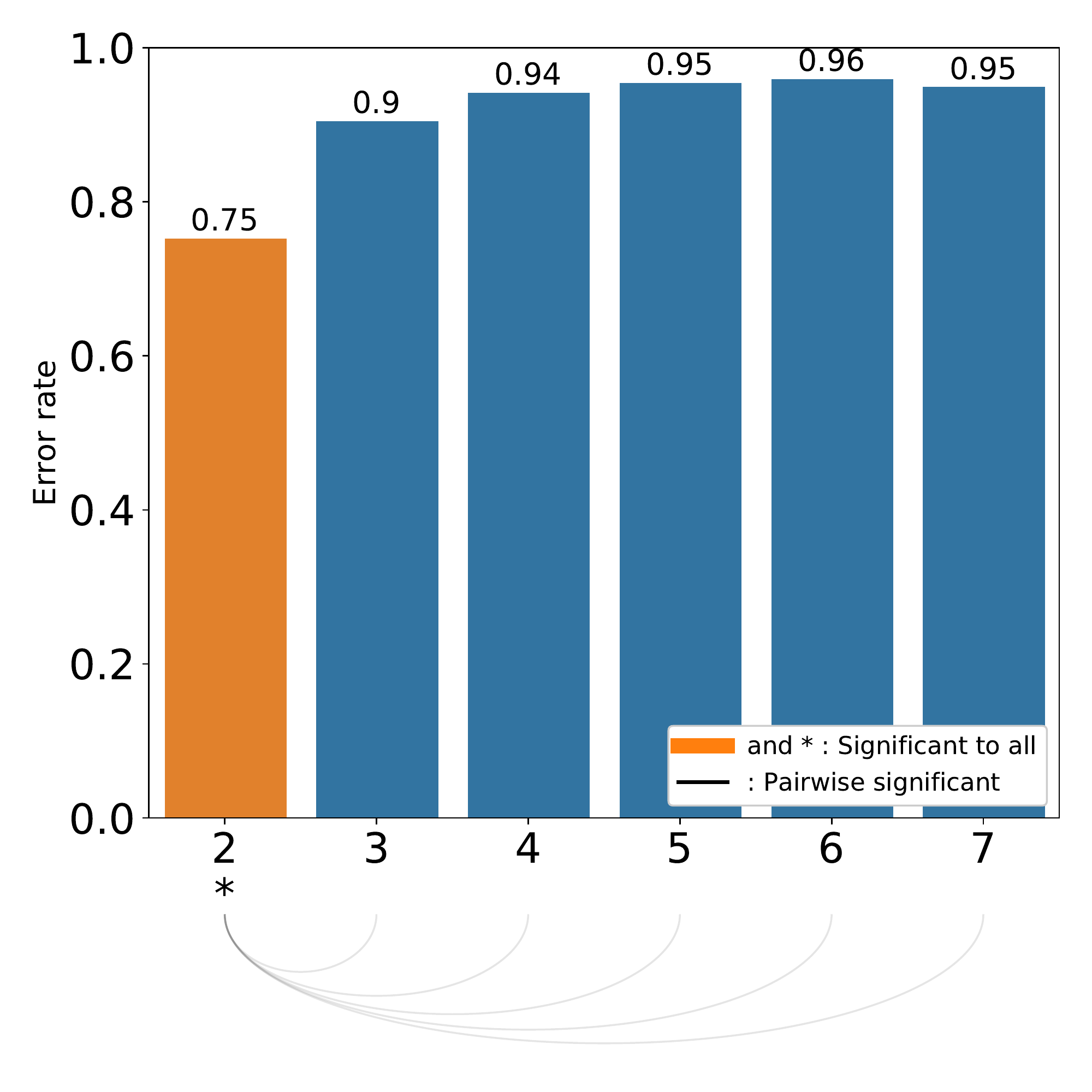}}&	
		{\includegraphics[align=c,width=.235\linewidth]{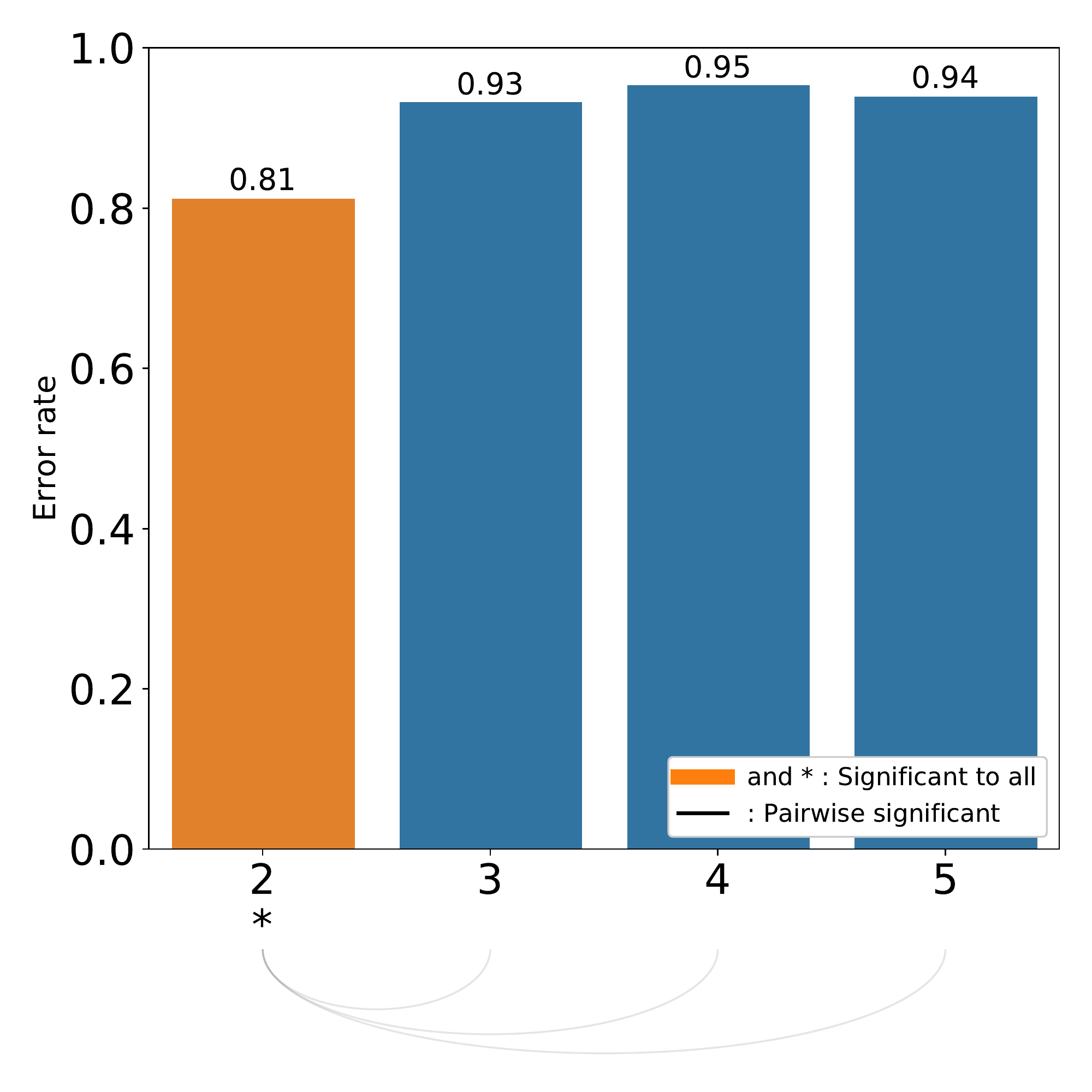}}&
		{\includegraphics[align=c,width=.235\linewidth]{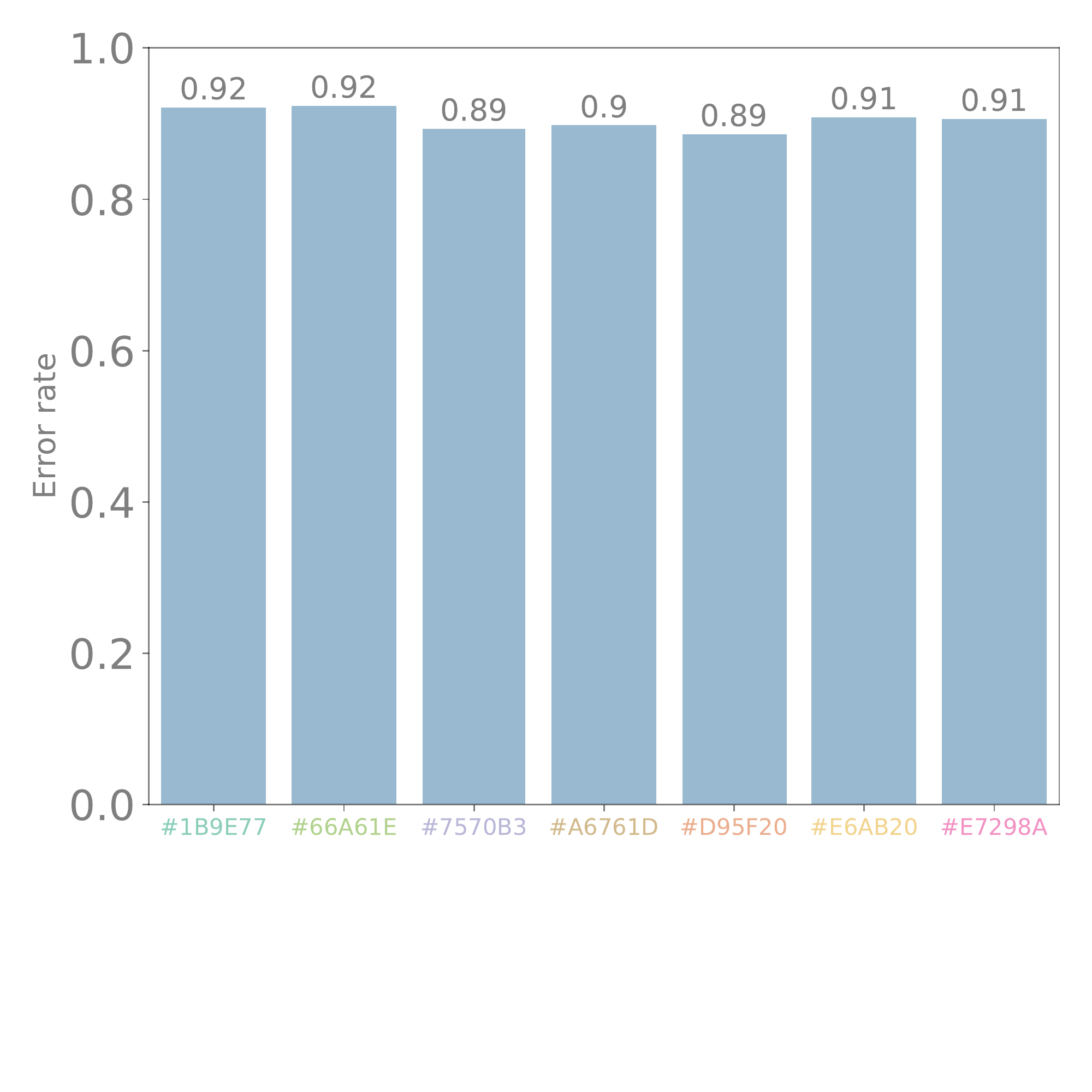}}\\	
    \hline
	\end{tabular}
  \caption{\label{fig:perf_resnet_param_value}Trained \resnet error rates (ER) on the \emph{test} set. The first row shows the overall parameters ERs, while the next rows present the parameters ERs by \textit{type}. A plot is faded if the ANOVA test on its given parameter and type aggregation failed; otherwise, it is opaque. An arc between two labels means that the pairwise comparison between the ERs of the two parameter values is significant according to a Wilcoxon rank-sum test. The significance threshold was $p{\text -}value < 0.05$ for the ANOVA and pairwise tests in the \textit{overall} studies, while it was $p{\text -}value < 0.025$ in the \textit{per type} studies. A reading example is given in the caption of \autoref{fig:perf_resnet_type}.}
\end{figure*}

At the end of the learning phase, the best epoch accuracy rates on the \emph{validation} and \emph{test} sets reached 74\% and 76\%, respectively, showing that the model was not overfitted and was able to generalize. A \emph{Matthews correlation coefficient}~\cite{baldi2000mcc} of $0.754$ on the \emph{test} set confirmed that the model learned to solve the task. Thus, we can expect that incorrect predictions are not due to hazards but rather combinations of parameter values from the data.

A Kruskal-Wallis ANOVA test \cite{kruskal1952wallis} was run on each parameter value prediction sequence to verify whether they had significant effects on the performances (overall or on a specific type value). For the parameters found to have significant effects (opaque plots in \autoref{fig:perf_resnet_param_value}), pairwise Wilcoxon rank-sum tests were run to check if their values led to significantly different performances. The significance level for the overall studies was $\alpha = 0.05$. When splitting the data \textit{by type}, a Bonferroni correction was applied, reducing the significance level to $\alpha = 0.025$. All parameters showed significance under at least one condition, except \emph{outlier shape,} which will not be studied further in this section.

For the remaining parameters, Figures~\ref{fig:perf_resnet_type} and \ref{fig:perf_resnet_param_value} present the trained model error rates (ERs) on the \emph{test} set. Next, we describe the main insights we can learn from these results.

\textbf{Type:} As we could expect, \emph{type} is a key parameter relative to the task difficulty, as shown by the large differences between the ERs in \autoref{fig:perf_resnet_type}. The \typeconj type led to significantly more errors than other values. The \typecol type has been shown to be significantly harder than both \typeshape and \typered; these last two not being significantly different to each other. The significant gaps between the \textit{type} values motivated the study of other parameters for each \emph{type} value separately and extended the statement from the visual search literature: ``the more parameter values the outlier shares with distractors, the more difficult the task is to solve"~\cite{quinlan1987combination}. \autoref{fig:perf_resnet_type} also shows that experimental objects of the \typecol type led to a significantly higher ER than those of the \typeshape type, which is surprising in view of the visual search literature and will be discussed in \autoref{sec:resnet_limitations}.

\textbf{Number of colors:} Overall, the ER almost linearly increases as \nbCols increases, as shown in \autoref{fig:perf_resnet_param_value}. A significant shift in performance between 1 and 2 \nbCols (and basically, between 1 and any other value) can be observed. This shift was probably induced by a bias in our data generation process, which will be discussed in \autoref{sec:resnet_limitations}. The increase in difficulty is even more severe with experimental objects of the \typecol type, where color is the only relevant dimension for identifying the outlier. When color is not a relevant dimension (type \typeshape), \nbCols does not have any significant effect on the task difficulty. Finally, for experimental objects of type \typeconj, \nbCols only has a significant impact on the performances obtained with 2 and other (higher) values. Beyond 2 colors, it seems that the task is already so hard to solve that further increasing the number of colors does not make the task significantly harder.

\textbf{Number of shapes:} \autoref{fig:perf_resnet_param_value} shows that there is a significant ER shift between 1 and other values of \nbShapes, as was observed with \nbCols. Again, this will be discussed in \autoref{sec:resnet_limitations}. This value is set aside, and there remains only one significant difference between the other \nbShapes values. Hence, we can assume that, overall, increasing the number of shapes does not significantly increase the task difficulty.
When shape is an irrelevant dimension for identifying the outlier (type \typecol), the ANOVA test passes, meaning that \nbShapes does have a significant effect. However, the only significant effect among the \nbShapes values is that 5 \nbShapes is easier than 2, 3 and 4 \nbShapes. This counterintuitive result could mean that as shape is an irrelevant dimension in the \typecol type images, its heterogeneity (\textit{i.e.}, \nbShapes) has no direct effect on the task difficulty. When shape is the only relevant dimension (type \typeshape), the ANOVA test passes for the \nbShapes parameter. However, no value is found to be significantly easier than any other, and the error rates remain under 1\%. Finally, the ER of \nbShapes follows the same trend as that of \nbCols for experimental objects of the \typeconj type.

\textbf{Outlier color:} The \emph{outlier color}, unlike \emph{outlier shape}, does have an impact on the task difficulty. Overall, there are three ``harder to find" colors (\green, \brown and \orange) and three ``easier to find" colors (\bluegreen, \yellow and \pink), with \purple in between. However, the ER differences between them remain small.
When color is a relevant dimension (type \typecol), this trend becomes stronger. It is noteworthy that all three ``easier to find" colors are among the 4 most saturated colors, which could be expected since saturated colors tend to capture attention~\cite{camgoz2002saturation,camgoz2004saturation}. When color is not a relevant dimension, the outlier color has no effect on the task difficulty. Finally, the \textit{outlier color} has no significant effect on \typeconj. Since we already know that the \textit{outlier shape} was not significant either, this means that neither the color nor the shape of the outlier matters for this type value. We can assume that \nbCols and \nbShapes are more impactful for the task difficulty than the outlier properties, and this extends the findings of \cite{duncan1989search,treisman1977multidim} about the harmfulness of \emph{heterogeneity} in a representation.\\

In this section, we did not study parameter value effects on the experimental objects of the \typered type. As we can see in \autoref{fig:perf_resnet_param_value}, no parameter had any effect on the experimental objects of this type. The overall ER of the \typered type is 1\%, and all \nbCols, \nbShapes and \textit{outlier color} ER values are under 1\%. We conclude that there is no univariate condition that affects experimental objects of type \typered.

\subsection{Interpretation}

\subsubsection{Limitations}
\label{sec:resnet_limitations}

As mentioned in \autoref{sec:methodology}, there most likely exist parameter configurations for which a metric diverges from what humans would effectively be capable of visualizing. However, since this study is built on DNN model predictions, other limitations must be considered as well.

The first limitation concerns the visual encoding of the experimental objects and the CNN model architecture. The model was able to learn to solve the task, but it is possible that a slight modification of its architecture or a slightly different visual encoding (\textit{e.g.}, different color or shape sets, wider stimuli), could have led to different trends in the predictions. This limitation is not necessarily exclusive to a CNN but rather a common issue in the definition of an experimental design space. Evaluation results are only true within the scope of their experiment, which is designed to be as generalizable as possible to scenarios outside the experiment.

Another limitation comes from the difficulty metric results, which showed that the model predictions were much more capable of identifying shapes than colors. We suspect that the model learned a strategy that would not correlate to human behaviors. It is possible that when the outlier was identifiable by its shape, the model simply counted the number of colored pixels in each cell (regardless of the color itself) and predicted the only colored pixel counts that did not appear at least twice.

Finally, as stated in \autoref{sec:dataset_gen}, some parameter configurations could not be generated. The bars corresponding to the error rates for both 1 \nbCols and 1 \nbShapes were only computed from a specific type value. In the overall-\nbCols-ER plot in \autoref{fig:perf_resnet_param_value}, it is a biased view to interpret that ``1" has a lower ER than other values since it is only composed of experimental objects of the \typeshape type, whereas other values are computed from experimental objects of all type values.

\subsubsection{Hypotheses}
\label{sec:autom_hypothesis}

These results allow us to propose several hypotheses:

\Htype: \textbf{The type difficulty should follow the order (easiest to hardest): \typered, \typeshape and \typecol, \typeconj.} This refers to the contribution of Quinlan and Humphreys~\cite{quinlan1987combination}, who showed that the more features the target shares with the nontargets, the more difficult the task is.

\Hconj: \textbf{The search task on experimental objects of type \typeconj is the hardest among all types. The task difficulty increases for both \nbCols and \nbShapes and quickly caps} (\textit{i.e.}, the difficulty no longer increases when there are more than 2 shapes and colors). The fact that conjunction search is harder than feature search was confirmed by prior work~\cite{duncan1989search,treisman1980theory} in a perception context.

\Hred: \textbf{The search task on experimental objects of type \typered is the easiest among all types. The task difficulty is not affected by either \nbCols or \nbShapes}, as shown by Nothelfer \textit{et al.}~\cite{nothelfer2017redundant}.

\Hcol: \textbf{When color is the only relevant dimension, the task difficulty increases with \nbCols, whereas \nbCols has no effect when color is not a relevant dimension.}

\Hshape: \textbf{When shape is the only relevant dimension, the task is easy and is not affected by \nbShapes. When it is not a relevant dimension, the task becomes significantly easier when there are 5 \nbShapes}, although this seems counterintuitive and might be induced by the non-significance of heterogeneity in irrelevant dimensions. 

The first three hypotheses extend the findings of the visual search community that we expect to observe in the context of this experiment. It is noteworthy that the \resnet predictions validate \Hconj and \Hred and that \Htype would be validated if the difference between the \typeshape and \typered types was found to be significant. With the last two hypotheses, we will study the different impacts of the shape and color dimensions on the resulting performances when related to outlier detection tasks with respect to our representations. More precisely, we will focus on the capacity limits of these dimensions (\emph{i.e.}, at which threshold value of \nbShapes or \nbCols does the task difficulty increase or become capped?) in an information visualization-like context. Whereas the visual search literature studies how the human brain processes stimuli, we expect to observe to what extent these theories stand.

\subsubsection{Parameter Space Reduction}
\label{sec:autom_parameters_space}

As we consider several parameters that have relatively high numbers of different values, testing all of their combinations would not be possible during a standard user evaluation. Moreover, as a large number of \textit{experimental trials} may bias the results due to loss of attention or tiredness, it is necessary to keep the completion time of the entire evaluation reasonable~\cite{purchase2012hci}. We call a \textit{trial} an experimental object on which a subject resolves the task.

We use the model results as a difficulty metric to guide the user evaluation parameter space reduction process.
The \textit{type}, \nbShapes and \nbCols distributions should remain uniform since they are the main conditions upon which the hypotheses are built. Each value of each parameter should also occur more than once. The values of other parameters are distributed as uniformly as possible within the selected combinations of \emph{types}, \nbShapes, and \nbCols. Following this condition, we can reduce the number of trials to $124$. That is still too many trials, considering the task is not ``easy" and its completion time will probably be measured in tens of seconds. The parameter space needs to be reduced even more so that the experiment is short enough to prevent subject tiredness.

To further reduce the parameter space, some \nbShapes and \nbCols values are removed from the study. First, the value ``1" is removed from both the \nbShapes and \nbCols parameters. As previously seen (\autoref{sec:resnet_limitations}), the value ``1" leads to \textit{type} distribution imbalances that may bias the results. For \nbCols, the values of 3 and 6 are removed. These are middle-end values in line with the observed ERs and significance of their contiguous values. For example, no significant difference can be observed between the error rates obtained with 3 and 4 \nbCols, except with the \typecol type where there is no significant difference between 2 and 3 \nbCols. With the same reasoning, the \nbShapes value of 4 is removed as well.

\begin{figure}[!bt]
	\centering
	\begin{subfigure}[b]{.45\linewidth}
		\includegraphics[align=c,width=\linewidth]{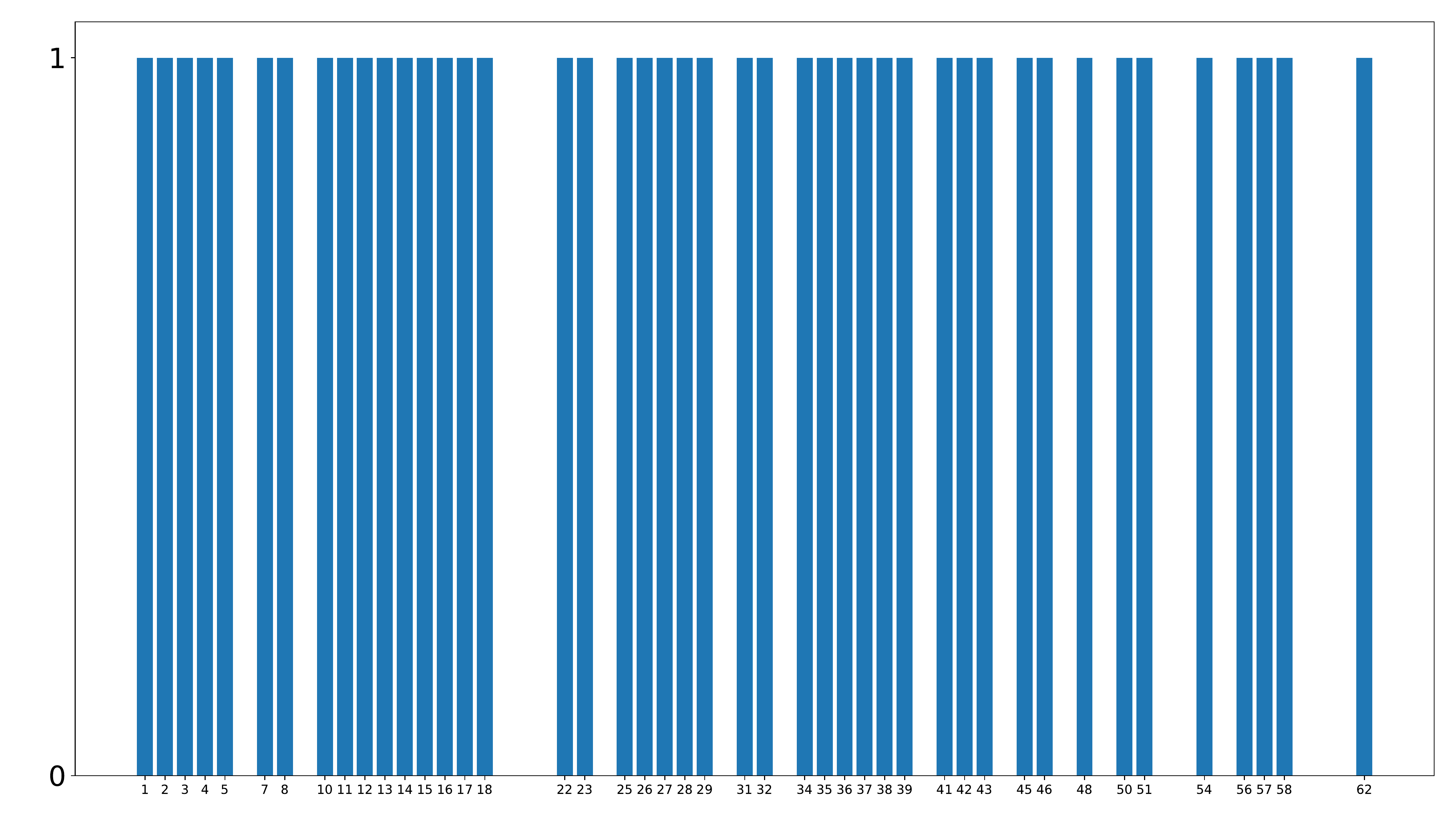}
		\caption{\label{fig:eval_outpos_distrib}\textit{Outlier position} distribution}
	\end{subfigure}
	\begin{subfigure}[b]{.45\linewidth}
		\includegraphics[align=c,width=\linewidth]{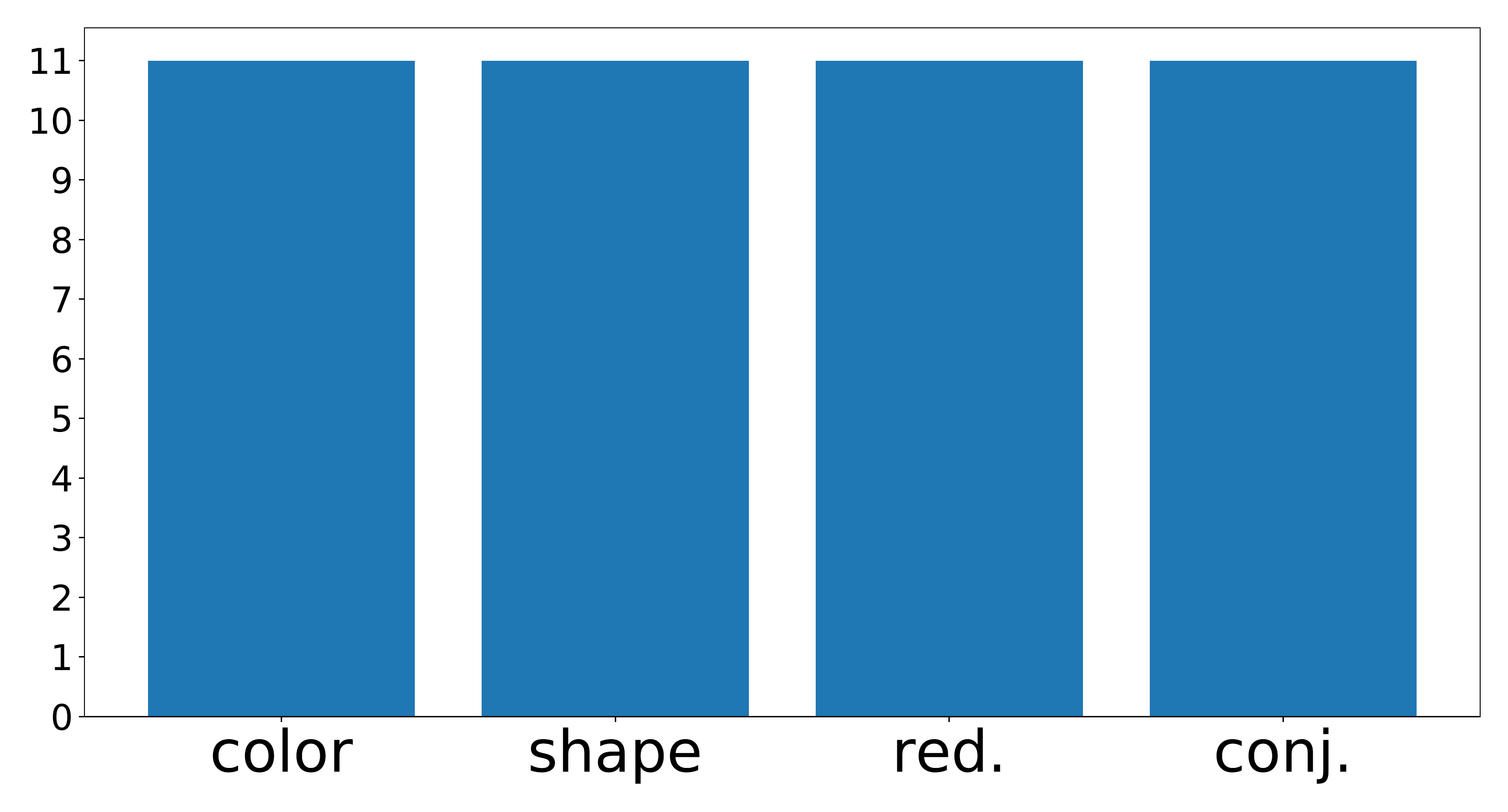}
		\caption{\textit{Type} distribution}
	\end{subfigure}
	\begin{subfigure}[b]{.45\linewidth}
		\includegraphics[align=c,width=\linewidth]{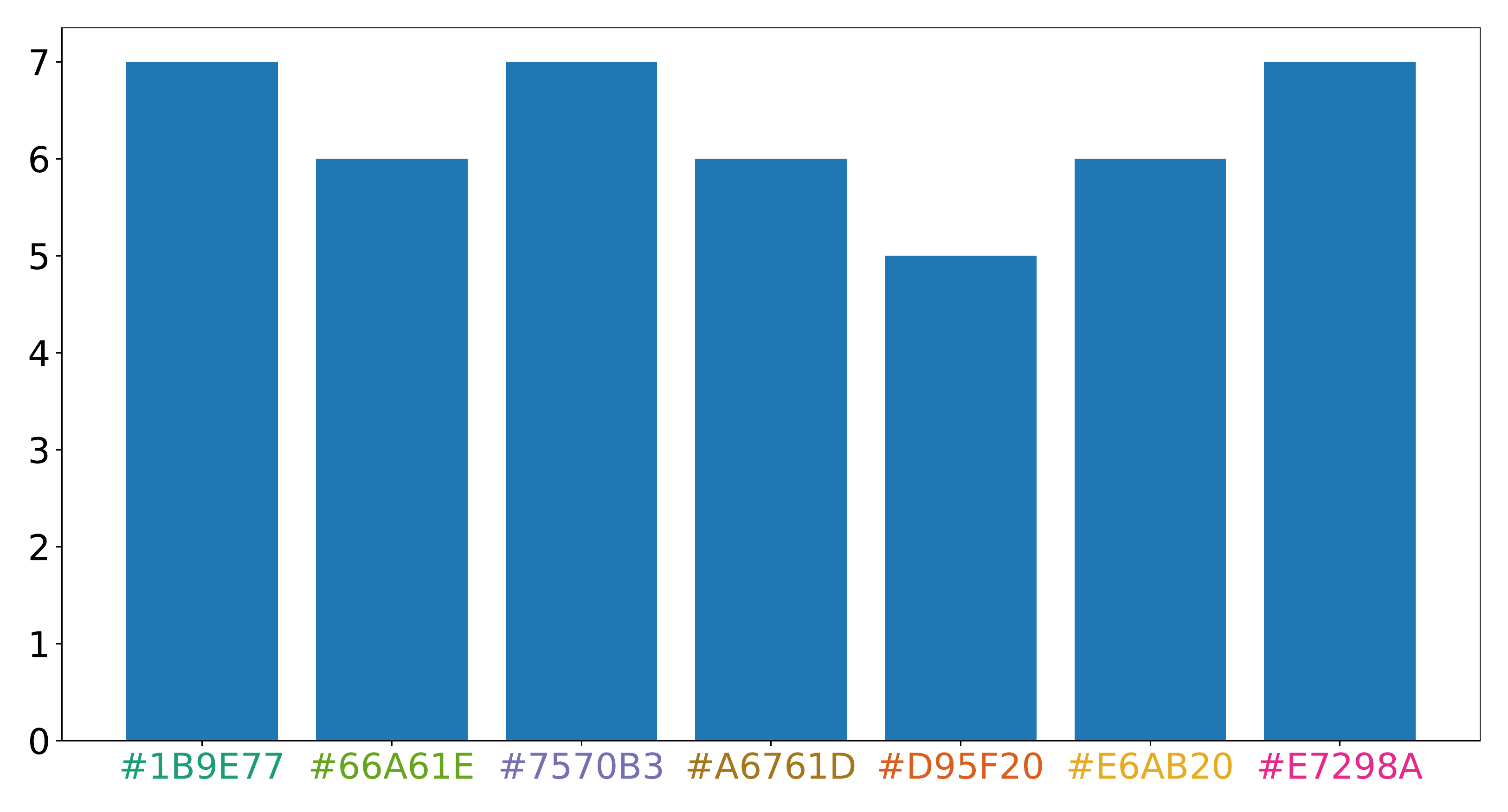}
		\caption{\textit{Outlier color} distribution}
	\end{subfigure}
	\begin{subfigure}[b]{.45\linewidth}
		\includegraphics[align=c,width=\linewidth]{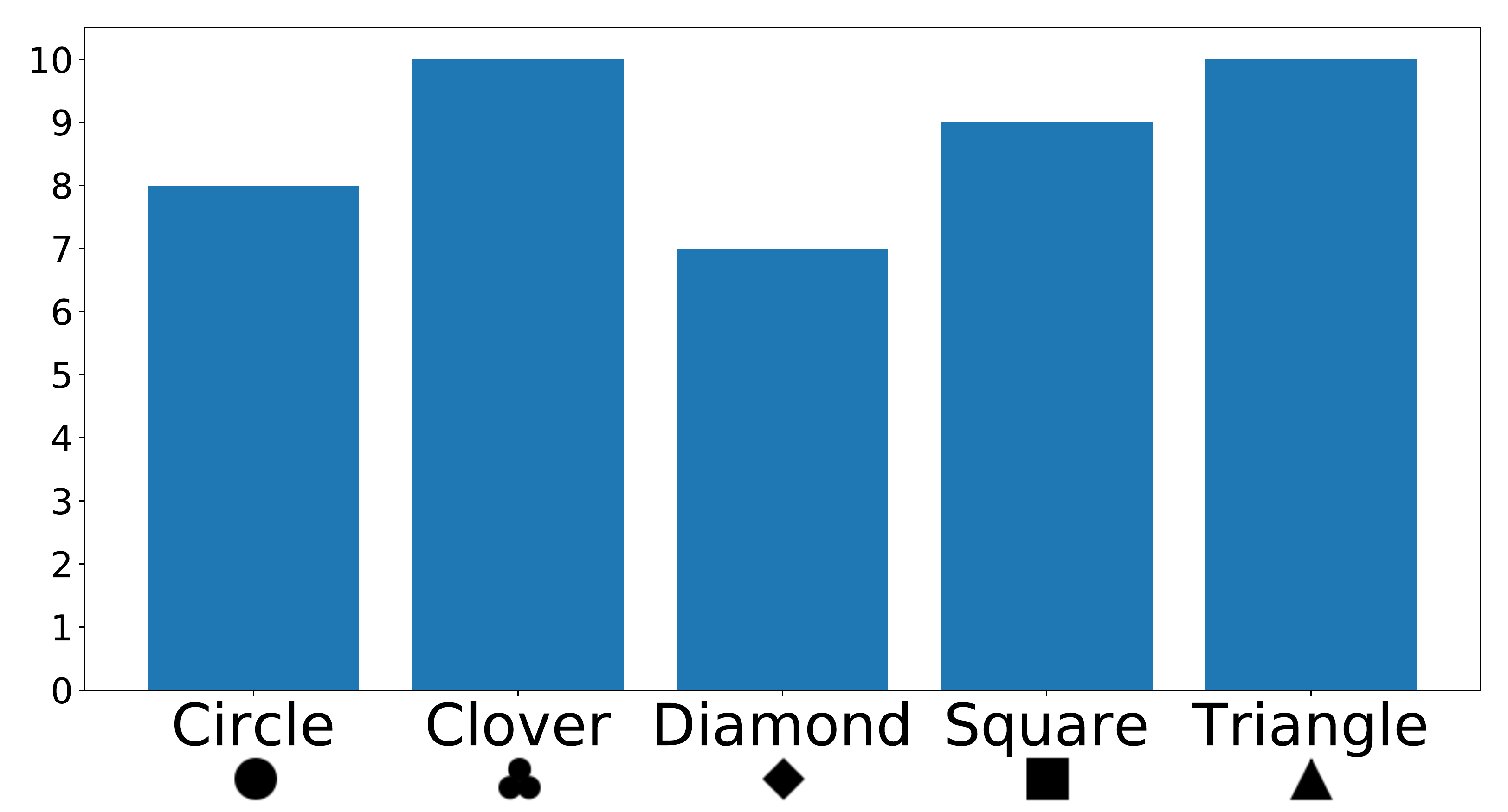}
		\caption{\textit{Outlier shape} distribution}
	\end{subfigure}
	\begin{subfigure}[b]{.45\linewidth}
		\includegraphics[align=c,width=\linewidth]{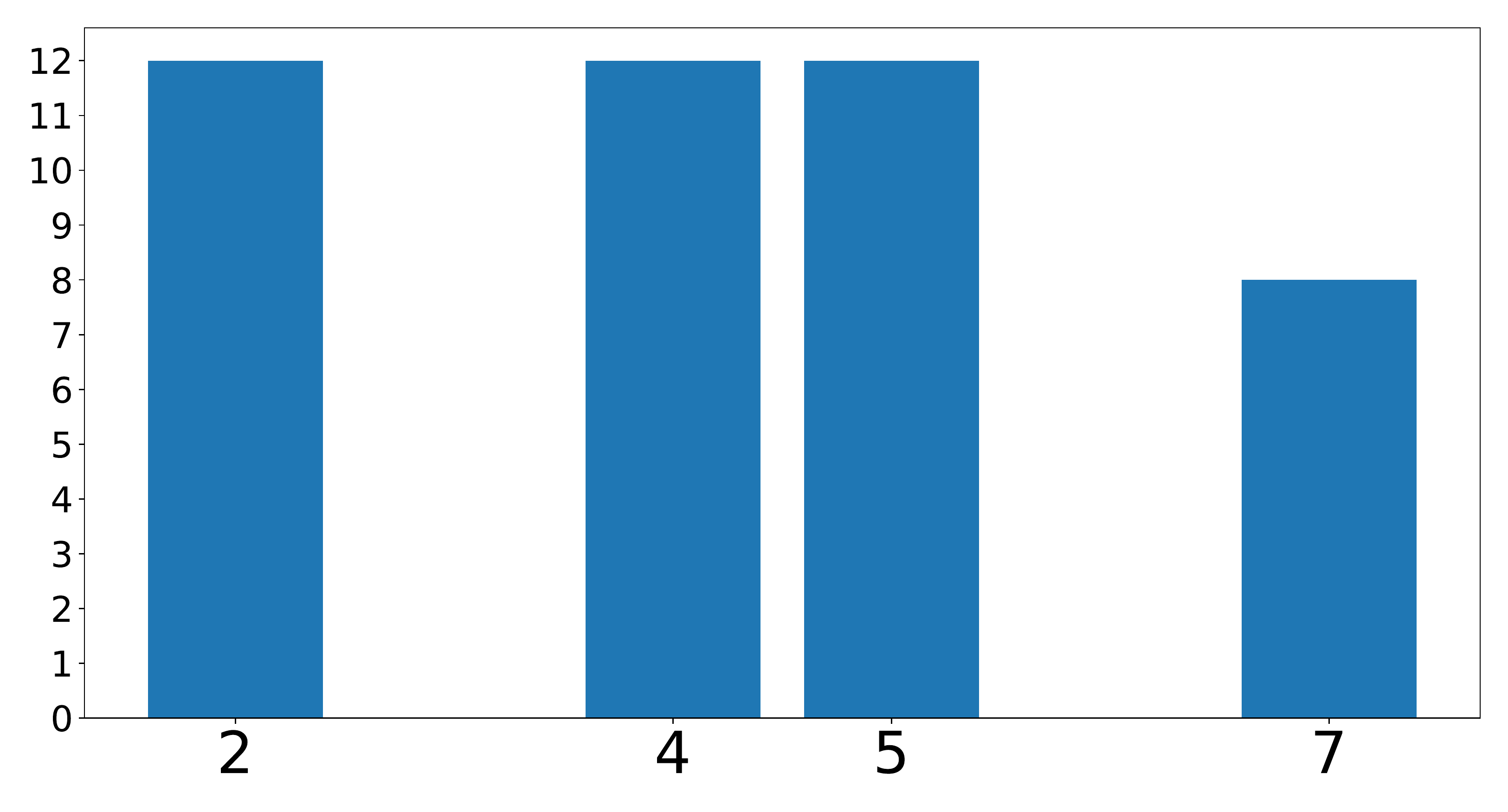}
		\caption{\nbCols distribution}
	\end{subfigure}
	\begin{subfigure}[b]{.45\linewidth}
		\includegraphics[align=c,width=\linewidth]{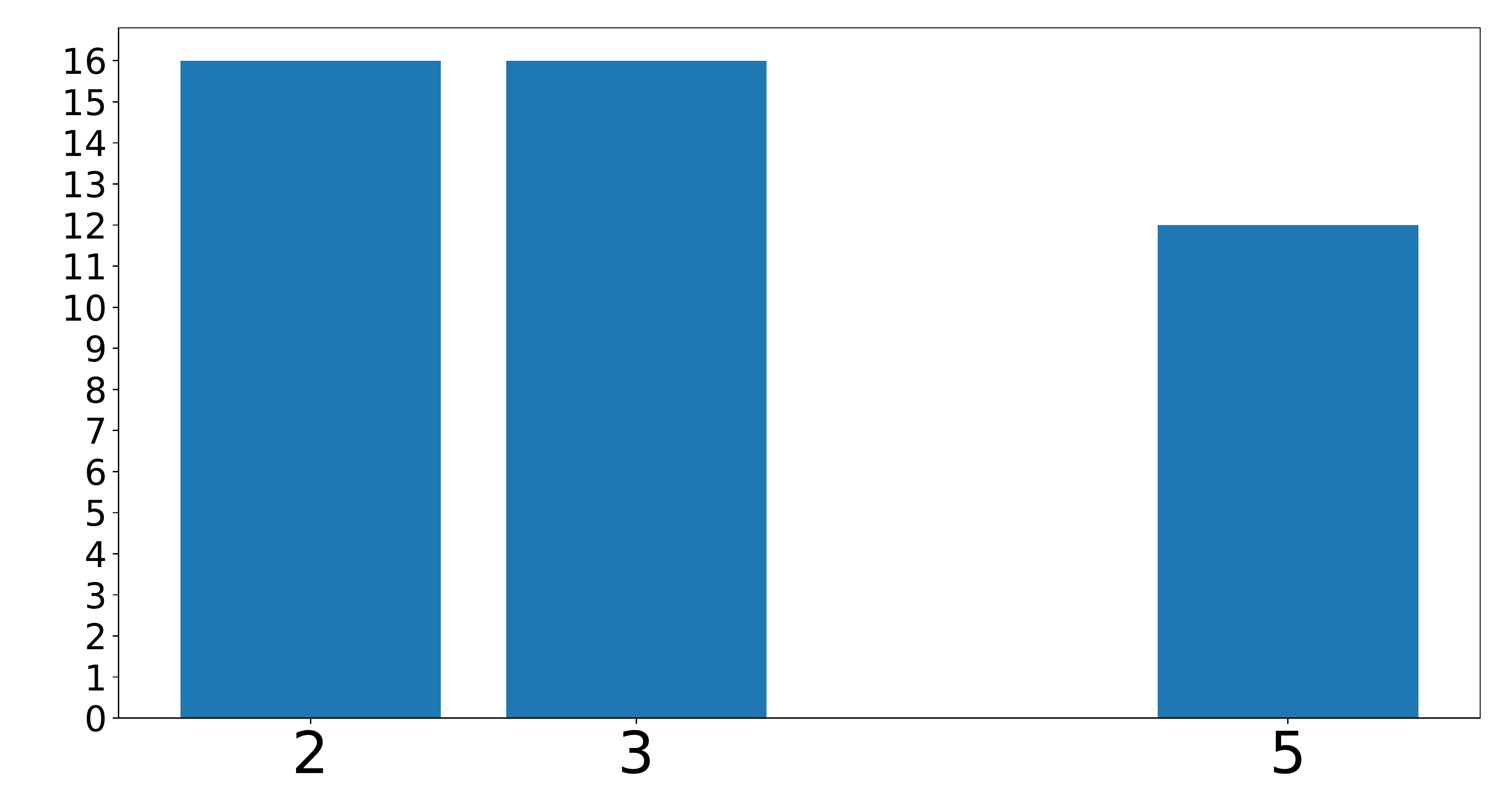}
		\caption{\nbShapes distribution}
	\end{subfigure}
	\caption{\label{fig:endEvalDataDistrib}Parameter value distributions in the \num{44} selected trials of the end-user evaluation.}
\end{figure}

The \nbCols values are then reduced from \{1, 2, 3, 4, 5, 6, 7\} to \{2, 4, 5, 7\}, and the \nbShapes values are reduced from \{1, 2, 3, 4, 5\} to \{2, 3, 5\}. Still excluding the combination (7; 5) for (\nbCols; \nbShapes), we end up with $(|\nbCols| * |\nbShapes| -1) * |\emph{type}| = 11 * 4 = 44$ trials for the user evaluation. The parameter value distributions within these 44 trials are shown in \autoref{fig:endEvalDataDistrib}.

\section{User Evaluation}
\label{sec:user_eval}
This section presents the setup, choices, constraints and results of the user evaluation. The preliminary hypotheses for the evaluation were defined in \autoref{sec:autom_hypothesis} to study the effects of the variations in the \nbCols, \nbShapes and \textit{types} on the task difficulty.

\subsection{Experimental Setup}

\subsubsection{The Task}

As mentioned in \autoref{sec:task}, the task consists of identifying an outlier stimulus in an $8 \times 8$ grid of distractor stimuli. A time limit for each trial is included to encourage subjects to solve the task as quickly as possible. Based on the pilot experiments, this time limit is set to 30 seconds and leads to a good compromise between the evaluation completion time and the error rates.

\subsubsection{Dataset and Order}

The data used in this experiment are randomly extracted from the \textit{test} set defined in \autoref{sec:dataset_gen} to fit the reduced parameter space defined in \autoref{sec:autom_parameters_space}.

The trials order is randomly set, and every subject runs the trials following this order but with a random shift so that they do not all start with the same trial.

\subsubsection{Evaluation Protocol}

The subjects are first asked to read and understand the task statements. These statements present all the colors, shapes and types that can occur during the experiment and provide a grid example. Subjects are free to ask any question, and we, in return, make sure they understand the task. Then, the subjects have to follow an 8-trials tutorial. The first 4 trials are shown already solved, along with information about their parameters. Each of them represents a different \textit{type} value. In the next 4 tutorial trials, subjects are asked to solve the task without a time limit and are given feedback about the correctness of their answers. Again, the 4 trials each represent a different \textit{type} value. Once a subject has completed the tutorial, he/she can replay it or start the evaluation.

Following the recommendations of Purchase~\cite{purchase2012hci}, we designed an additional 8 trials for practice before starting the 44 evaluated trials. The subjects are not aware that there are practice trials, and we do not consider them during the results study. This ensures that all subjects are at peak performance when the real evaluation (with monitored trials) starts. The 44 experimental trials are then displayed one after another with a three seconds break between each trial (either validated or skipped due to the time limit). During the three seconds break, the space reserved for the trial images is filled with white (the background color). For each trial, the subject response times and answers are recorded. After the $26^{th}$ trial, a one minute long pause is given to the subjects, with the possibility of resuming the evaluation before the pause period expires. When the subjects complete all the trials, they are asked to fill out a questionnaire about what, according to them, made the task easier (or harder) to solve. The whole protocol lasts approximately 20 to 30 minutes for each subject.

\subsubsection{Evaluation User Interface}

The evaluation tool consists of a website specifically implemented for this study. The website is displayed in a full-screen browser on a $1920 \times 1080$ resolution monitor. Every trial image is displayed with a 1:1 ratio ($256 \times 256$ pixels) in the middle of the screen, with a black border to bring it out of the white background. The task statements and the advancement of the evaluation are succinctly written above the trial image. Below the image, the remaining time for the current trial and a validation button are displayed. To solve the task, subjects have to select their answer by clicking directly on the corresponding stimulus on the image, which surrounds it with a black border. An answer can then be validated by clicking the validation button. The validation button is set wide enough so that it does not require any specific focus to be clicked on.

\subsubsection{Involved Subjects}

The subjects of this experiment are 18 men and 6 women, all of whom are undergraduate students, research staff or engineers in computer science. All subjects are between 21 and 50 years old with an average age of 24.8. They all reported having a perfect or corrected-to-perfect visual acuity, and none reported suffering from colorblindness.

\subsection{Results}
\label{sec:human_res}

During the evaluation, subject response times (RTs) and answers are recorded. Next, subjects performances are studied with regard to their RTs and error rates (ERs). The results are computed for 21 out of the 24 subjects; after looking at the subjects performances and answers to the questionnaire, we removed 2 subjects for whom their RTs were lower than average by more than $1.5$ times the standard deviation. In addition, these two subjects had ERs lower than average and were therefore considered outliers. We also removed 1 subject for which both their ER and RT were higher than the average plus $1.5$ times the standard deviation. We also found evidence in his/her questionnaire answers that the task was either misunderstood or not seriously solved.

\begin{figure}[!bt]
\centering
\includegraphics[width=.95\linewidth]{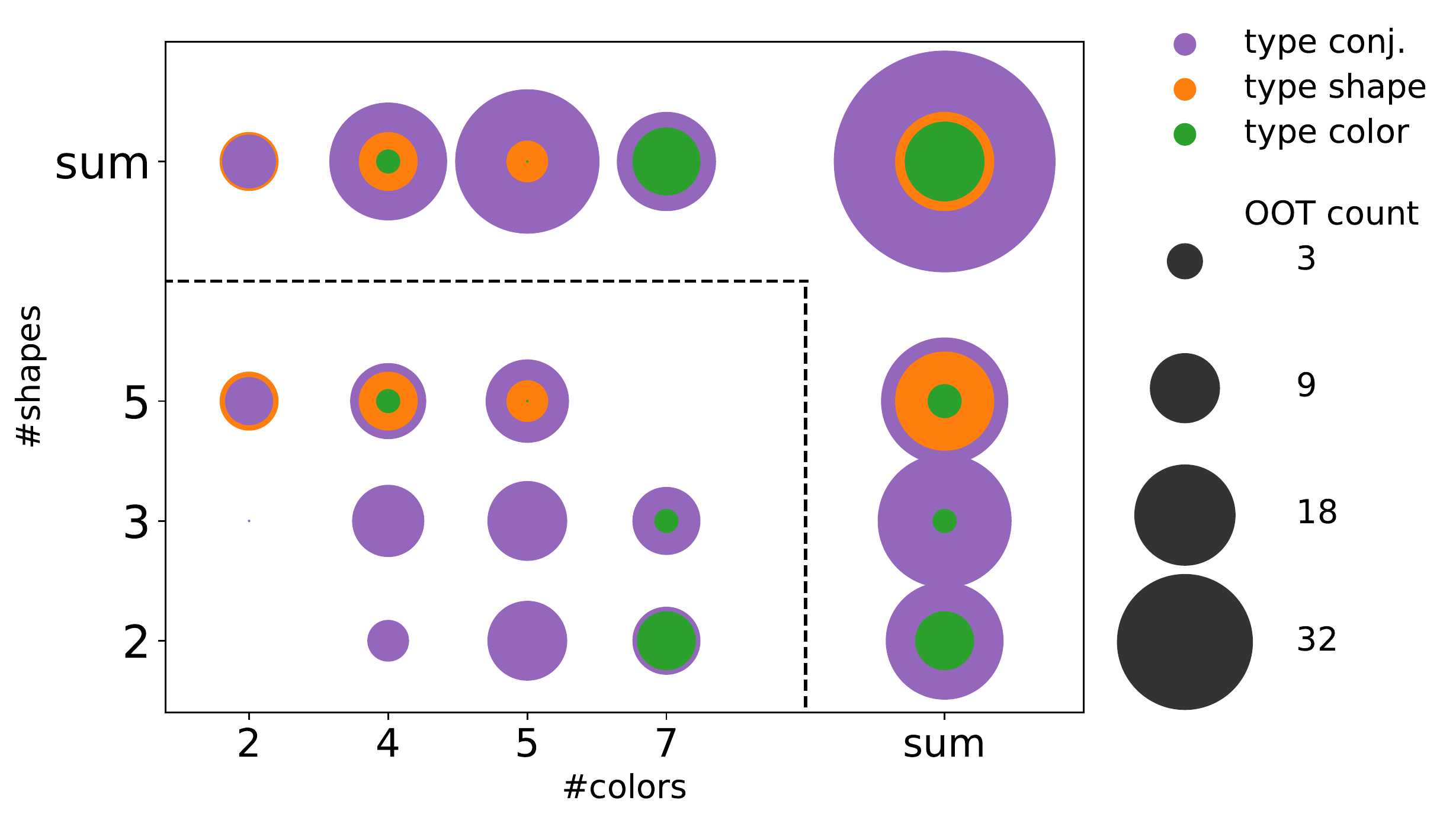}
\caption{Number of trials for which the subjects ran \textit{out of time} (OOT) per \textit{type}, \nbCols and \nbShapes. There are 116 OOT trials in total (5.5 per subject on average), 74\% of which are of type \typeconj, 16\% are of type \typeshape and 10\% are of type \typecol.}
\label{fig:OOT_distrib}
\end{figure}

\begin{figure}[!bt]
	\centering
	\begin{subfigure}[b]{.49\linewidth}
		\includegraphics[align=c,width=\linewidth]{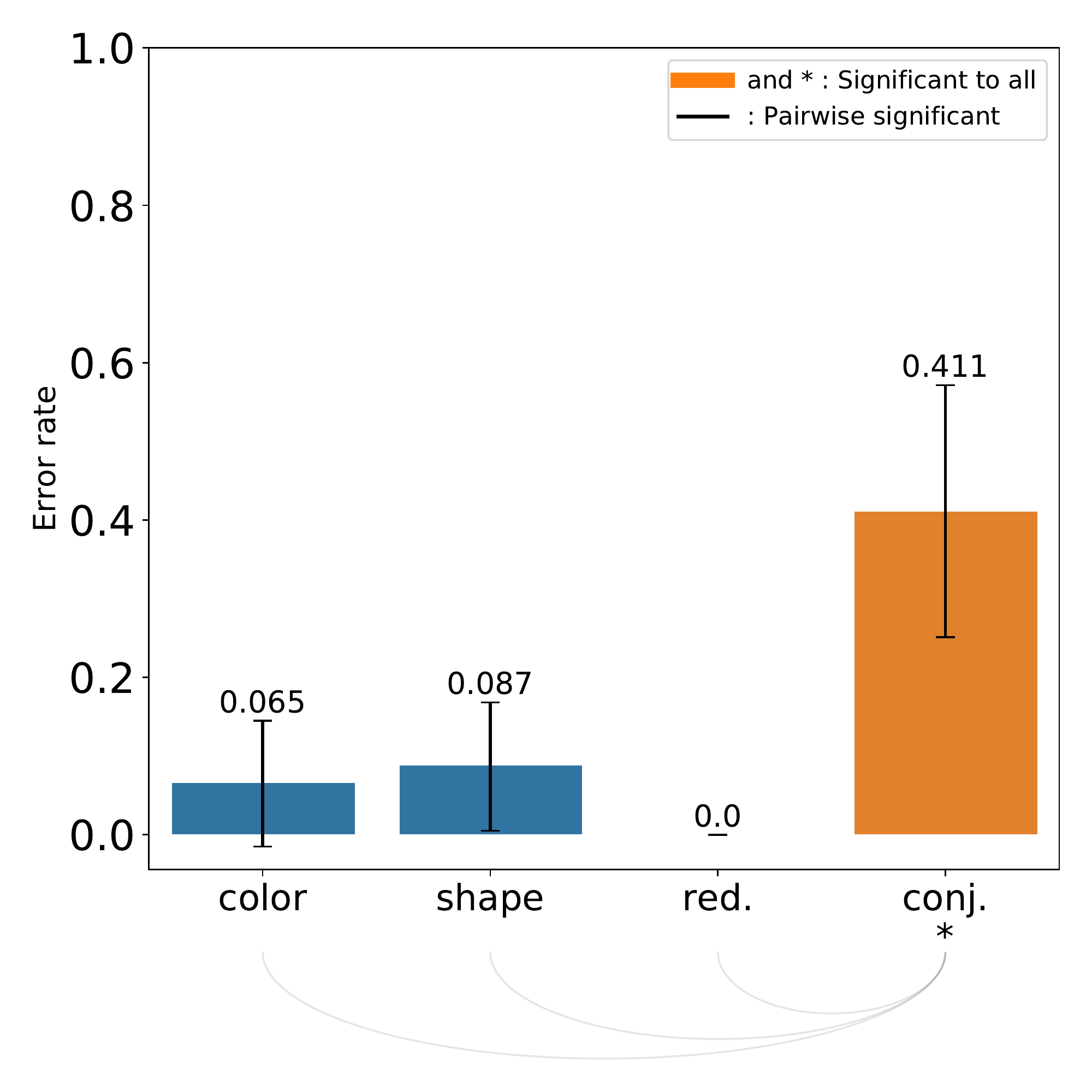}
		\caption{\label{fig:human_type_err}ER Type}
	\end{subfigure}
	\begin{subfigure}[b]{.49\linewidth}
		\includegraphics[align=c,width=\linewidth]{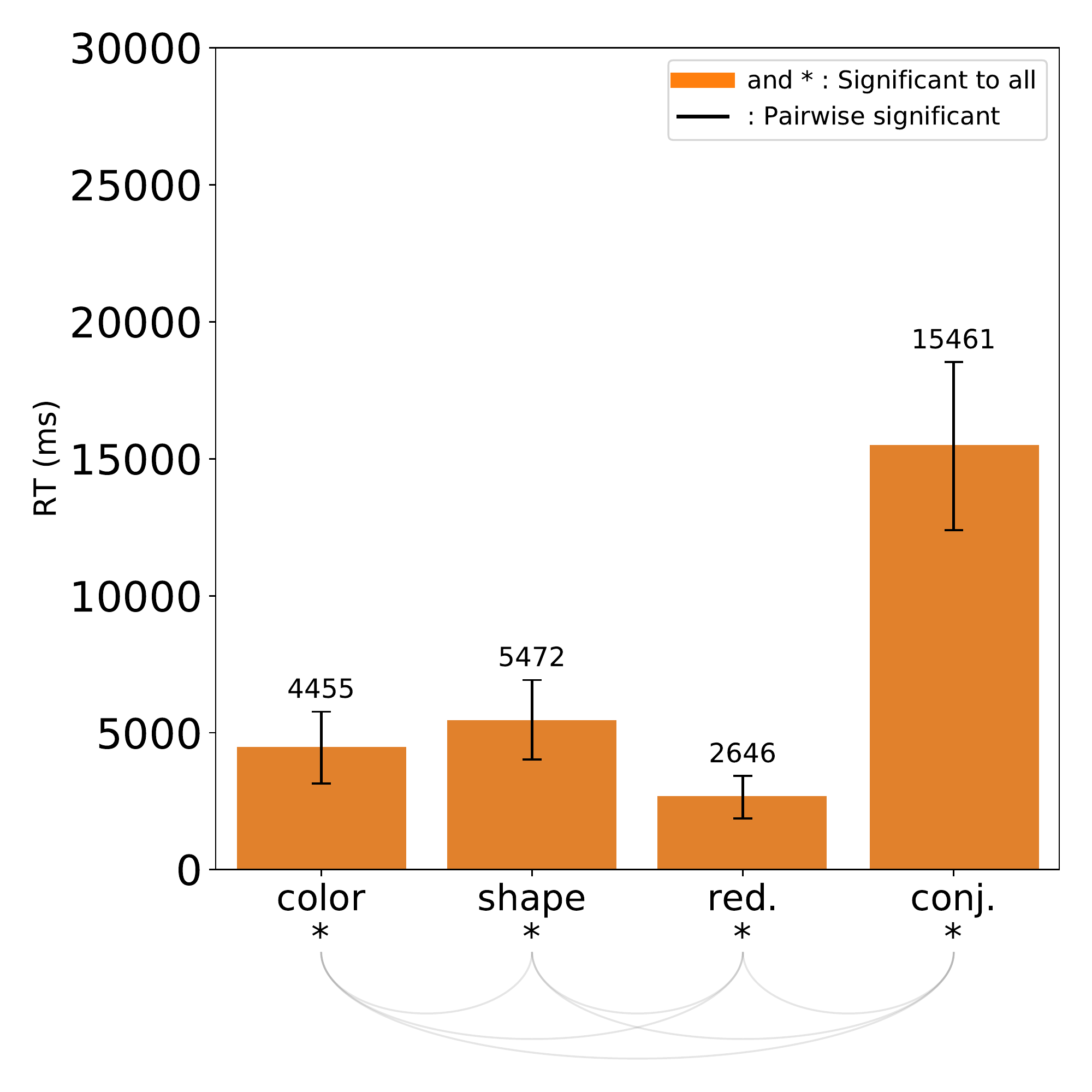}
		\caption{\label{fig:human_type_rt}RT Type }
	\end{subfigure}
	\caption{\label{fig:human_type_perf}Subjects ERs and mean RTs with standard deviation bars for \emph{type} values. ANOVA tests showed that \emph{type} had a significant effect on the ERs and RTs. An arc between two labels means that the pairwise comparison between the corresponding performance values is significant ($p{\text -}value < 0.05$) according to a Wilcoxon rank-sum test. \textit{Reading example}: the \typered type is the condition that is fastest to solve as its RT is significantly lower than other conditions. No pairwise significance test could be run for the \typered ER type since no errors were ever made on these trials.}
\end{figure}

\begin{figure*}
\centering
\begin{tabular}{|c|c|c||c|c|}
    \hline 
    & \multicolumn{2}{c||}{\nbCols} & \multicolumn{2}{c|}{\nbShapes} \\
    \cline{2-5}
    & \textbf{A} - Error rate & \textbf{B} - Response time & \textbf{C} - Error rate & \textbf{D} - Response time \\
    \hline
    {\shortstack{\begin{sideways}\hspace{-0.4cm}- Overall\end{sideways}}} \shortstack{\hspace{-0.26cm}\vspace{-0.8cm}\textbf{1}} &
    {\includegraphics[align=c,width=.215\linewidth]{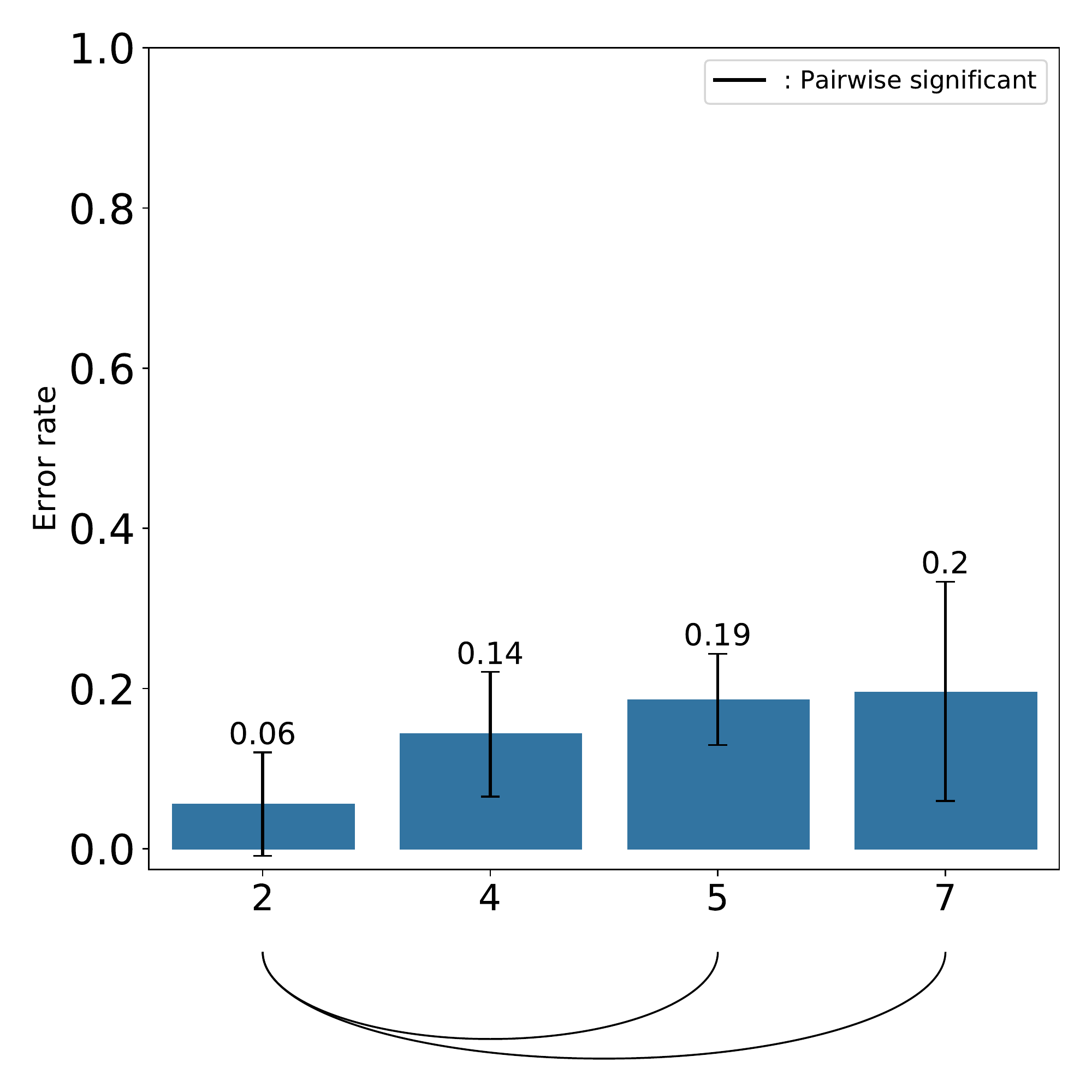}} &
    {\includegraphics[align=c,width=.215\linewidth]{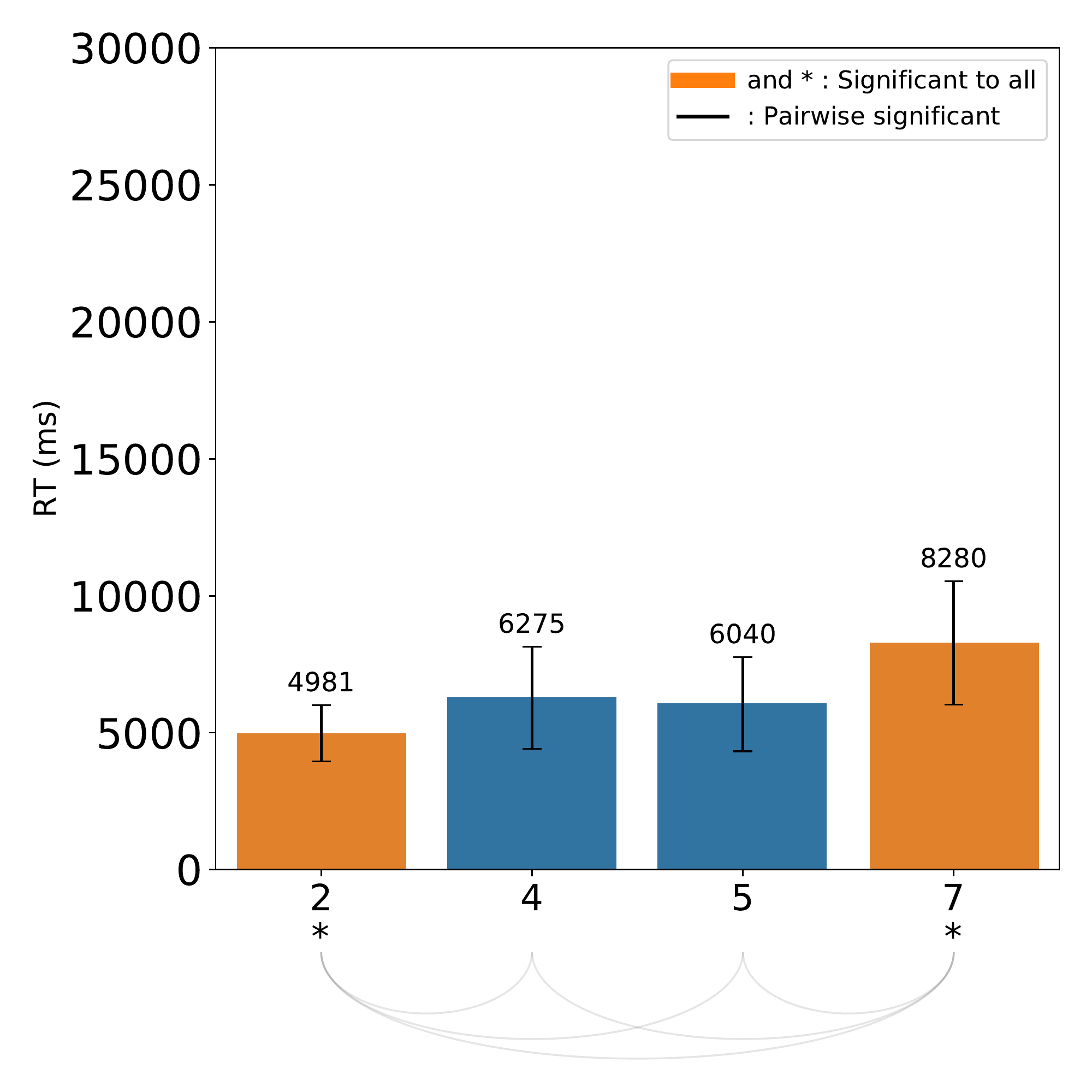}} &
    {\includegraphics[align=c,width=.215\linewidth]{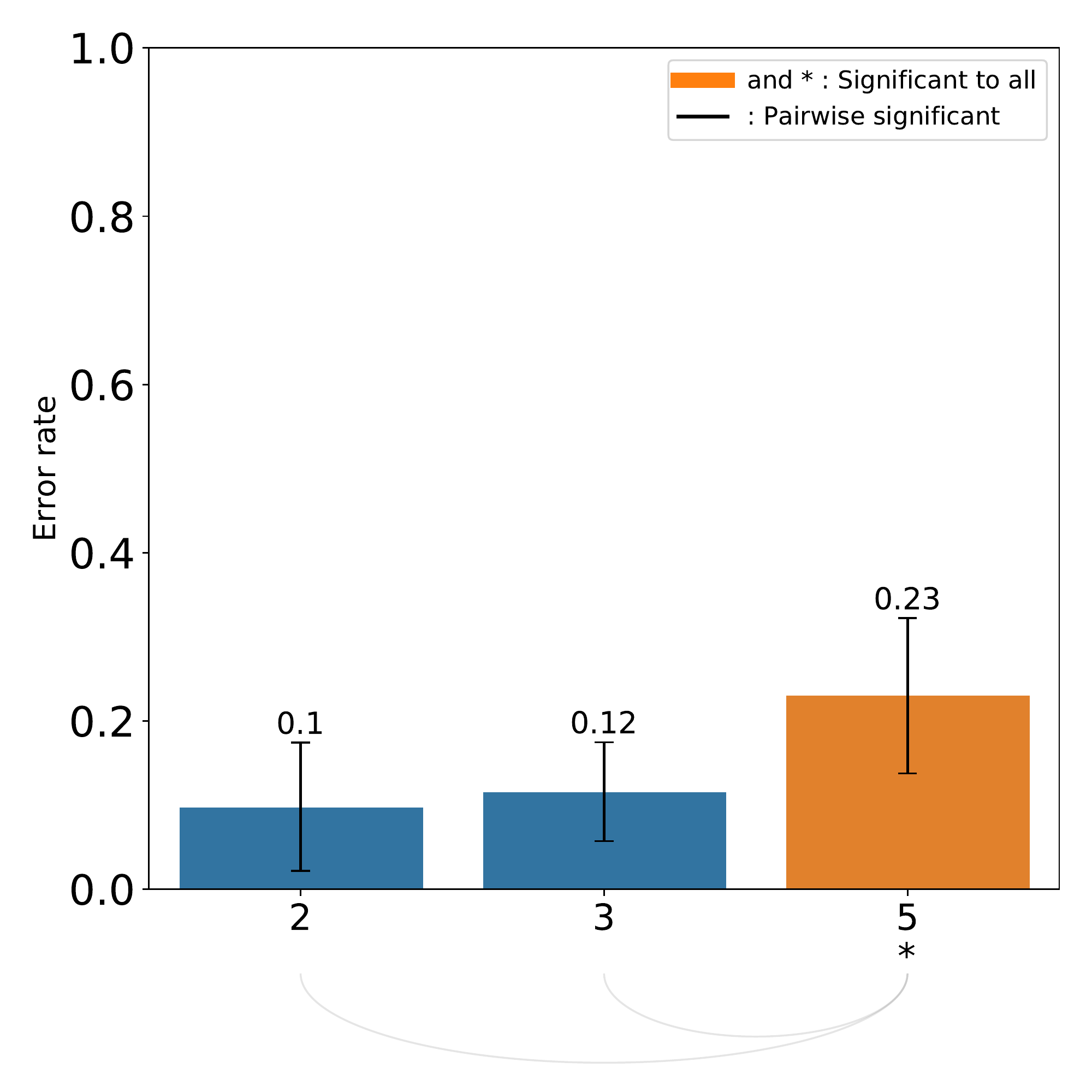}} &
    {\includegraphics[align=c,width=.215\linewidth]{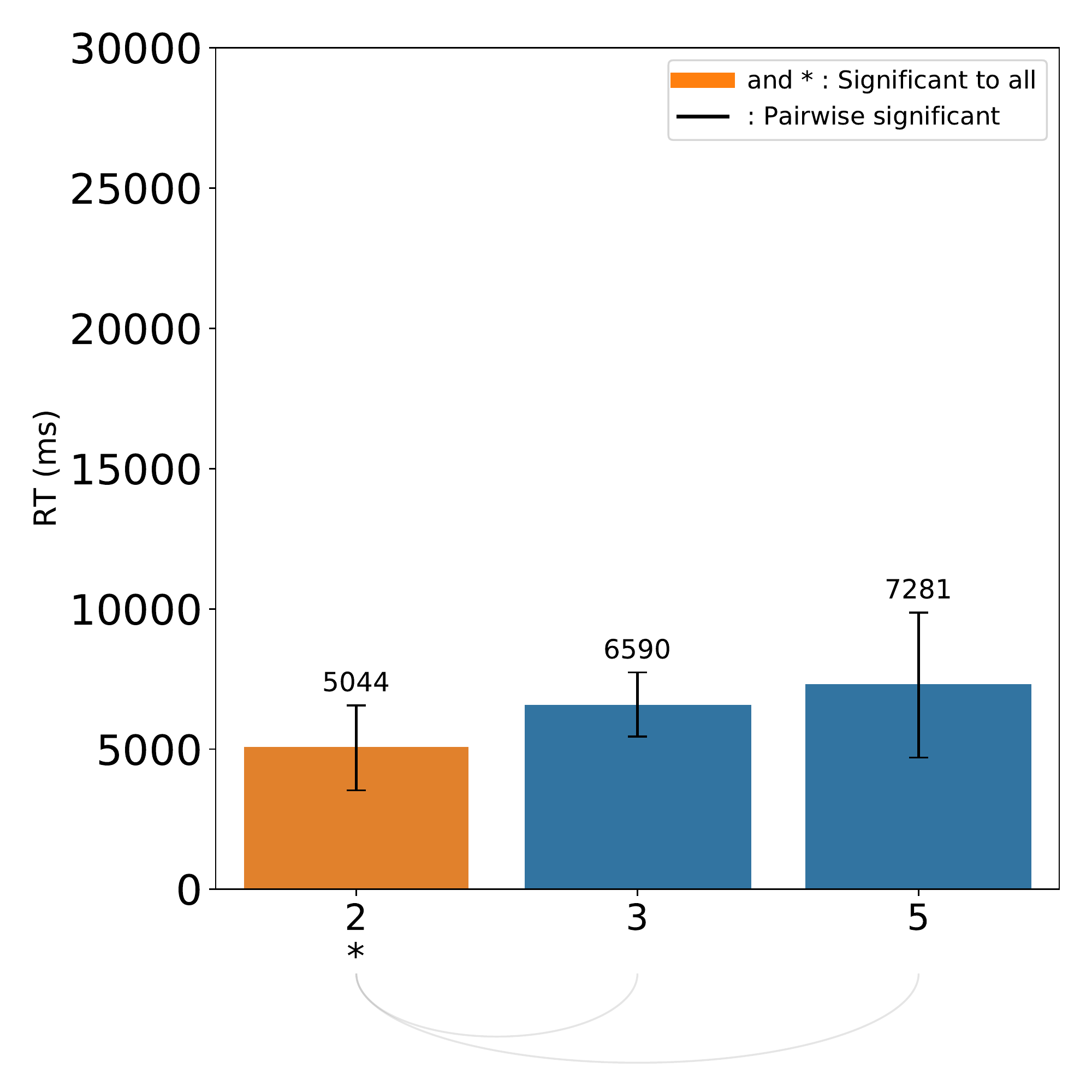}} \\
    \hline
    {\shortstack{\begin{sideways}\hspace{-0.5cm}- \emph{Color} type\end{sideways}}} \shortstack{\hspace{-0.31cm}\vspace{-0.9cm}\textbf{2}} &
    {\includegraphics[align=c,width=.215\linewidth]{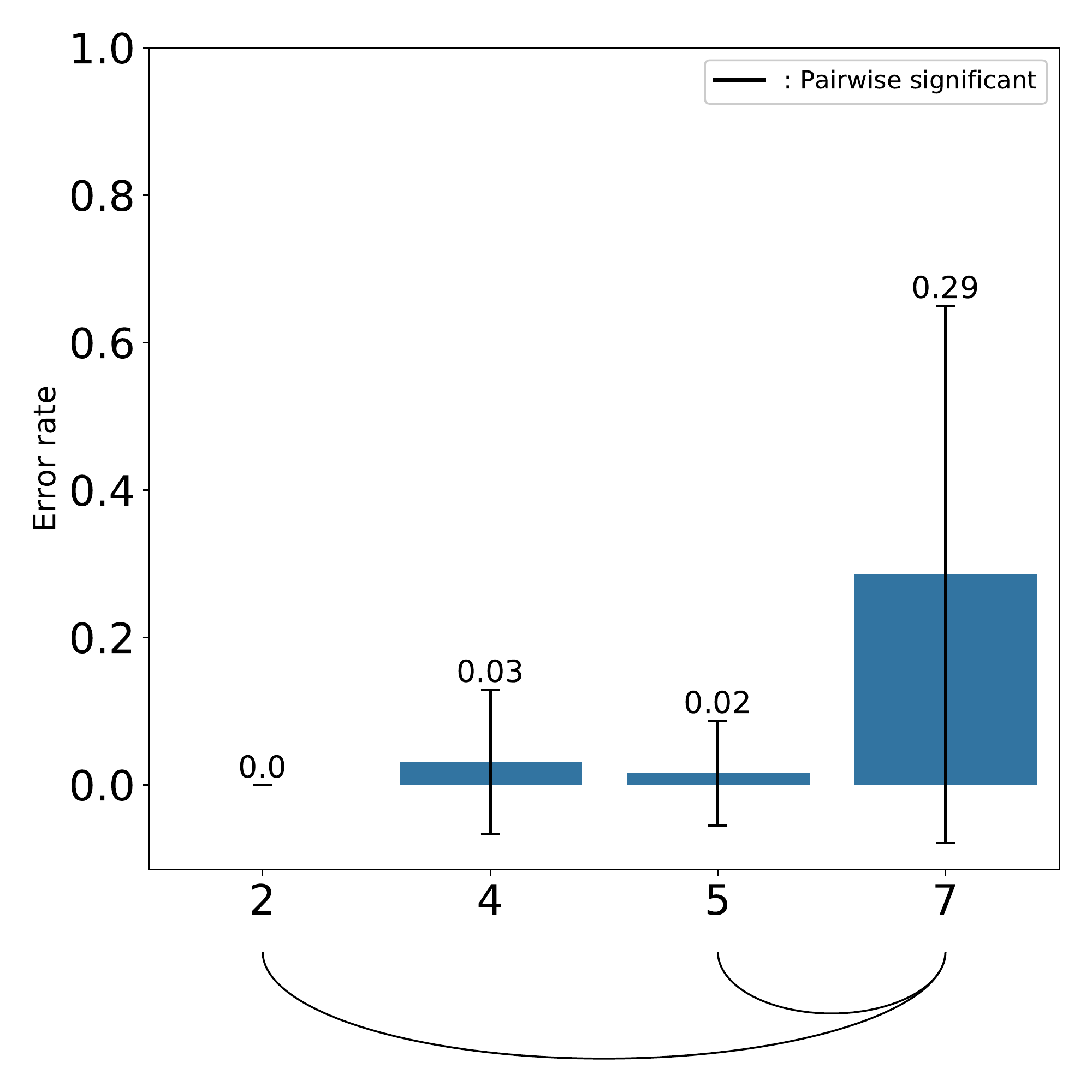}} &
    {\includegraphics[align=c,width=.215\linewidth]{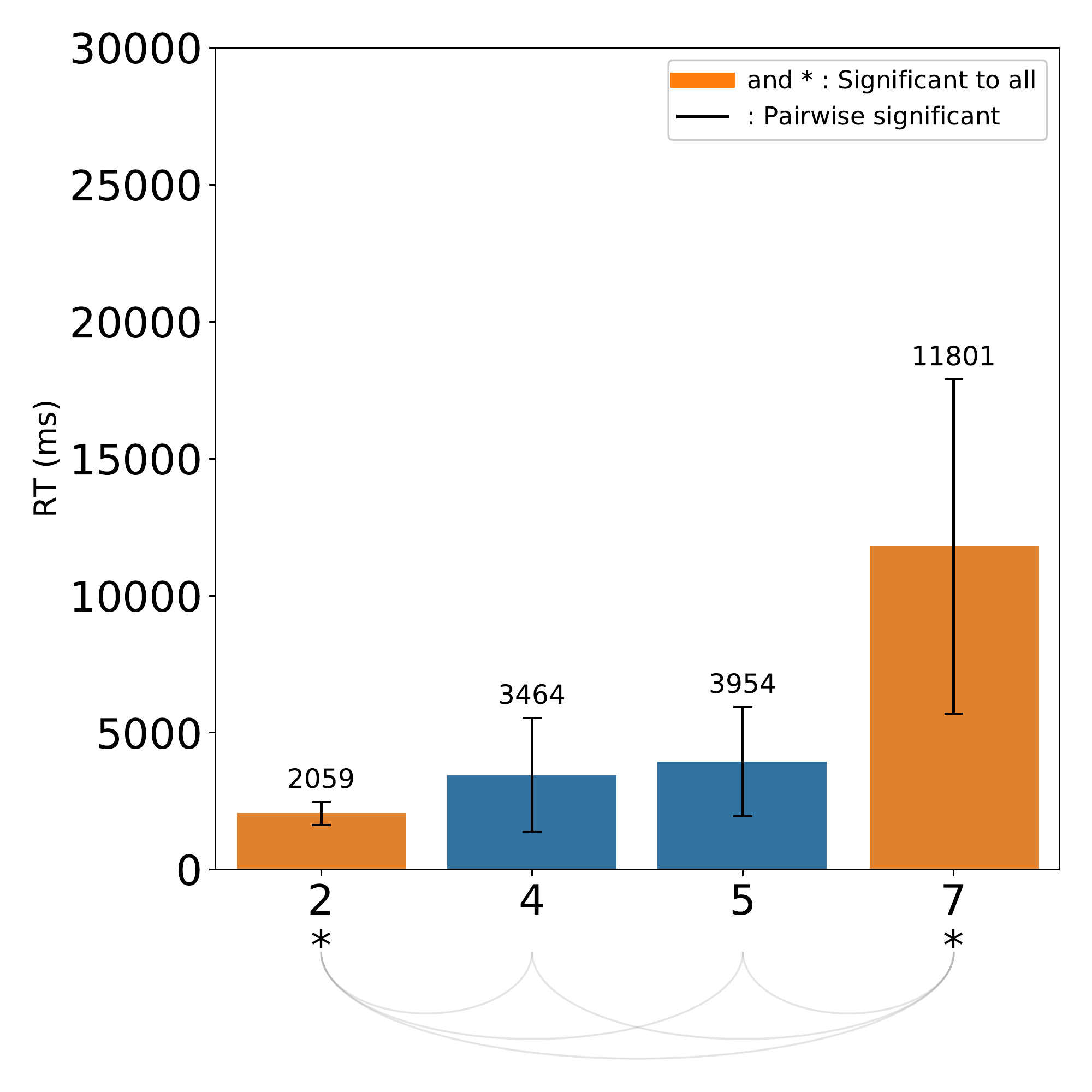}} &
    {\includegraphics[align=c,width=.215\linewidth]{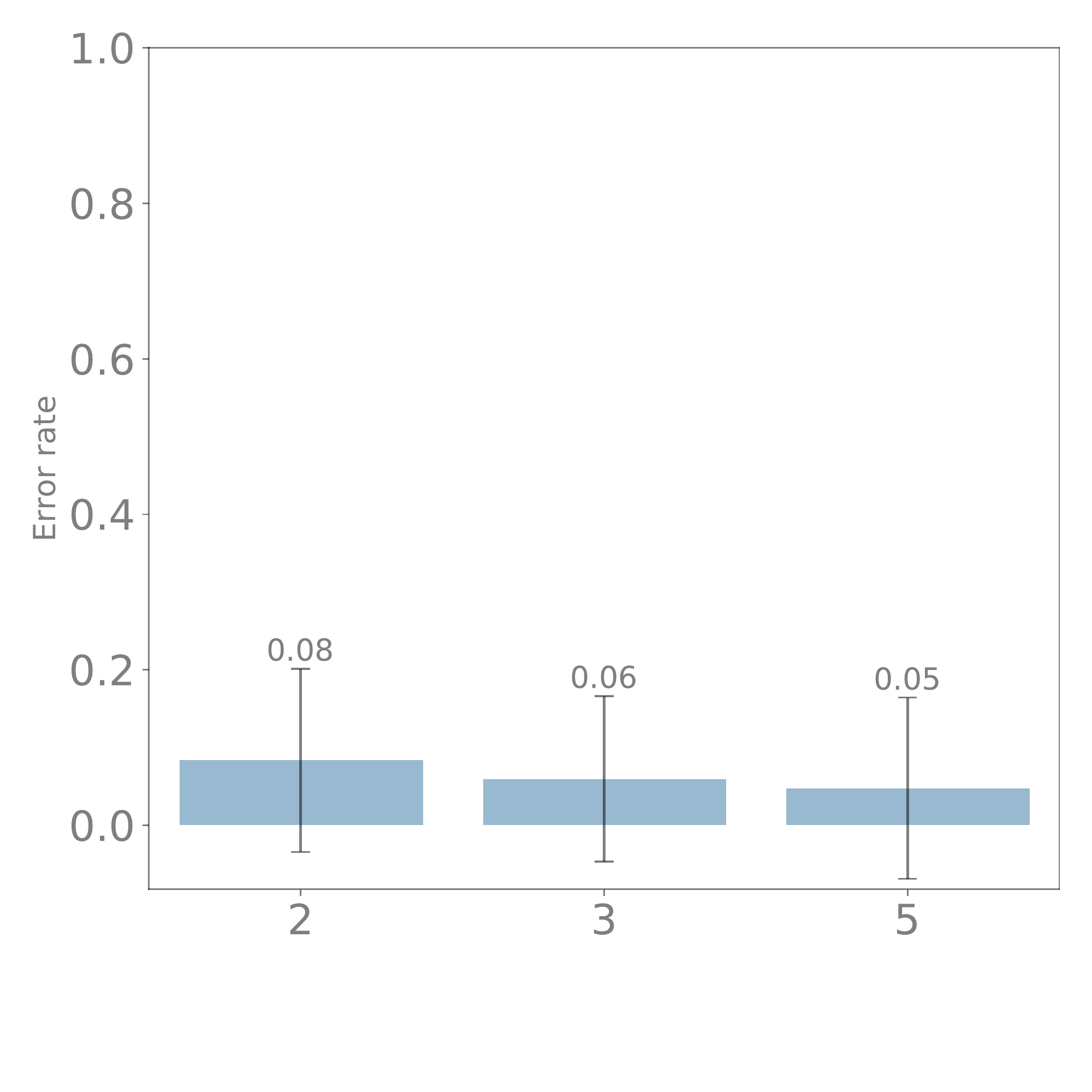}} &
    {\includegraphics[align=c,width=.215\linewidth]{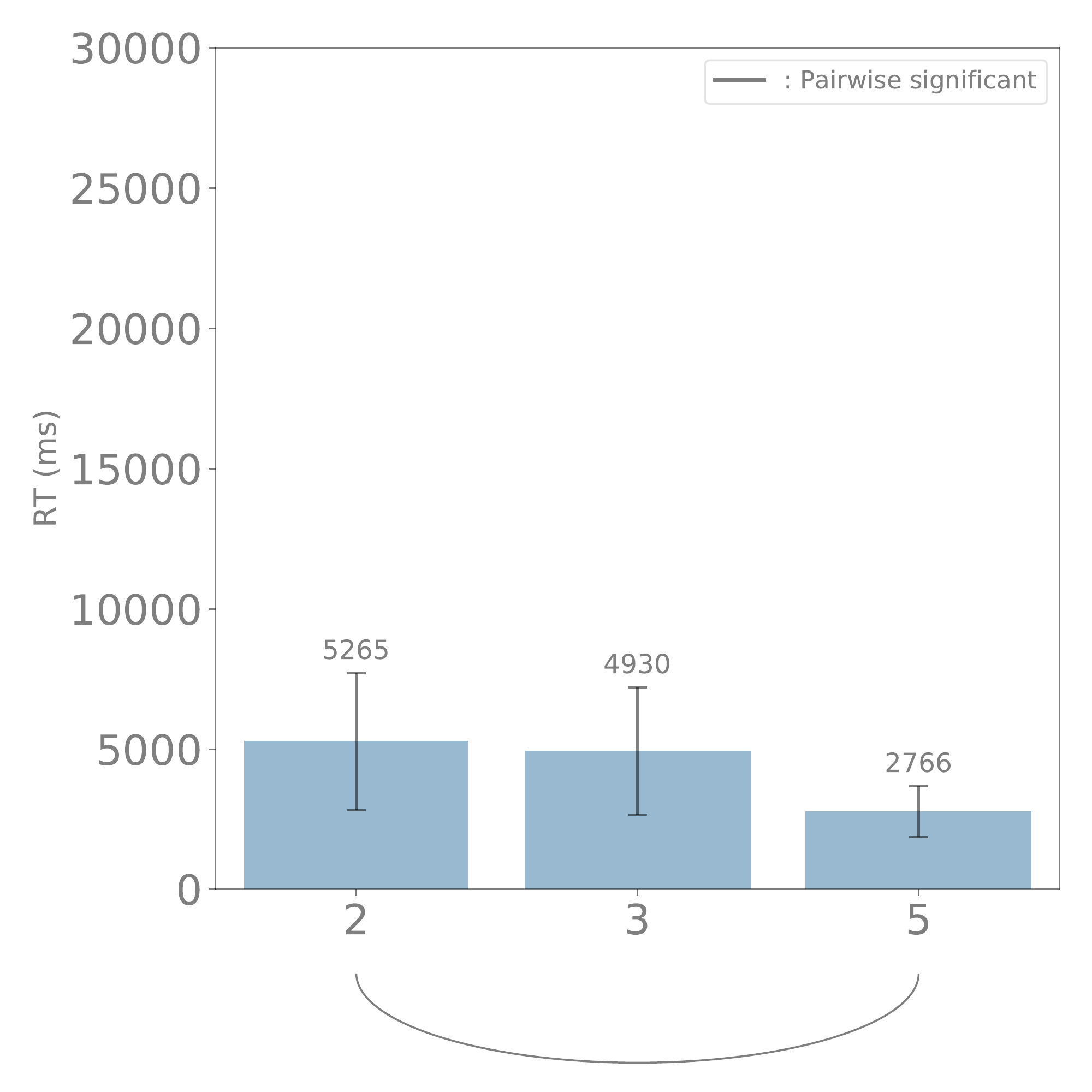}} \\
    \hline
    {\shortstack{\begin{sideways}\hspace{-0.5cm}- \emph{Shape} type\end{sideways}}} \shortstack{\hspace{-0.31cm}\vspace{-0.9cm}\textbf{3}} &
    {\includegraphics[align=c,width=.215\linewidth]{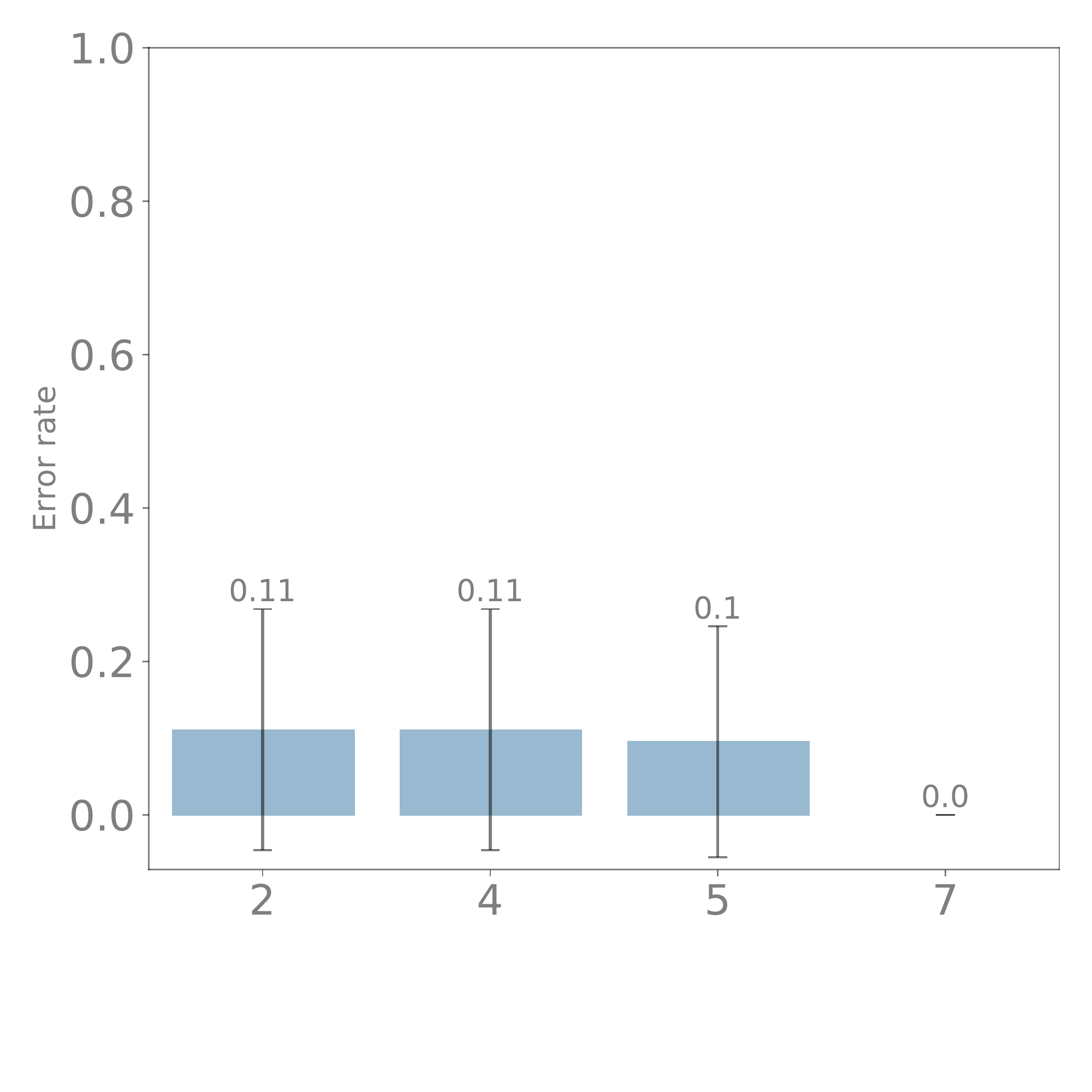}} &
    {\includegraphics[align=c,width=.215\linewidth]{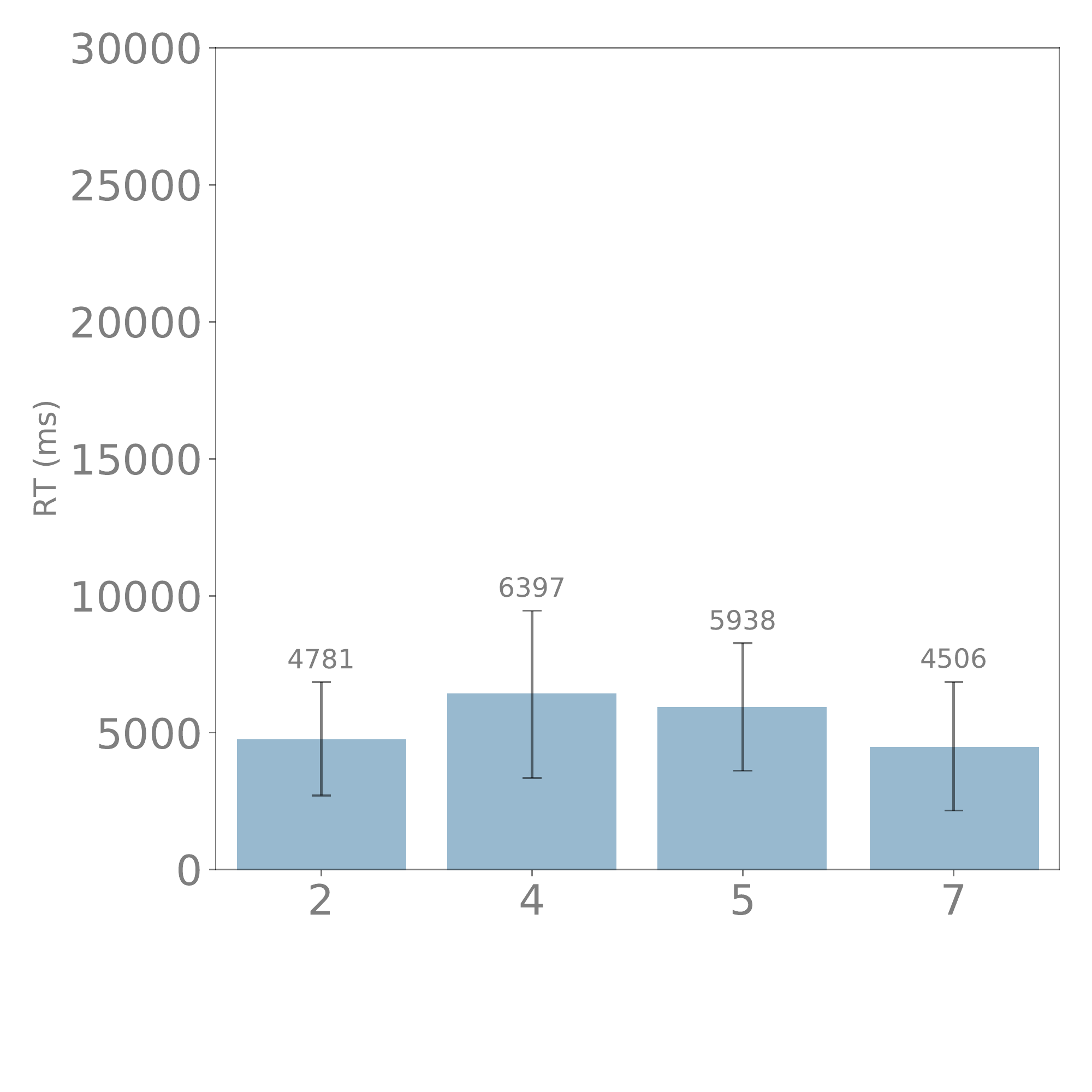}} &
    {\includegraphics[align=c,width=.215\linewidth]{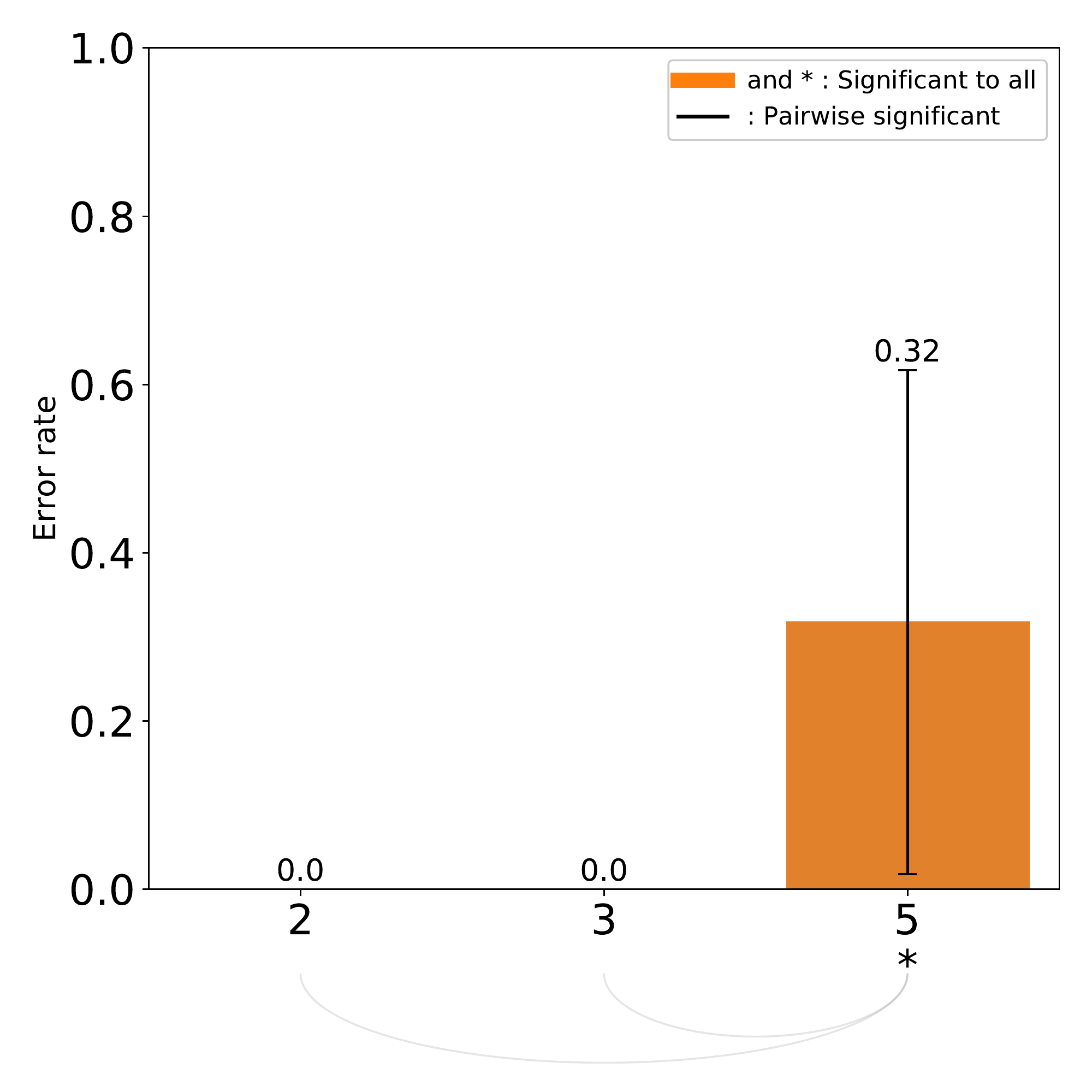}} &
    {\includegraphics[align=c,width=.215\linewidth]{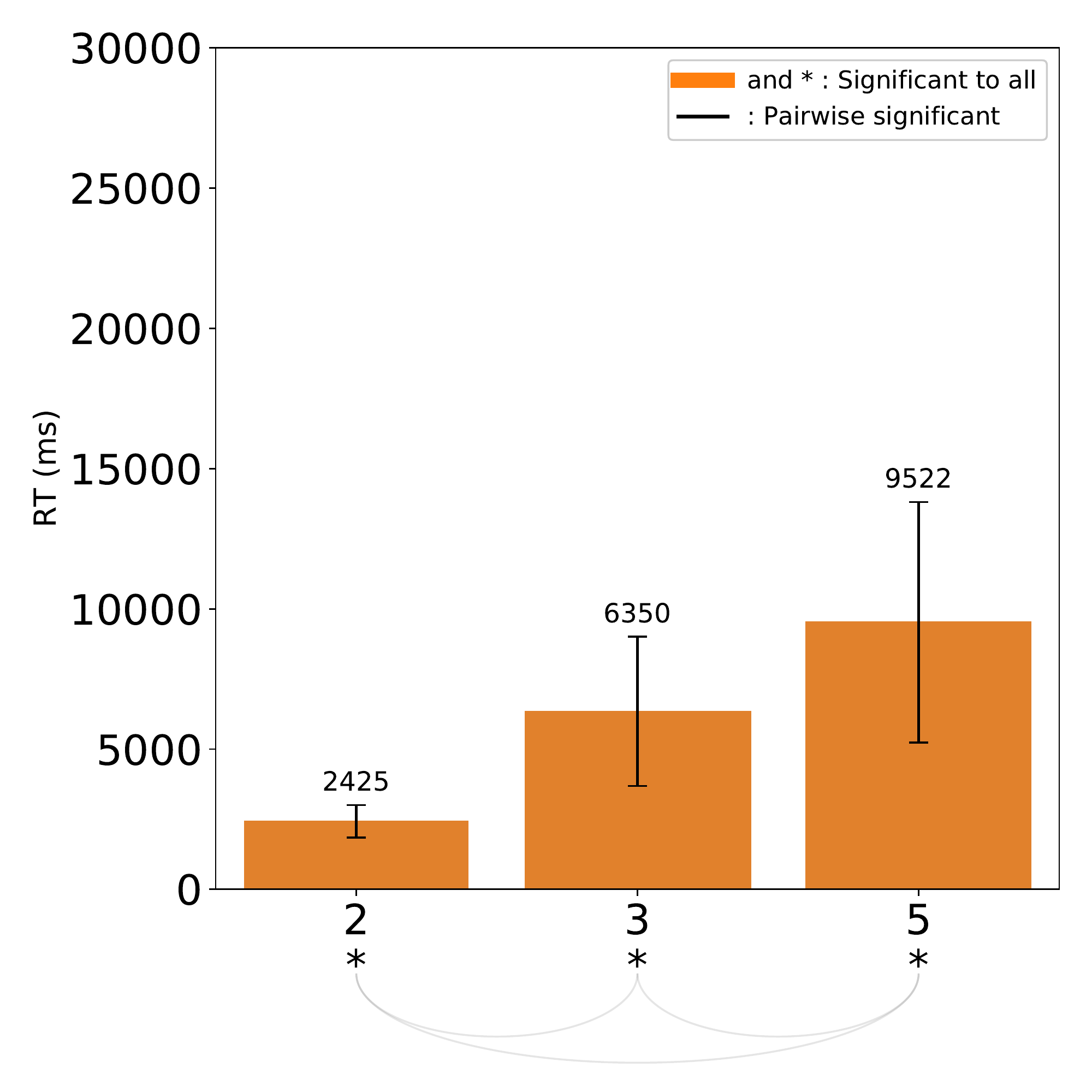}} \\
    \hline
    {\shortstack{\begin{sideways}\hspace{-0.5cm}- \emph{Red.} type\end{sideways}}} \shortstack{\hspace{-0.31cm}\vspace{-0.9cm}\textbf{4}} &
    {\includegraphics[align=c,width=.215\linewidth]{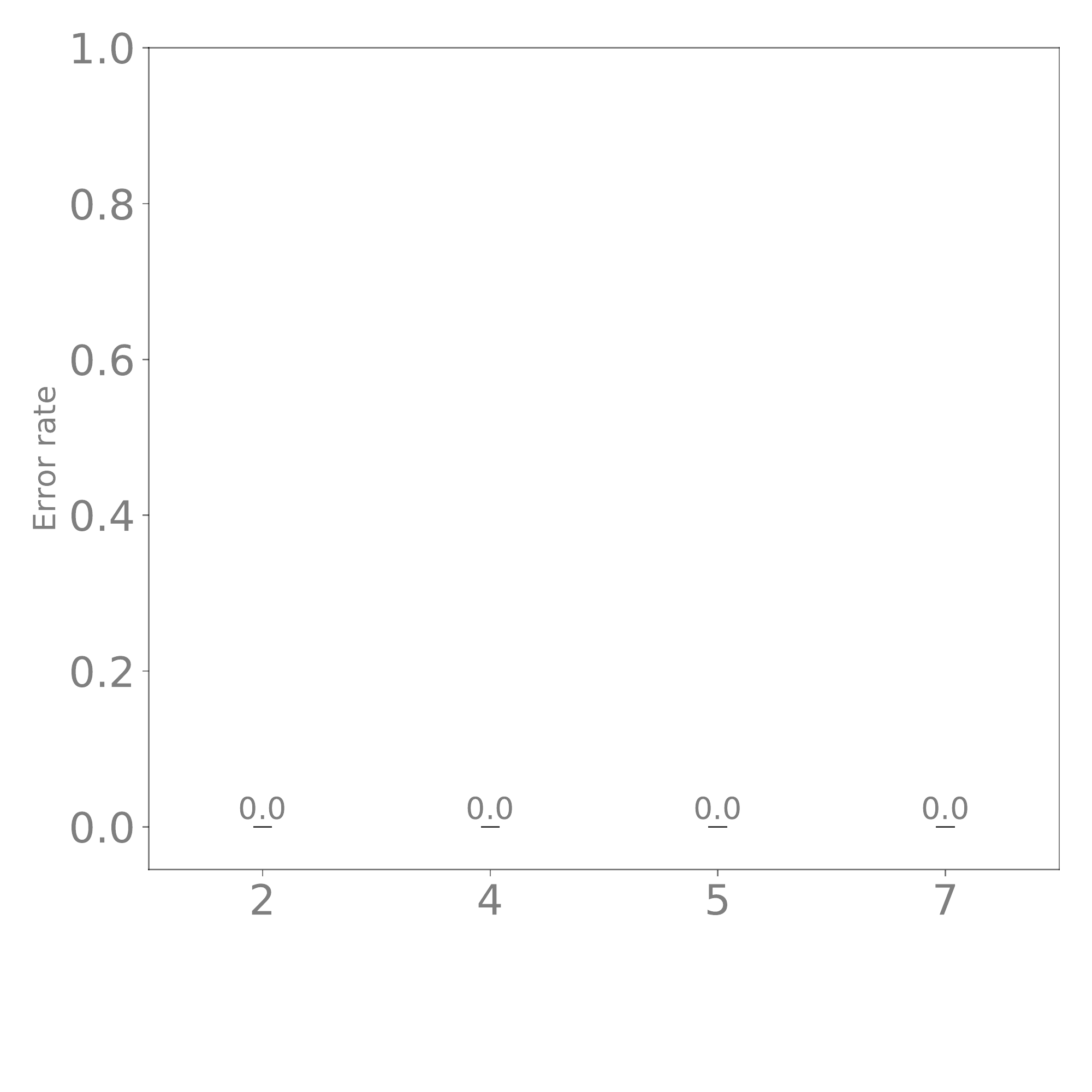}} &
    {\includegraphics[align=c,width=.215\linewidth]{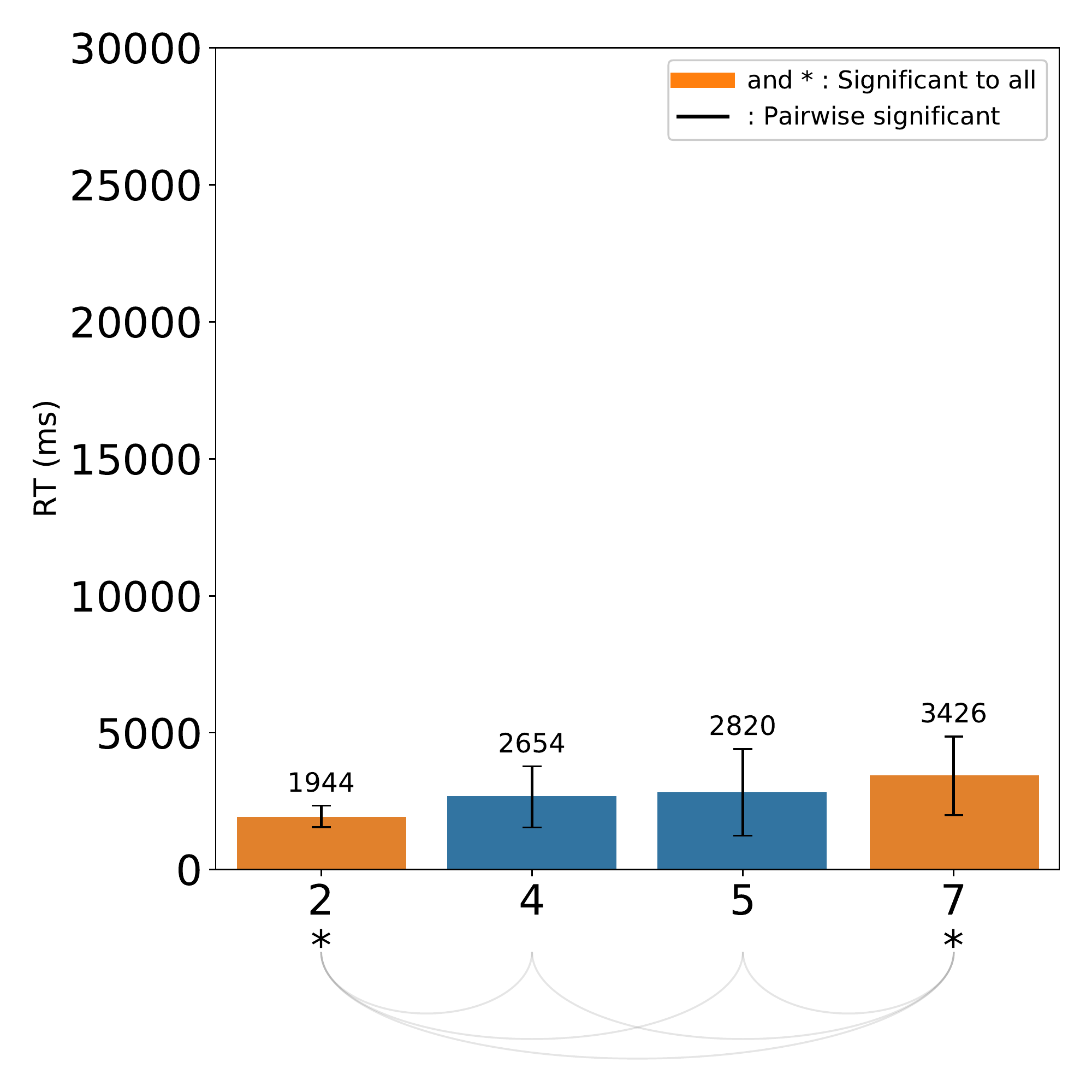}} &
    {\includegraphics[align=c,width=.215\linewidth]{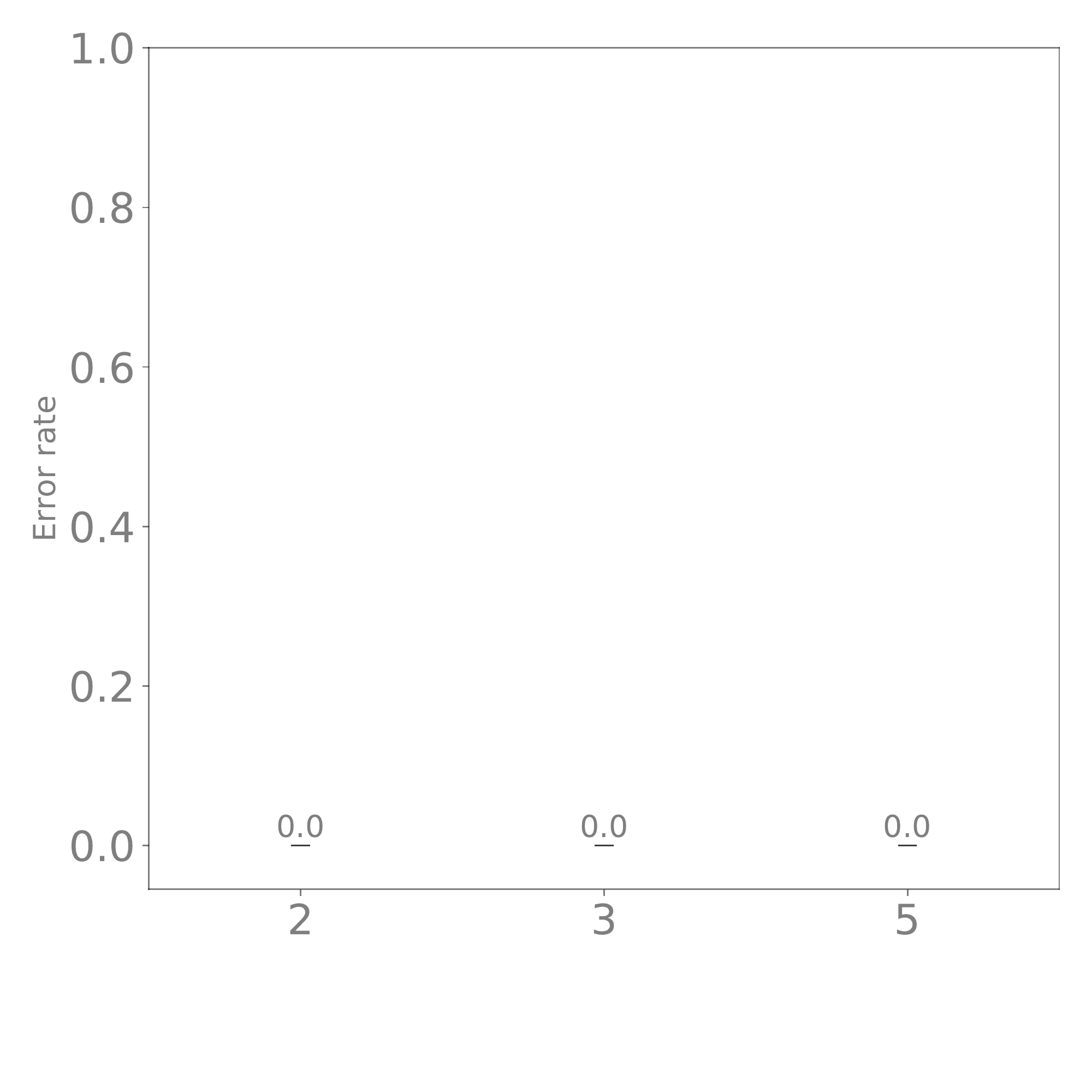}} &
    {\includegraphics[align=c,width=.215\linewidth]{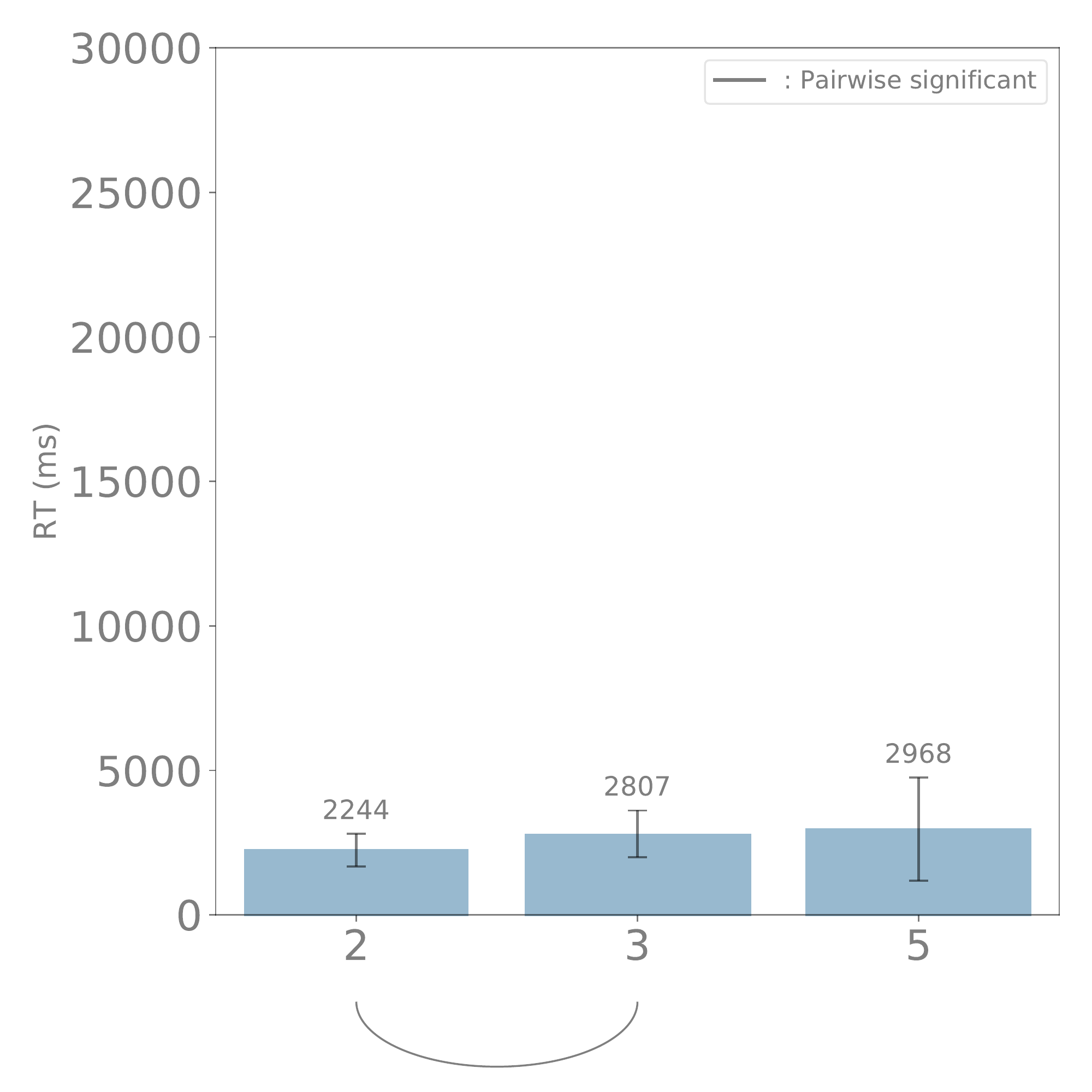}} \\
    \hline
    {\shortstack{\begin{sideways}\hspace{-0.5cm}- \emph{Conj.} type\end{sideways}}} \shortstack{\hspace{-0.31cm}\vspace{-0.9cm}\textbf{5}} &
    {\includegraphics[align=c,width=.215\linewidth]{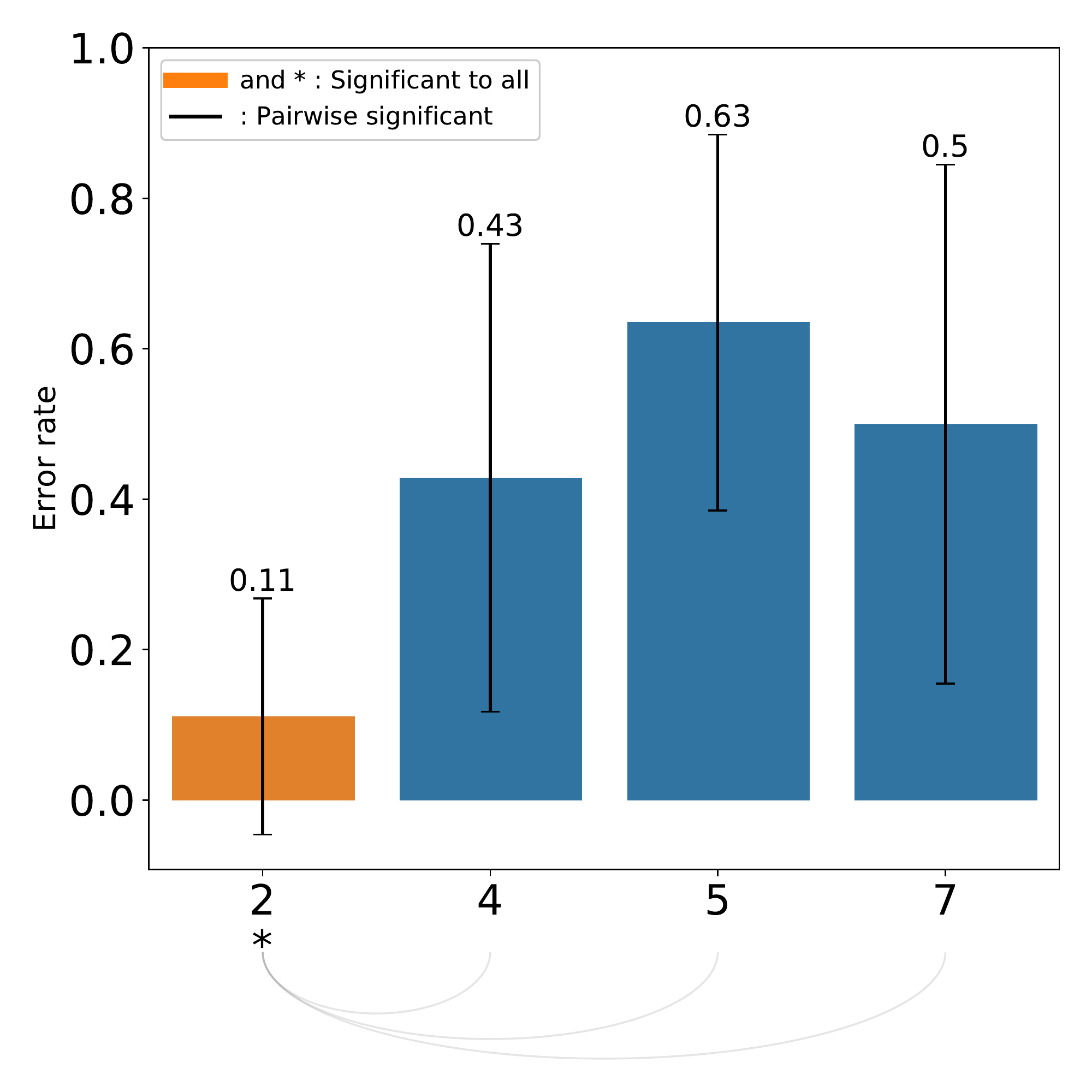}} &
    {\includegraphics[align=c,width=.215\linewidth]{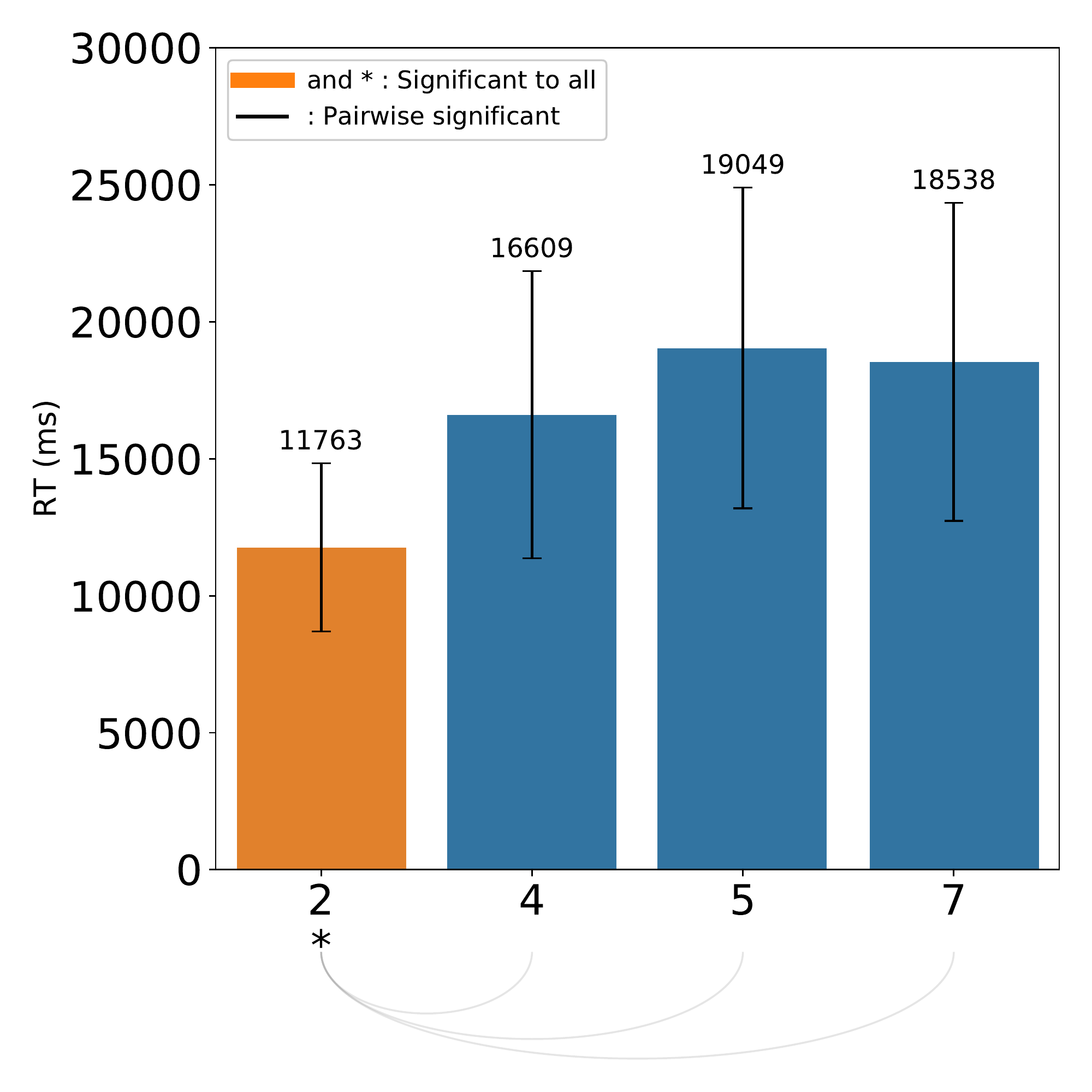}} &
    {\includegraphics[align=c,width=.215\linewidth]{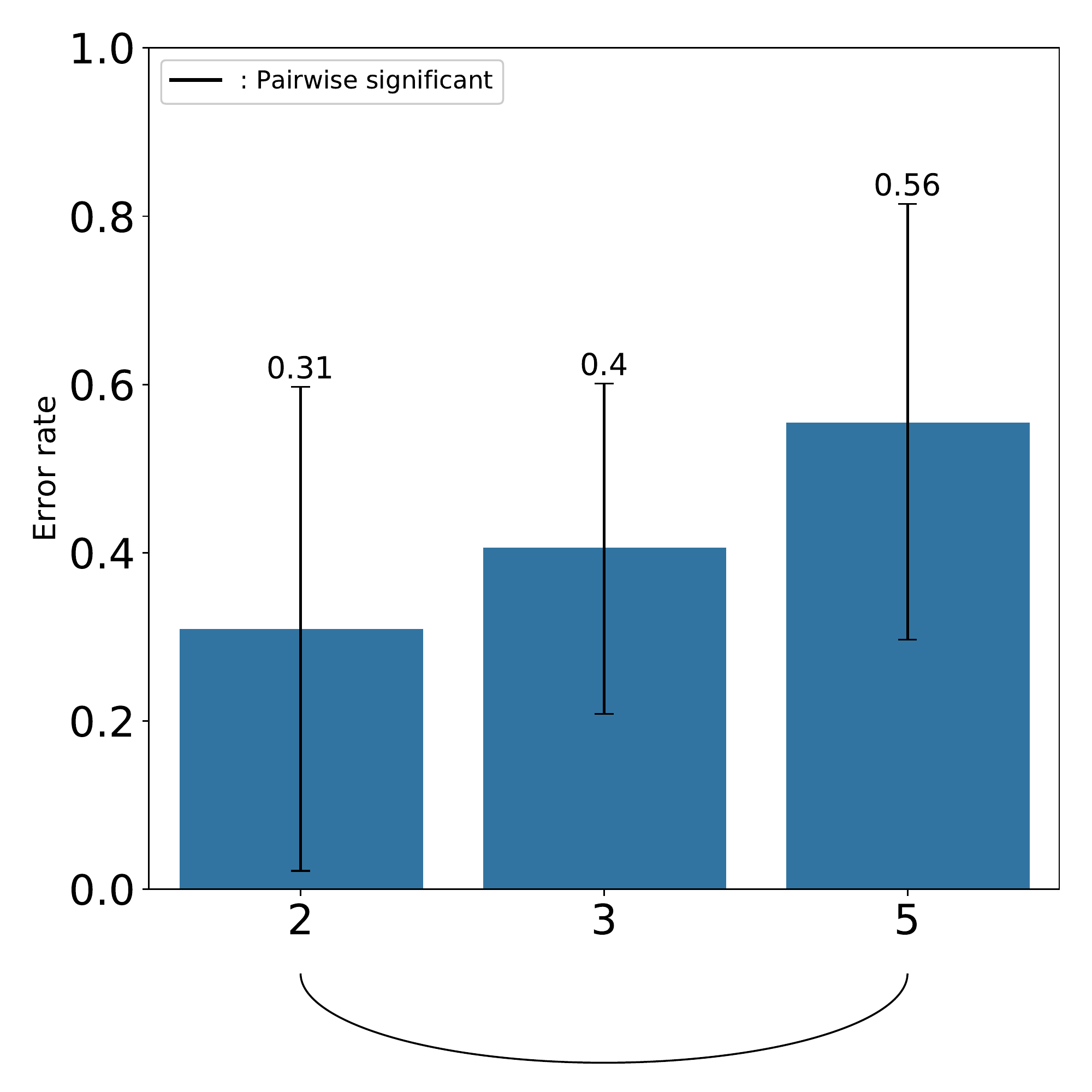}} &
    {\includegraphics[align=c,width=.215\linewidth]{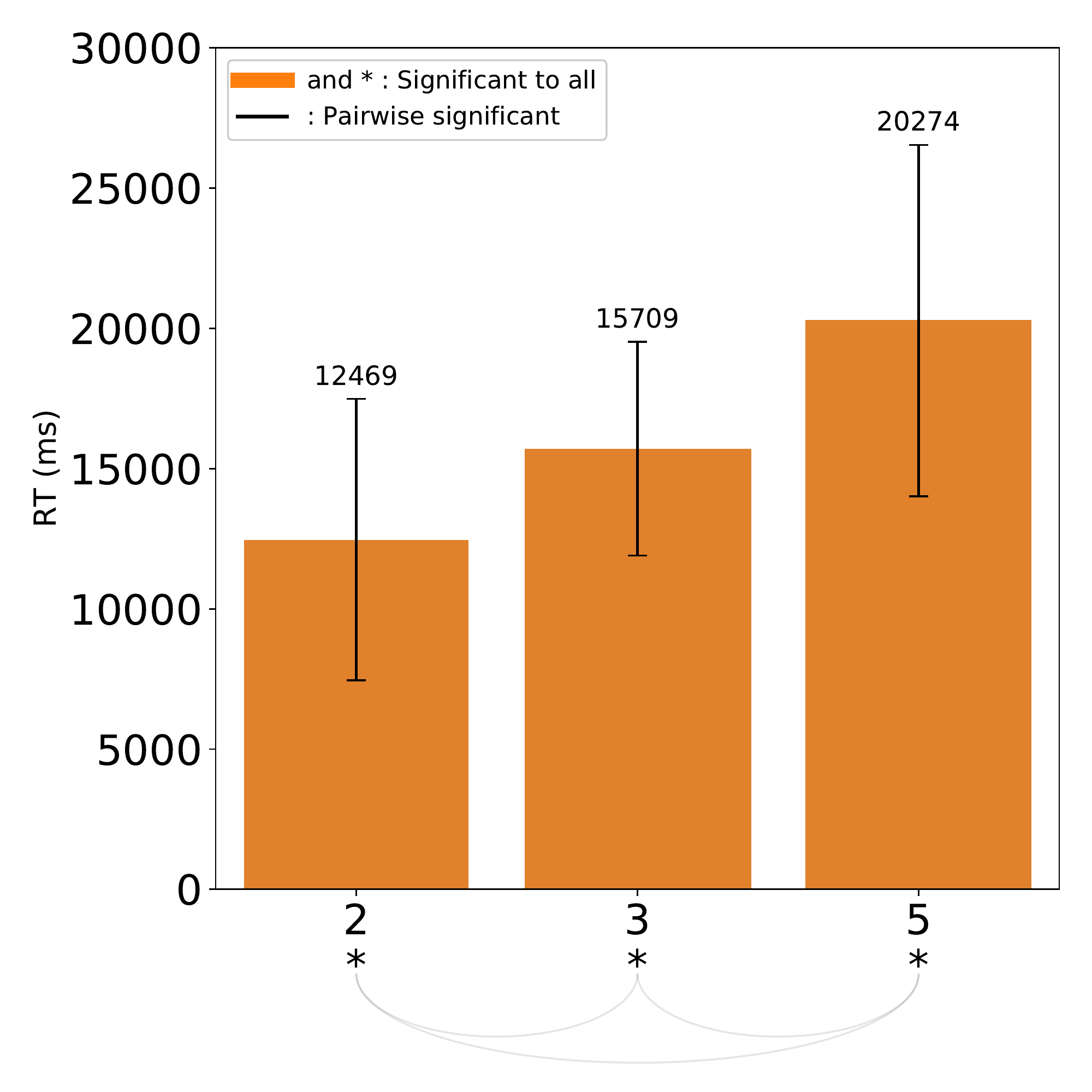}} \\
    \hline
\end{tabular}
\caption{Subjects ERs and mean RTs with standard deviation bars measured during the evaluation. The first row shows the overall performances in terms of the parameter values, while the next rows present the parameter values achieved by \textit{type}. A plot is faded if the ANOVA test on its given parameter and type aggregation failed; otherwise, it is opaque. An arc between two labels means that the pairwise comparison between the corresponding performance values is significant according to a Wilcoxon rank-sum test. The significance threshold is $p{\text -}value < 0.05$ for the ANOVA and pairwise tests in \textit{Overall} studies, while it is $p{\text -}value < 0.025$ in the \textit{per type} studies. A reading example is given in the caption of \autoref{fig:human_type_perf}. In this table of plots, columns are given letters and rows are given numbers for ease of reference.}
\label{fig:err_rt_human_per_param_value}
\end{figure*}

\subsubsection{Quantitative Results}
\label{sec:hypothesis_validation}

In the following, we describe the subject results and interpret them to confirm or refute the preliminary hypotheses defined in \autoref{sec:autom_hypothesis} that aim to study the effects of the variations in \textit{type}, \nbCols and \nbShapes on the task difficulty.

Similar to the DNN results analysis, we first run Kruskal-Wallis ANOVA tests\cite{kruskal1952wallis} on all considered parameters for the ER and RT measures. For the RTs, only validated answers are considered (\textit{i.e.}, subjects did not run \textit{out of time}, OOT). On average, OOT trials account for 5.5 trials out of the 44 of the evaluations per subject, 74\% of which are of type \typeconj, and their distribution among \textit{type}, \nbCols and \nbShapes is shown in \autoref{fig:OOT_distrib}. Again, results were studied overall and per type value. The significance level was the same as those of the DNN results analysis: $\alpha = 0.05$ for the overall studies and $\alpha = 0.025$ for the \textit{per type} studies. These tests showed that the three considered parameters presented significant effects on performance. The results are presented in Figures~\ref{fig:human_type_perf} and \ref{fig:err_rt_human_per_param_value}. \\

We \textbf{accept \Htype} since the results plainly corroborate its statement. We can see in \autoref{fig:human_type_perf} that the ER and RT performances show the same trend in terms of \textit{type} value difficulty, although there are fewer significant pairwise differences in the ERs than in the RT results. The \typered type led to the best performances, as all subjects answered correctly on all type \typered trials with a mean RT below 3 seconds. Although the \typecol and \typeshape type ERs are not significantly different (6.5\% and 8.7\%, respectively), type \typecol (4.5 s) trials were significantly faster to solve than \typeshape (5.5 s) trials, though such a small difference might not be relevant in an information visualization context. Finally, type \typeconj is hardest, with an ER of 41.1\% and an average RT of 15.5 s, and it accounts for 74\% of the OOT trials (see \autoref{fig:OOT_distrib}).

We also \textbf{accept \Hconj}, although the results are not straightforward to read. In \autoref{fig:human_type_perf}, the \typeconj type is the condition that significantly led to the worst performances in terms of both the ERs and RTs. In this regard, we validate that the \typeconj type is the hardest type value. In \autoref{fig:err_rt_human_per_param_value} A-5 and B-5, we see that \nbCols has a significant effect on type \typeconj trials with respect to the ERs and RTs. The \nbCols effect is only significant between 2 (11\% ER; 11.8 s RT) and the other values ($>$42\% ER; $>$16.6 s RT). This confirms that the task difficulty of type \typeconj trials increases with \nbCols but caps very quickly (\textit{i.e.}, the task difficulty no longer significantly increases starting from 4 \nbCols).
In \autoref{fig:err_rt_human_per_param_value} C-5 and D-5, no threshold effect on neither the ERs nor RTs can be directly observed as \nbShapes varies. As mentioned above, the RT results on OOT trials were not considered since it would have been incorrect to interpret them as \textit{wrong answers in 30 seconds}. Since 86 type \typeconj trials were OOT (see \autoref{fig:OOT_distrib}), 86 out of $11*21=231$ answers were not considered in the computation of the type \typeconj RT performances, and these 86 trials would have had RTs above 30 seconds. These missing points make the RT results look less ``poor" than they truly are and hide the threshold effect we expected to observe for \Hconj. The same interpretation can be made with the \typeconj type and \nbCols performances, and strengthens the threshold effect that can already be observed.

We \textbf{accept \Hred} with a restriction regarding the context in which it is considered. \autoref{fig:human_type_perf} shows that all answers are correct for type \typered trials (\textit{i.e.,} 0\% ER) and that they are significantly faster to solve than any other type value. Hence, the \typered type is the easiest type value. All answers being correct, we validate that \nbCols and \nbShapes variations do not affect the ERs. For the RT results, \nbCols is shown to have a significant effect on type \typered trials (see \autoref{fig:err_rt_human_per_param_value} B-4). The RT variations have small amplitudes, and the results remain between 1.9 and 3.4 seconds. Such RTs are more than acceptable, and their variations do not denote a significant loss of performance with respect to solving an outlier detection task in an information visualization context. \\

Globally, the experiment enables us to extend the findings of the literature. From here, we will study the results with regard to the two remaining hypotheses and later discuss the capacity limits of the color and shape dimensions that this experiment enables us to observe. \\

We \textbf{accept \Hcol} since \nbCols significantly affects the resulting performance when color is the only relevant dimension. \autoref{fig:err_rt_human_per_param_value} A-1 and B-1 show that, overall, \nbCols has significant effects on both the ERs and RTs. The only significant ER differences lie between 2 and 5 or 7 \nbCols. On the other hand, the RT performances show that the task difficulty significantly increases with \nbCols except between the values of 4 and 5. When color is the only relevant dimension, we can see in \autoref{fig:err_rt_human_per_param_value} A-2 that the ER remains low until 7 \nbCols, where it dramatically increases. The same behavior can be observed for the RTs (\autoref{fig:err_rt_human_per_param_value} B-2), although the growth seems linearly related to \nbCols until the value 7. When color is not a relevant dimension (\autoref{fig:err_rt_human_per_param_value} A-3 and B-3), it has no significant effect on performance.

We \textbf{reject \Hshape} since \nbShapes does affect the task difficulty when shape is the only relevant parameter. Overall, we see in \autoref{fig:err_rt_human_per_param_value} C-1 and D-1 that \nbShapes has a significant effect on performance. The error rates are significantly higher when there are 5 shapes in a trial, while the RTs are significantly lower when there are 2 shapes. When shape is not a relevant dimension (\textit{i.e.,} \typecol type, \autoref{fig:err_rt_human_per_param_value} C-2 and D-2), it has no effect on the performances. When it is the only relevant dimension (\autoref{fig:err_rt_human_per_param_value} C-3 and D-3), both the ERs and RTs are significantly impacted by variations in \nbShapes, but they do not follow the same trend. In fact, subjects never failed to solve the task when shape was the only relevant dimension and there were less than 5 \nbShapes. On the other hand, the RT performances seem to increase linearly with \nbShapes. However, \autoref{fig:OOT_distrib} shows that subjects ran out of time 18 times on trials of the \typeshape type, always with 5 \nbShapes. With the same reasoning as that of \Hconj, we know that the 5 \nbShapes column in \autoref{fig:err_rt_human_per_param_value} D-3 is missing 18 (out of $(|\nbCols|-1)*21 = 63$) data points that would have taken more than 30 seconds, and this is a significant number since it represents 28.6\% of the data points in that column.

\subsubsection{Qualitative Results}
\label{sec:qualitative_results}

The qualitative results of this evaluation are built upon the subjects answers to the questionnaire.

The first question was to order, from easiest to hardest, the different \emph{type} values. For this question, almost all subjects (20 out of 21) ranked the \typered type as the easiest and \typeconj as the hardest. More than half of the subjects (14 out of 21) ranked the \typecol type as easier than \typeshape. This result corroborates their performances (see \autoref{fig:human_type_perf}), as the \typeconj type has the highest ER and RT; the \typered type is significantly faster to answer than \typecol, which is also significantly faster to answer than \typeshape.

\begin{figure}[!bt]
\centering
\begin{tabular}{|c|c|c|}
\hline
 &  \textit{Outlier color} & \textit{Outlier shape}\\
 \hline
 \begin{sideways}\hspace{-0.6cm}Error rate\end{sideways} & \includegraphics[align=c,width=0.40\linewidth]{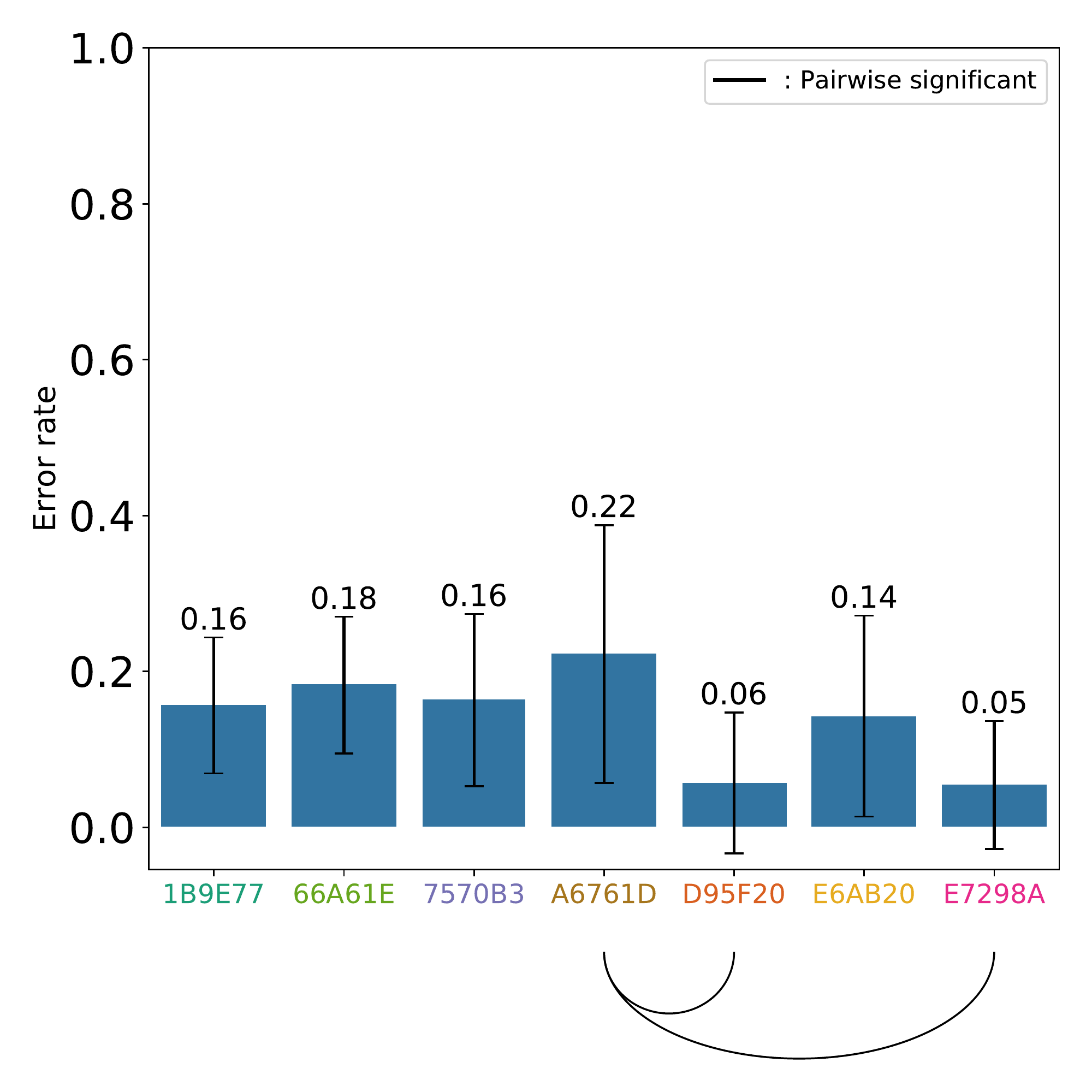} & \includegraphics[align=c,width=0.40\linewidth]{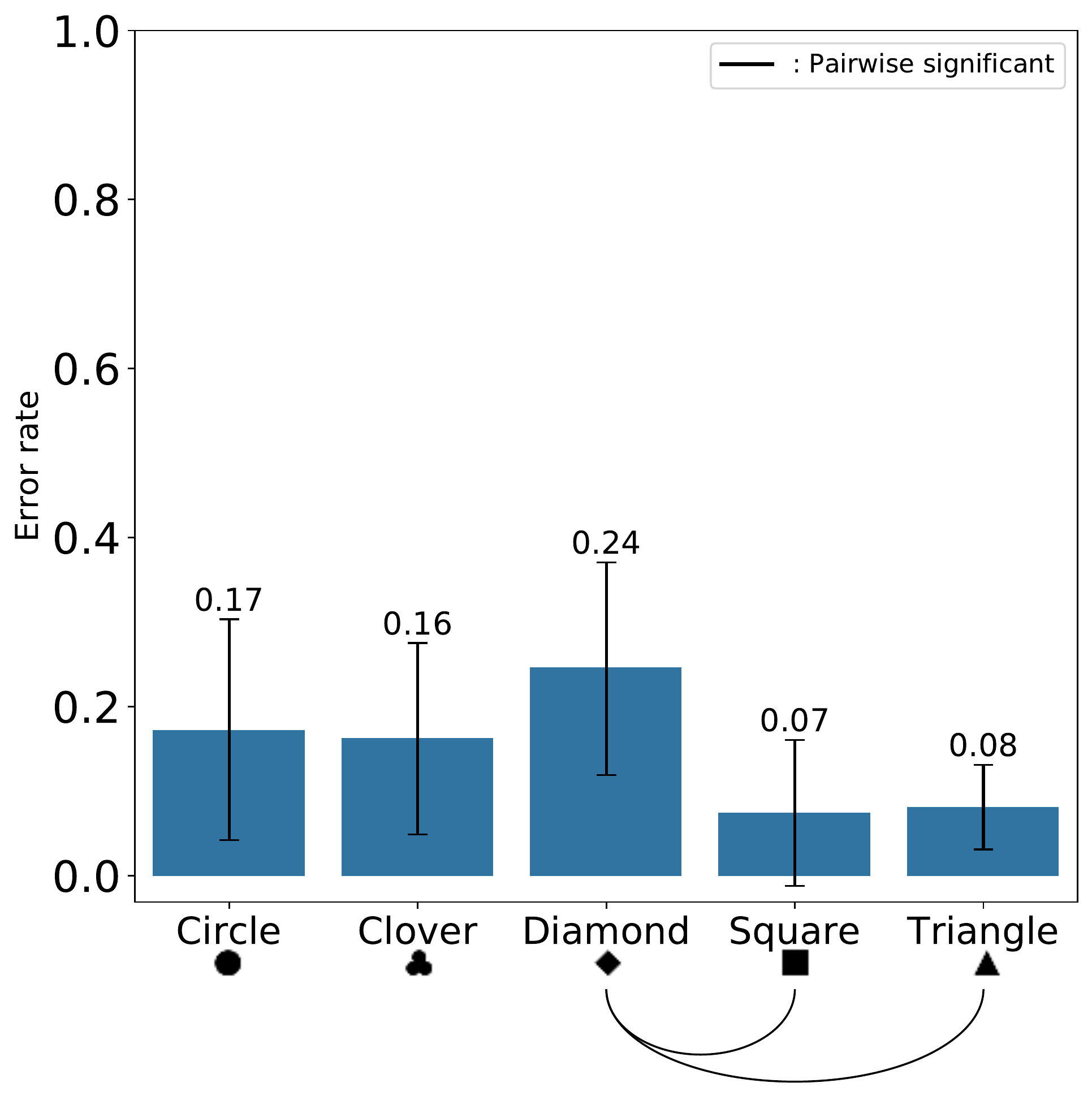} \\
 \hline
 \begin{sideways}\hspace{-0.8cm}Response time\end{sideways} & \includegraphics[align=c,width=0.40\linewidth]{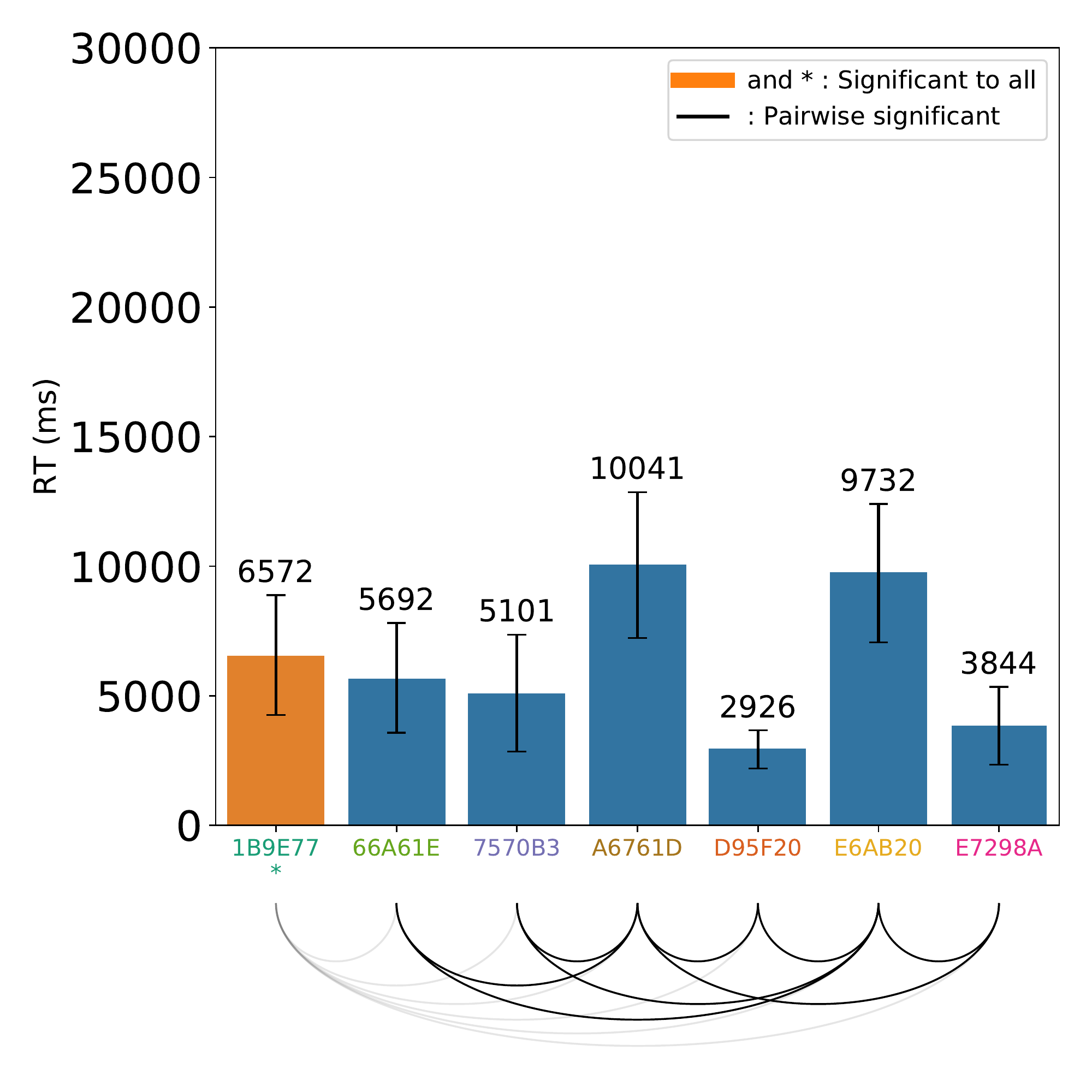} & \includegraphics[align=c,width=0.40\linewidth]{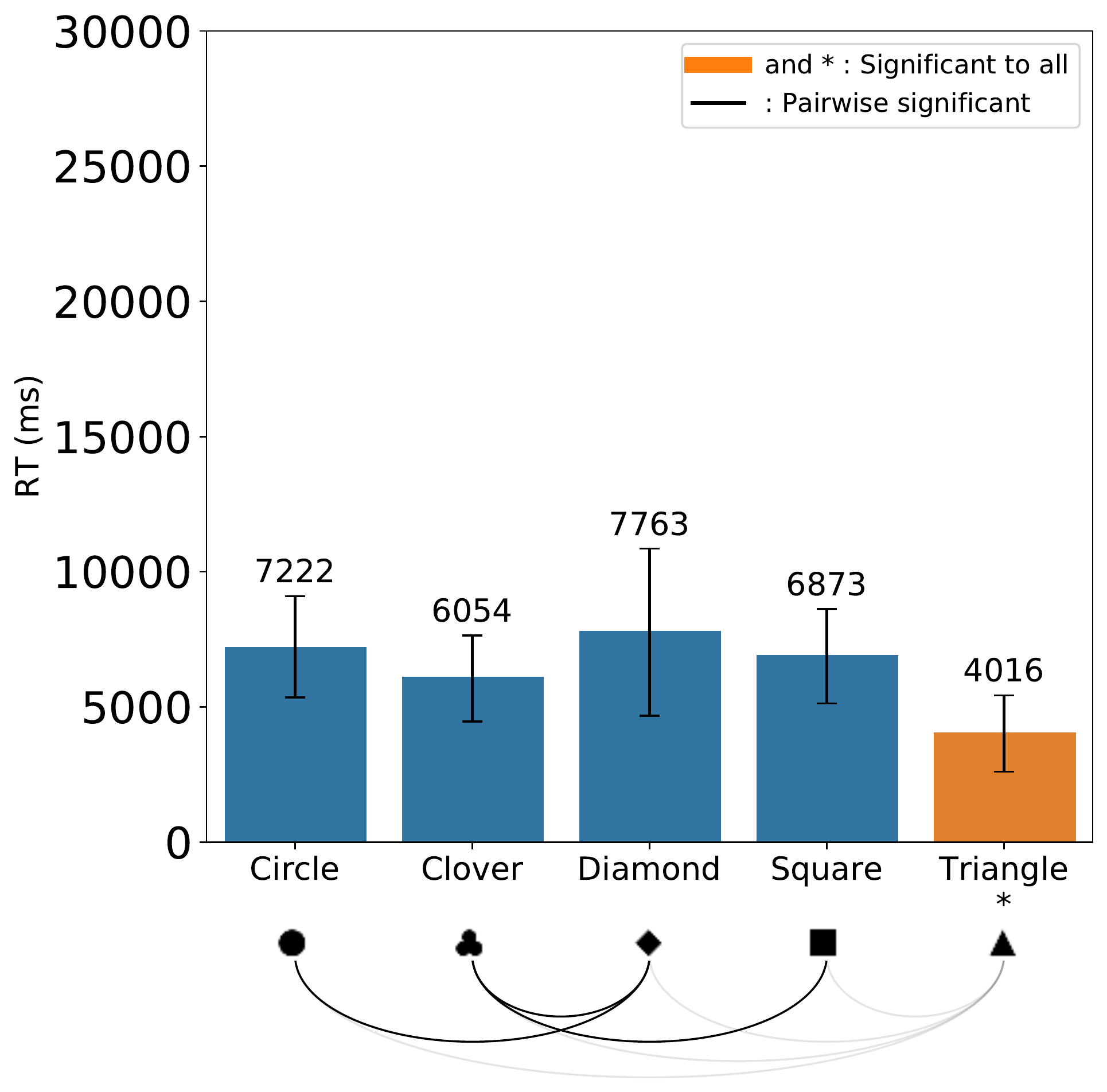} \\
 \hline
\end{tabular}
\caption{Subjects ERs and mean RTs with standard deviation bars for the \emph{outlier color} and \emph{outlier shape}. An ANOVA test showed that both had significant effects on the results. An arc between two labels means that the pairwise comparison between the corresponding performance values is significant ($p{\text -}value <~0.05$) according to a Wilcoxon rank-sum test.}
\label{fig:human_res_outlier_shape_color}
\end{figure}

The second question was to report whether an \emph{outlier color} was easier to find than others. Ten subjects clearly identified \pink as easier than others, and few subjects found that \green and \bluegreen made the task more difficult to solve when both were present in a trial. This feeling partially reflects their RT performances, though the ERs only show significant differences that would lead one to think that \brown is the most difficult color (see \autoref{fig:human_res_outlier_shape_color}). Only a few subjects reported \orange to be an easy color, and neither \brown nor \yellow were cited as hard-to-find colors despite their low performances, as reported in \autoref{fig:human_res_outlier_shape_color}. The latter point can be explained by the fact that \brown and \yellow primarily occurred as \textit{outlier colors} in type \typeconj trials. In these complex trials, subjects probably did not remember the outlier colors and shapes since the complexity of the display mostly came from its heterogeneity (the \nbCols and \nbShapes parameters).
Finally, some subjects reported that the outlier color did not matter as long as the contrast between stimuli colors remained high enough. Although our experiment did not aim to measure the impact of outlier saturation on the task, it would be an interesting extension of the works by Camgöz \textit{et al.}~\cite{camgoz2002saturation,camgoz2004saturation} about the effects of saturation on attention.

Then, subjects were asked the same question about shapes. The answers were more varied than those for the color question, and only the \square and \circle were commonly reported to be easier than average. This feeling can hardly be observed in the corresponding results (see \autoref{fig:human_res_outlier_shape_color}). The task was not specifically faster to solve when the outlier shape was a square or circle, although they both had low ERs. On the other hand, the \diamond was never reported as an easy shape, which suggests that subjects found it harder to find. This assumption is confirmed by the results: the diamond has the highest ER and RT. More than half of the subjects reported that the difficulty of finding the outlier using its shape dimension was dependent on the trial distractors. Many answers were similar to the following example: ``triangle among circles is pretty easy to find, whereas triangle among diamonds is hard".

We then asked the subjects to provide estimations of the \nbShapes and \nbCols values for trials during which they found the task hard. The answers were spread out between 3 and 6 colors and between 2 and 5 shapes, meaning that the perception of difficulty truly varied from one subject to another. A majority of answers reported a \nbCols value right above that of \nbShapes (\textit{e.g.}, (3-2), (4-3) or even (5-4)), which allows us to think that the capacity limit of color is higher than that of shape. It is important to note subjects were not provided the \nbCols and \nbShapes values of the trials and also did not know either that we preemptively removed some parameter values (\textit{e.g.}, 3 colors or 4 shapes, see \autoref{sec:autom_parameters_space}), and this explains why they sometimes answered with values they never saw but that still represented their feelings.

The penultimate question asked the subjects to report their strategy for solving the task. The main reported strategy was: first, observe if an outlier pops out of the display. If not, identify the colors and shapes in the trial; then, for each color, browse all the shapes of that color to find if there are two occurrences of the same stimulus. Finally, repeat the process until the outlier is found. This strategy is a typical behavior in visual search tasks, where the display is first processed preattentively, then by sections (\textit{i.e.}, texture segregation~\cite{callaghan1986textureSegr}), and if no ``match" has still been found, the display is ultimately processed serially.

Ultimately, subjects were asked what, overall, made a trial hard to solve. Two main factors came out of their answers. The first is the similarity between all the distractors in a grid. This corroborates the Duncan and Humphreys~\cite{duncan1989search} theory about target-nontarget and nontarget-nontarget similarity (see \autoref{sec:related_work}). The second reason is the direct neighborhood of the outlier in a trial. This may refer to the feeling that illusory conjunctions arose when the outlier was visually \textit{close} to its distractor neighbors.

\subsubsection{Capacity Limits of the Color and Shape Dimensions}
\label{sec:capacity_limits}

As a reminder, what we call the \textit{capacity limit} of color (or shape) is the maximum number of different features of that dimension that can be present in a display before a visual search task for an outlier becomes too arduous. Hence, we want to answer the following question: ``how many colors can one use in a representation before it becomes too complex to find an outlier?" when univariate or bivariate data are encoded with the color and shape dimensions.

We observed with \Hcol and \Hshape that \nbCols and \nbShapes do not have significant effects on performance when their respective dimensions are not relevant. Next, a dimension capacity limit will only be considered when the dimension is relevant with regard to identifying the outlier. With \Hred, we saw that type \typered trials were not affected by \nbShapes and that \nbCols variations led to small RT fluctuations that are not significant in an information visualization context. We assume that either the maximum values of these parameters in this experiment (7 \nbCols or 5 \nbShapes) are too small to observe a loss of performance or that the visual search process does not suffer from the noise induced by distractor heterogeneity when data are redundantly encoded. We believe that the latter assumption is correct since redundant encoding has been shown to make visual search tasks significantly easier~\cite{nothelfer2017redundant}.

\textbf{Color:} When color is the only relevant dimension for identifying the outlier (type \typecol), we observe (see \autoref{fig:err_rt_human_per_param_value} A-2 and B-2) that all \nbCols values led to ERs of less than 3\% and RTs of less than 4 seconds on average, except for the value 7, which reached a 29\% ER in over 11 seconds. We believe this kind of shift in performance is the consequence of exceeding the capacity limit of the color dimension in this experiment. Hence, when color is the only relevant dimension for representing data, its capacity limit is \textbf{strictly less than 7}. In conjunction search (type \typeconj), the capacity limit is \textbf{strictly less than 4} since we can see that the performances do not worsen as \nbCols increases from a value of 4.

\textbf{Shape:} When shape is the only relevant dimension (type \typeshape), the subjects did not make any errors when there were less than 5 shapes. At 5 shapes, the ER reached 32\%, which is close to the ER obtained for 7 \nbCols when color was a relevant dimension. As observed in \autoref{sec:hypothesis_validation}, the RT measures do show significant variations between every pair of values, though these differences remain linear and no threshold effect can directly be observed. However, we also observed in \autoref{fig:OOT_distrib} that there were several \textit{OOT} trials when there were 5 shapes in the type \typeshape trials, and these represent 28.6\% of the total trials of this variety. We conclude that the shape capacity limit is \textbf{strictly less than~5} when it is a relevant dimension.
For conjunction search (type \typeconj), no threshold effect can be observed as \nbShapes varies. As the number of OOT trials on type \typeconj trials is balanced among the \nbShapes values (see \autoref{fig:OOT_distrib}), there is no ``hidden" threshold effect due to missing data points either. Finally, no information about the capacity limit of shape can be determined from the results for the type \typeconj trials. Either 2 \nbShapes is already over the capacity limit, 5 \nbShapes is still under the limit, or the search task for an outlier in this case is actually linearly related to \nbShapes. \\

Therefore, the capacity limit of color is found to be higher than that of shape. This assumption reflects what subjects reported in the questionnaire (see \autoref{sec:qualitative_results}). Moreover, we can see in \autoref{fig:err_rt_human_per_param_value} that 5 \nbCols in type \typecol (\emph{1.5\% ER; 4 s RT}) trials led to significantly better performances than those obtained with5 \nbShapes in type \typeshape trials (\emph{31\% ER; 9.5 s RT}). Such a difference confirms that the shape dimension is more sensitive to heterogeneity than color when each is the relevant dimension for identifying an outlier.

\subsection{Results Sensitivity}
\label{sec:human_res_sensivity}

\begin{figure}[!bt]
	\centering
	\begin{subfigure}[b]{.325\linewidth}
		\includegraphics[align=c,width=\linewidth]{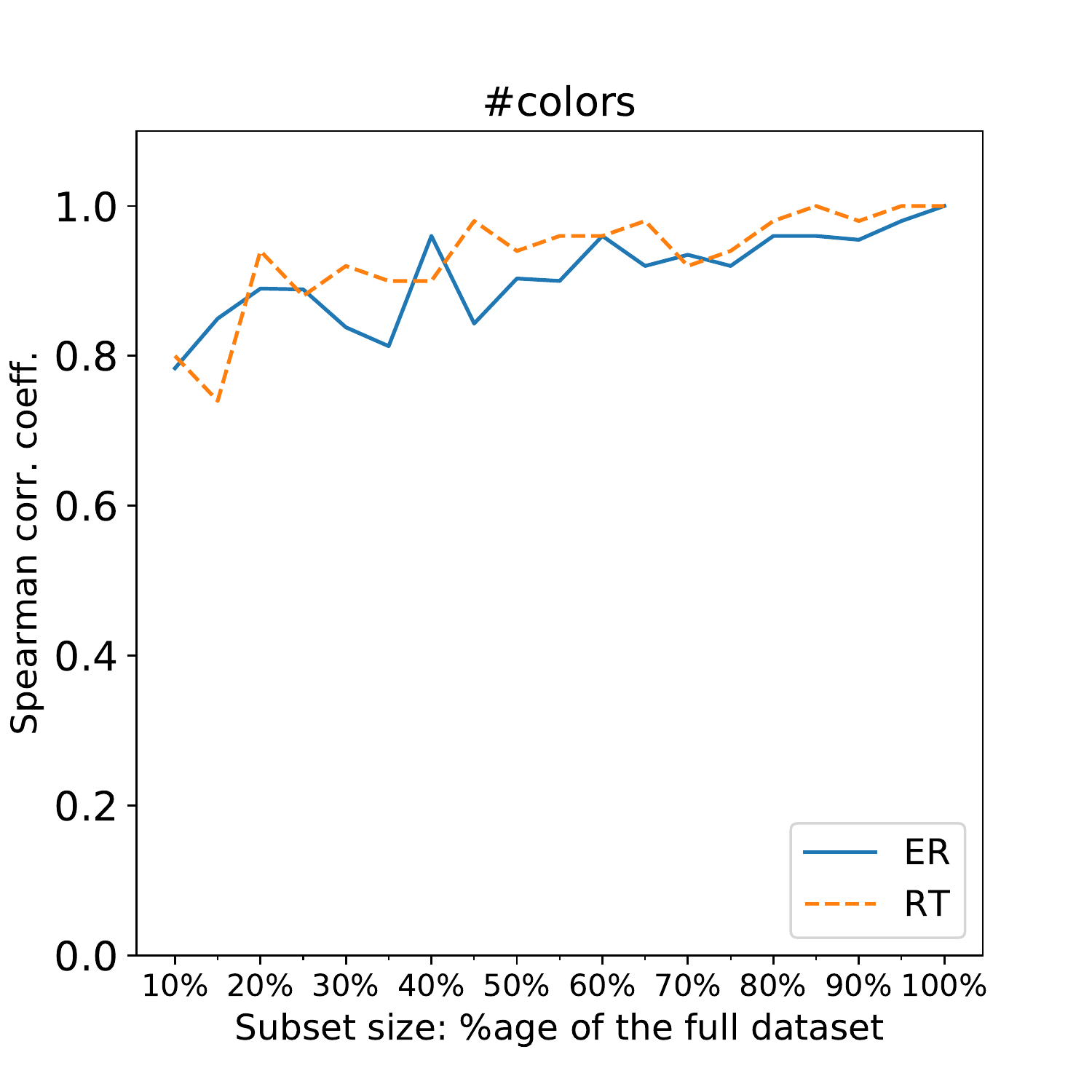}
	\end{subfigure}
	\begin{subfigure}[b]{.325\linewidth}
		\includegraphics[align=c,width=\linewidth]{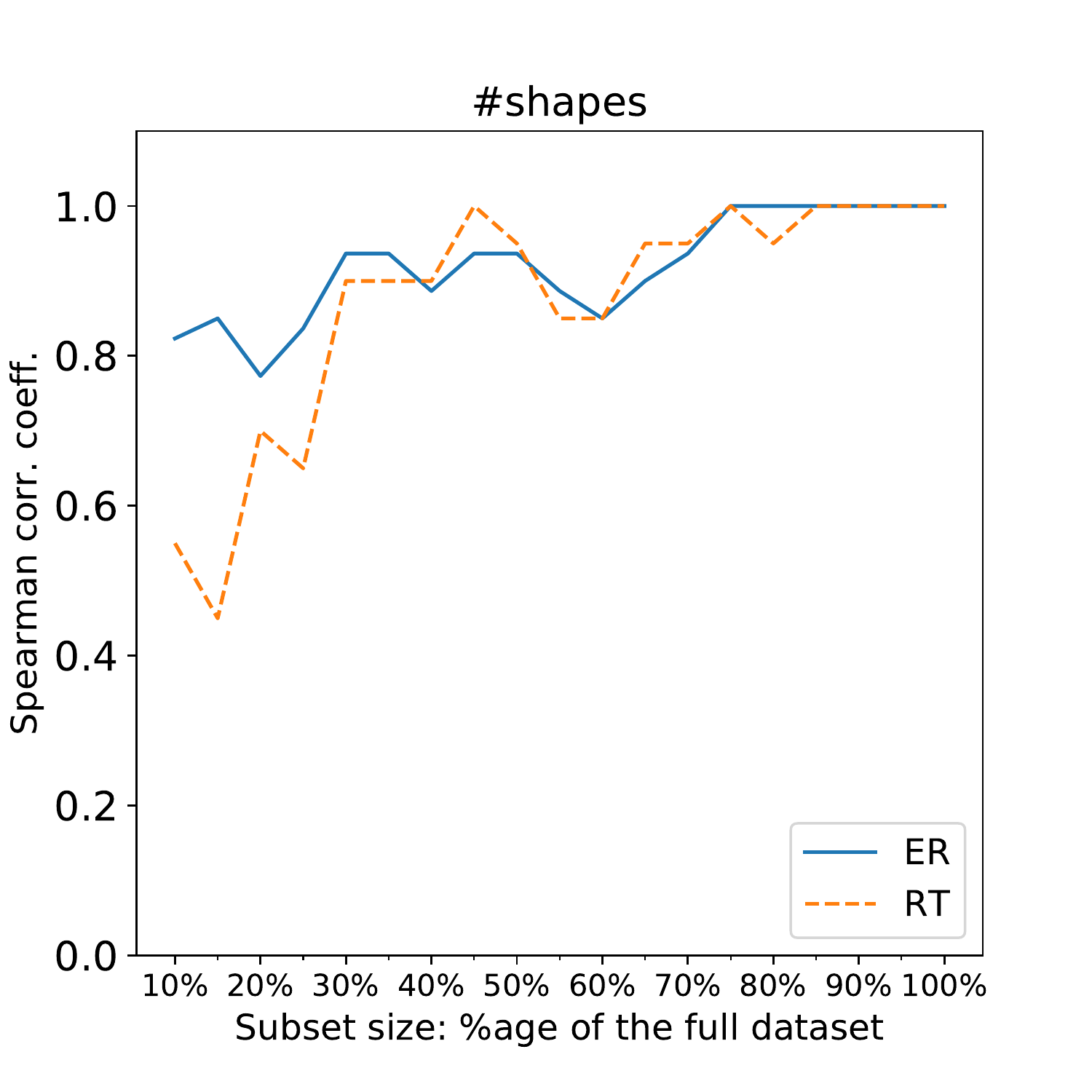}
	\end{subfigure}
		\begin{subfigure}[b]{.325\linewidth}
		\includegraphics[align=c,width=\linewidth]{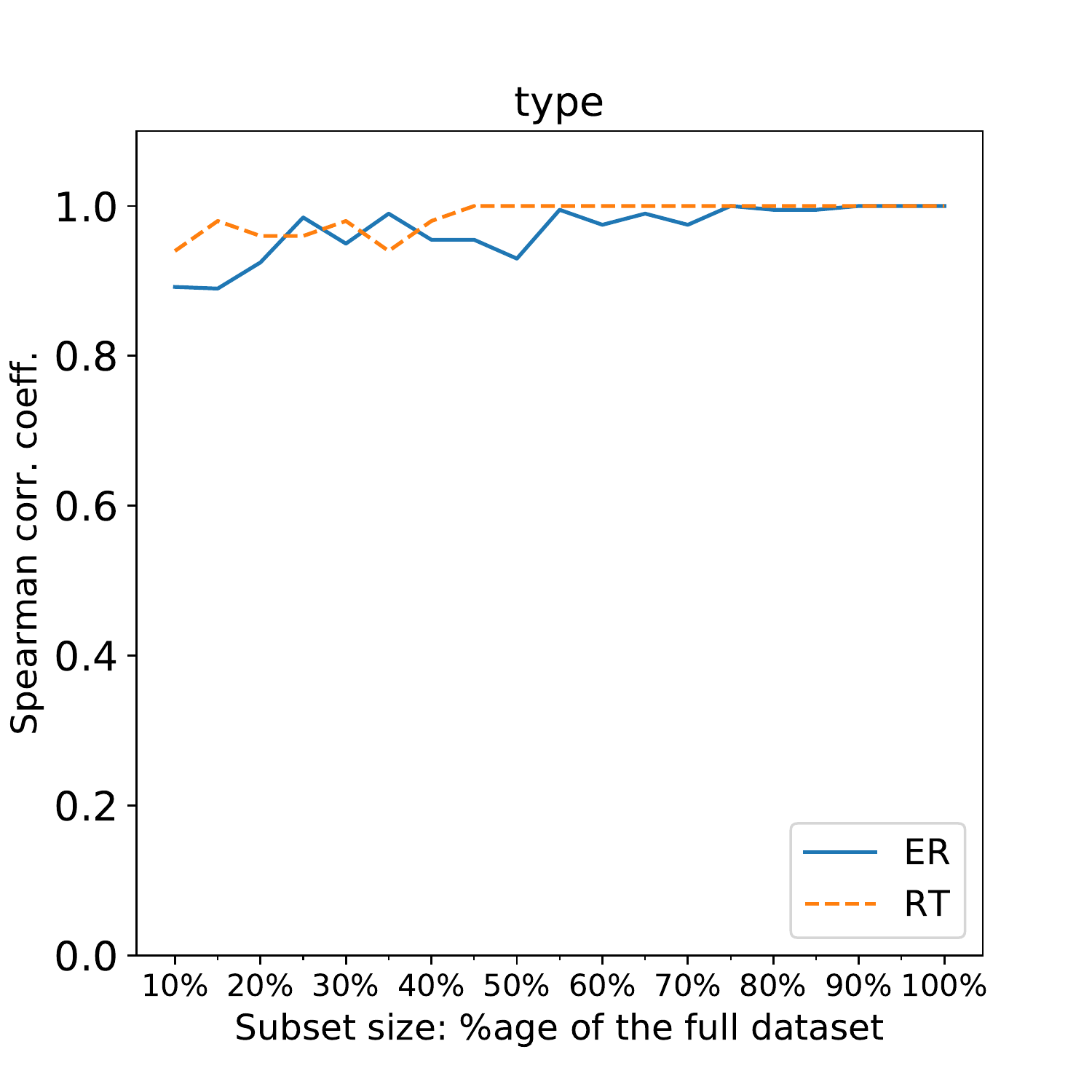}
	\end{subfigure}
	\begin{subfigure}[b]{.325\linewidth}
		\includegraphics[align=c,width=\linewidth]{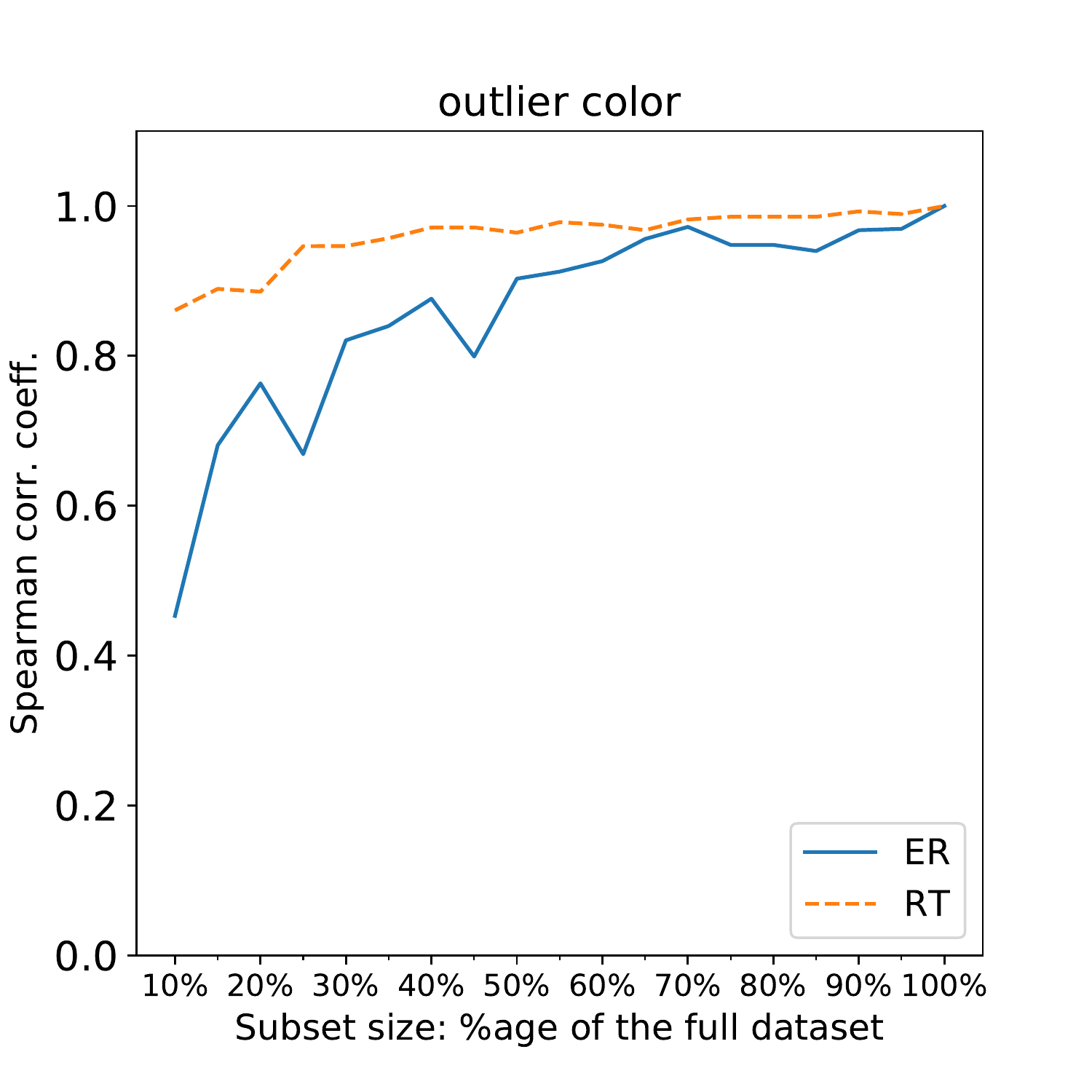}
	\end{subfigure}
	\begin{subfigure}[b]{.325\linewidth}
		\includegraphics[align=c,width=\linewidth]{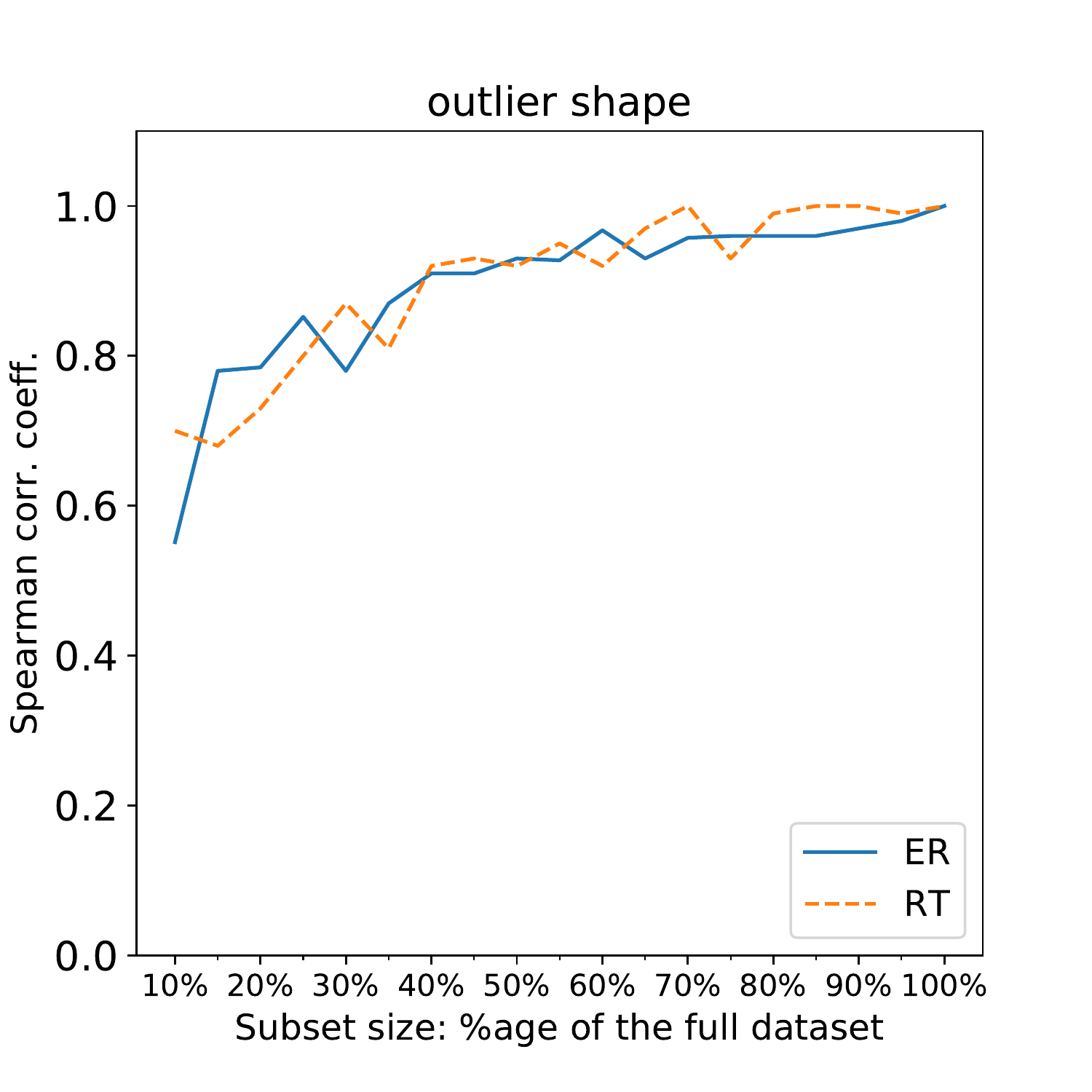}
	\end{subfigure}
	\begin{subfigure}[b]{.325\linewidth}
		\includegraphics[align=c,width=\linewidth]{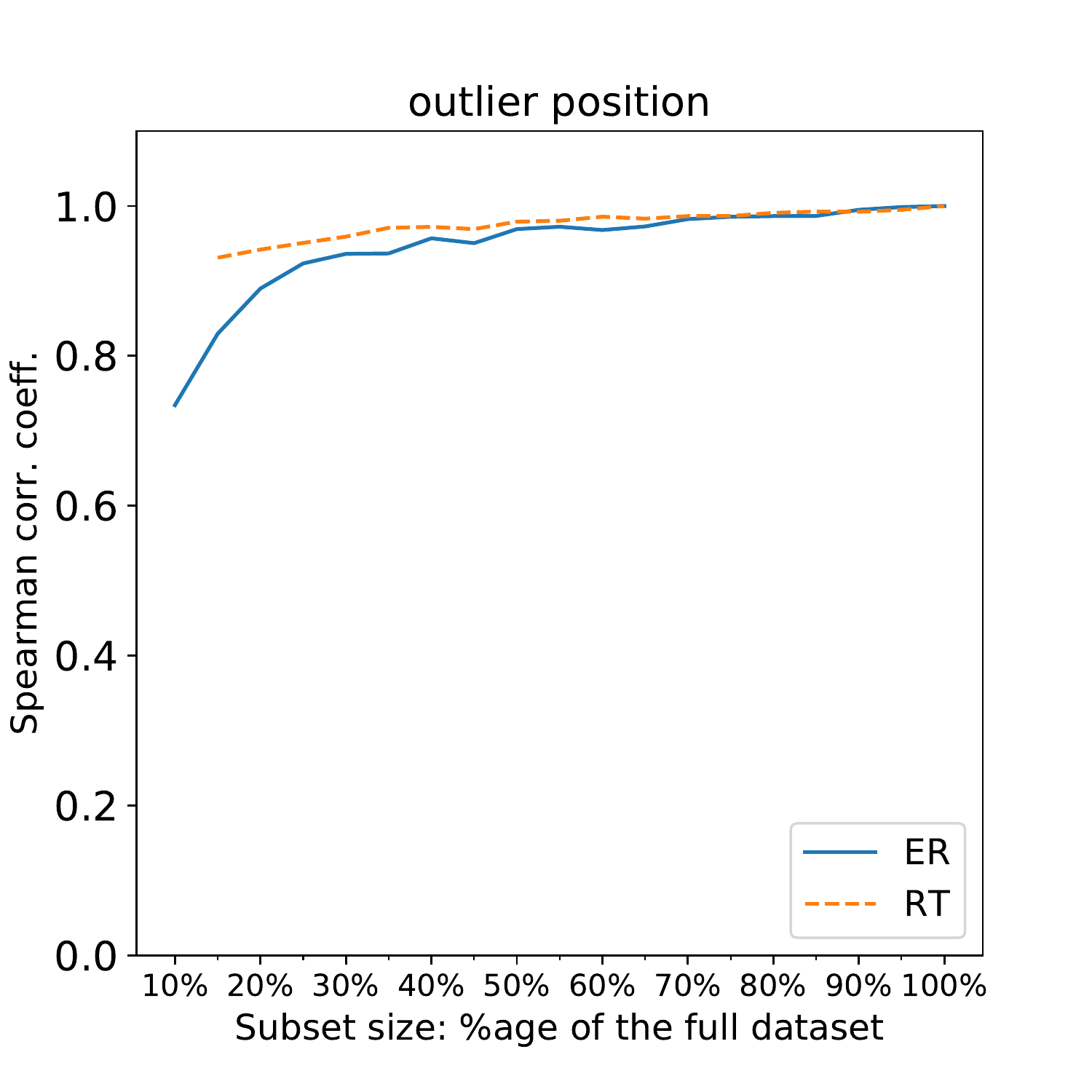}
	\end{subfigure}

	\caption{\label{fig:human_res_sensivity}Spearman correlation coefficients~\cite{zwillinger1999spearmanCorr} between random subsets of subjects (of different sizes) and all subject performances for each parameter. The X-axis corresponds to the size (percentage of the full dataset, from 10\% to 95\% with a 5\% step) of the subset of subjects for which their performances are compared to those of all subjects (100\%). Coefficients are averaged over 10 samplings for each subset size to minimize sampling-related uncertainties.}
\end{figure}

Similar to Demiralp \textit{et al.}~\cite{demiralp2014kernels}, we tested the robustness of our results to subject removal. We expect that more robust results in this sense have a greater ability to be generalized. To this end, we extracted random subsets of subjects of different sizes (from 10\% to 95\% of the full dataset) and computed the ER and RT performances from these subsets with each parameter. That is, for each parameter, we computed its mean ER and RT values on the subsets of subjects. Then, we used the Spearman correlation coefficient~\cite{zwillinger1999spearmanCorr} to quantify the correlation between the sequence of parameter performances values on a subset of subjects and the same sequence on the full dataset (\textit{i.e.}, all subjects). To reduce sampling-related uncertainties, each subset size was sampled 10 times so that the correlation coefficients reported in \autoref{fig:human_res_sensivity} were averaged over 10 computations. Spearman correlation coefficients range between -1 and 1, where 1 is a positive correlation, -1 is a negative correlation and 0 means no correlation. \autoref{fig:human_res_sensivity} shows that the correlation coefficients are over $0.5$ with 25\% of the subject data, indicating that some information is preserved. With 40\% to 50\% of the data, the correlation coefficients are all between 0.8 and 1, meaning that most of the information is preserved with only 50\% of the subjects.
Such high correlation coefficients with only half the subjects show that the results of these experiments are not very sensitive to subject removal. Hence, they are not biased by some subject-specific behaviors, and the number of subjects considered is sufficient for the results to be significant.

\subsection{Limitations}
\label{sec:human_limitations}

For any experiment, its results have to be considered within the limitations induced by its protocol.

First, our subjects are taken from a noticeably educated population in computer science. Hence, they are used to computer devices and are familiar with visualizations (as users or experts). The results of this experiment must be used with caution if one is to generalize them to a more generic population.

In the experimental design, we selected our color set from a color palette provider to make it categorical (see \autoref{sec:dataset_gen}) so that all colors could be as distinguishable as possible. Since the experiment aimed to study the capacity limit of color (and not to find which color set is best for representing categorical data), we mitigated the effect of the color set on the performances. However, some subjects found that the differences in saturation between some colors had impacts on their performances.

Another limitation is that the observed \nbCols performance variations are limited by an aspect of the experimental design. When generating an image with $5$ colors, these are not taken from a designed set of 5 categorical colors but are picked from the $7$ color set defined in \autoref{sec:data_space}. It is then important to note that the performances observed on trials with less than $7$ colors were not measured from trials with an optimal color set fitted to their \nbCols. The same limitation could be stated for shapes, although the notion of distance between shapes is more complex. Nevertheless, it was necessary to define a static set of colors and shapes so that we could keep track of the \textit{outlier color} distribution and be able to aggregate them to study our results. Having a different and optimal set for each \nbCols value would also have made the subjects able to infer properties of the display from the color set of a trial, and this could bias the ways in which they built their strategies.

Finally, as stated in \autoref{sec:data_space}, we balanced our shape sub-features (\emph{i.e.}, straight horizontal and vertical lines, curves, tilted lines) to try not to favor any particular kind. However, such a definition leads to imbalances between shape areas. The \triangle has a smaller area than the \square, so the colors they are filled with are not represented by the same number of pixels. Thus, when color is a relevant dimension for identifying the outlier, it may be harder to solve the task if the outlier shape is a \triangle than if it is a \square. Although our study does not enable us to validate this assumption, it would be interesting to experiment whether, when considering a large number of colored shapes, the shape areas are (or become) more important than the shapes themselves.

\section{Discussion}
\label{sec:discussion}
This experiment enabled us to study some hypotheses as well as the capacity limits of color and shape in an information visualization context. Nevertheless, there remain some interesting outcomes that we did not address, and we discuss them in this section.

\subsection{Number of Visually Different Stimuli}

\begin{figure}[!bt]
\centering
\begin{tabular}{|c|c|c|}
    \hline 
    & \multicolumn{2}{c|}{Effects of the \nbDiffStimuli meta-parameter on the task}  \\
    \cline{2-3}
    & Error rate & Response time \\
    \hline
    {\shortstack{\begin{sideways}\hspace{-0.5cm}\emph{Color} type\end{sideways}}} &
    {\includegraphics[align=c,width=.40\linewidth]{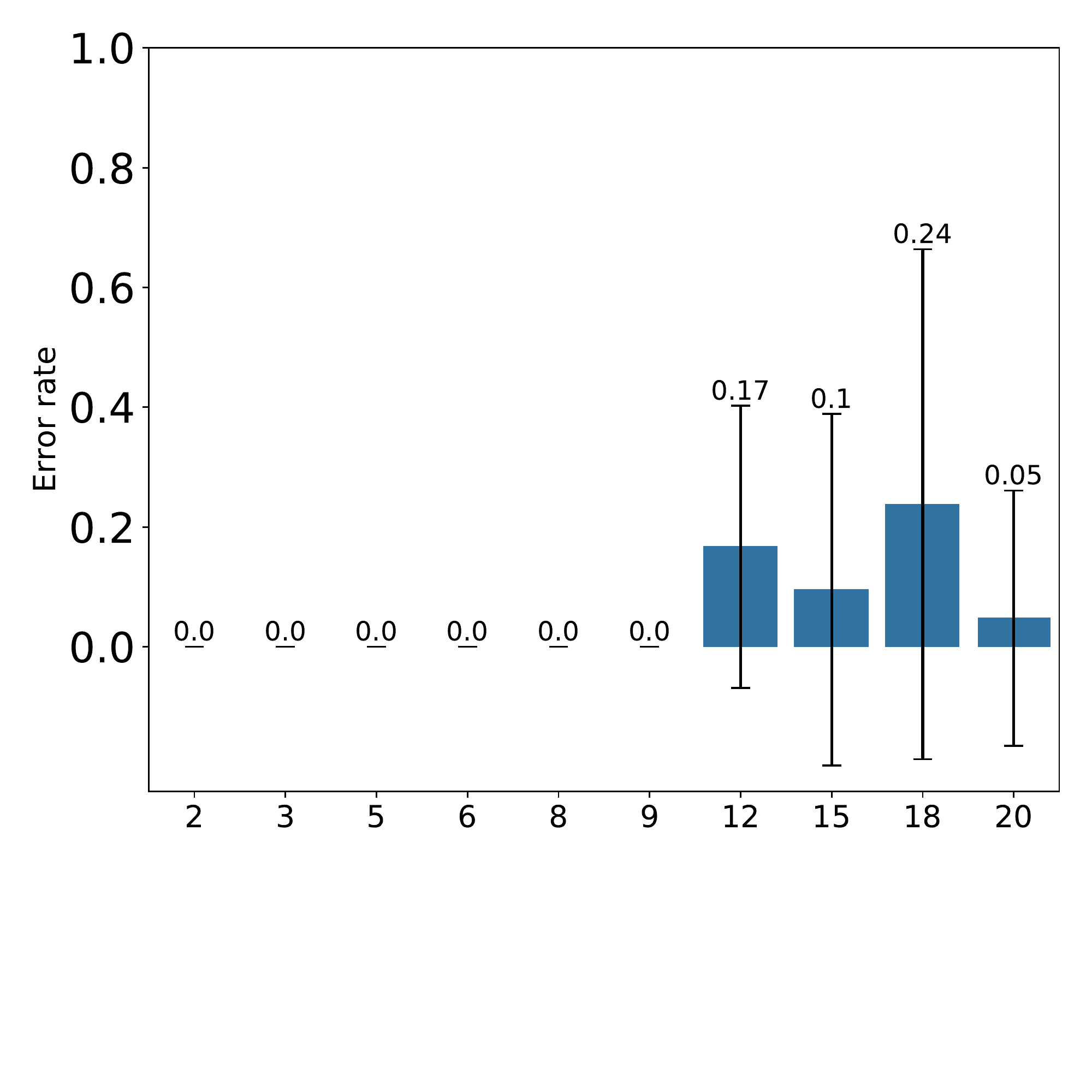}} &
    {\includegraphics[align=c,width=.40\linewidth]{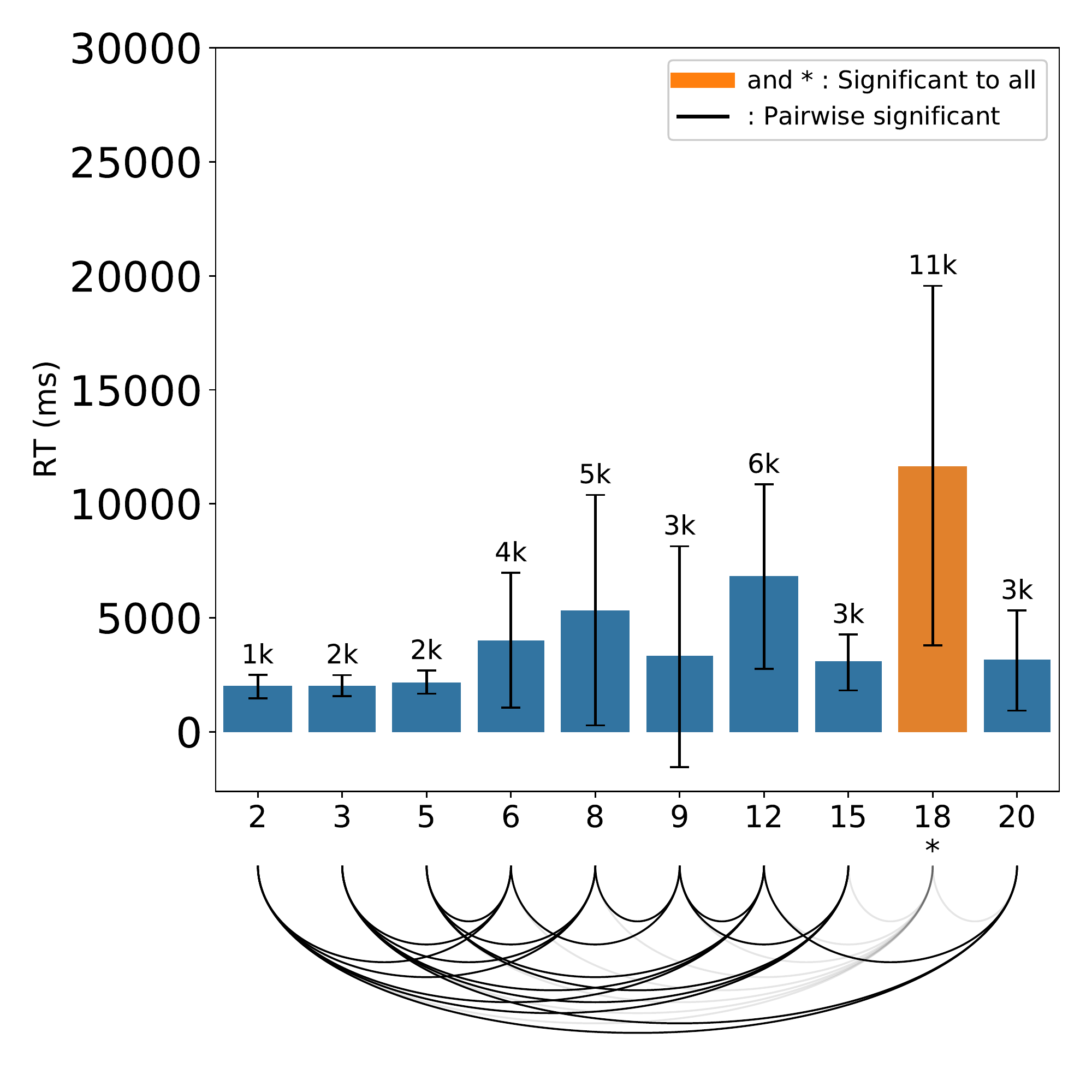}} \\
    \hline
    {\shortstack{\begin{sideways}\hspace{-0.5cm} \emph{Shape} type\end{sideways}}}&
    {\includegraphics[align=c,width=.40\linewidth]{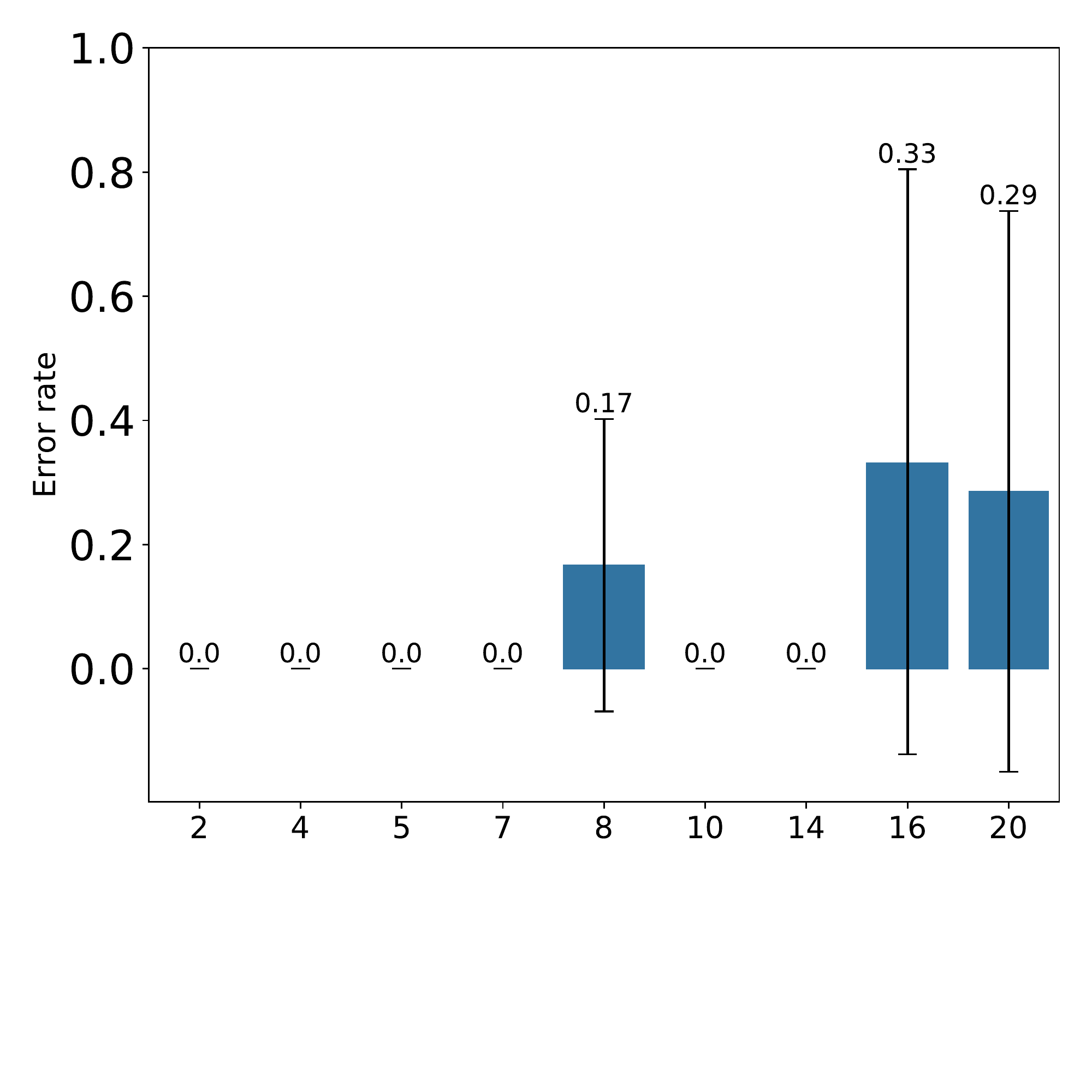}} &
    {\includegraphics[align=c,width=.40\linewidth]{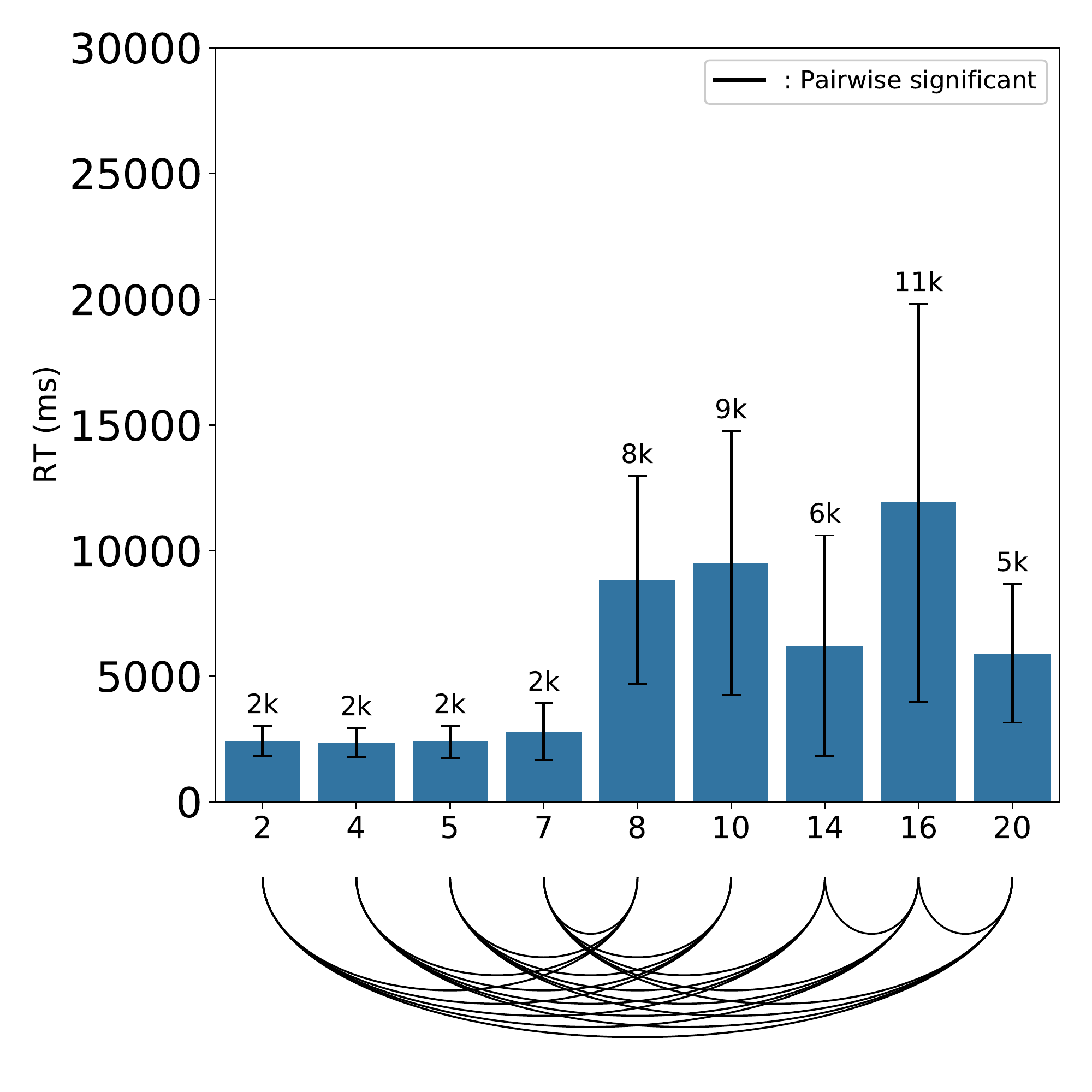}} \\
    \hline
    {\shortstack{\begin{sideways}\hspace{-0.5cm} \emph{Red.} type\end{sideways}}} &
    {\includegraphics[align=c,width=.40\linewidth]{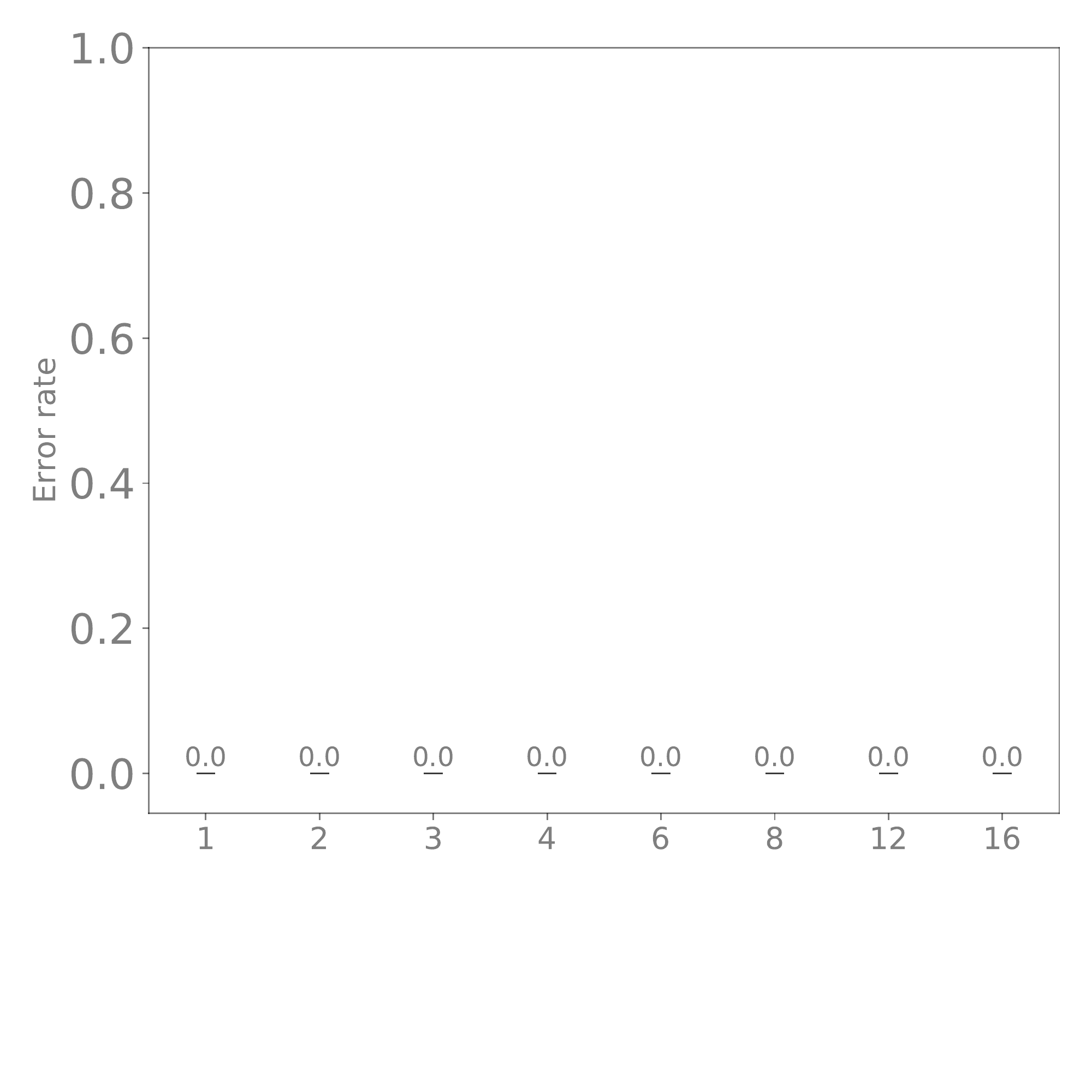}} &
    {\includegraphics[align=c,width=.40\linewidth]{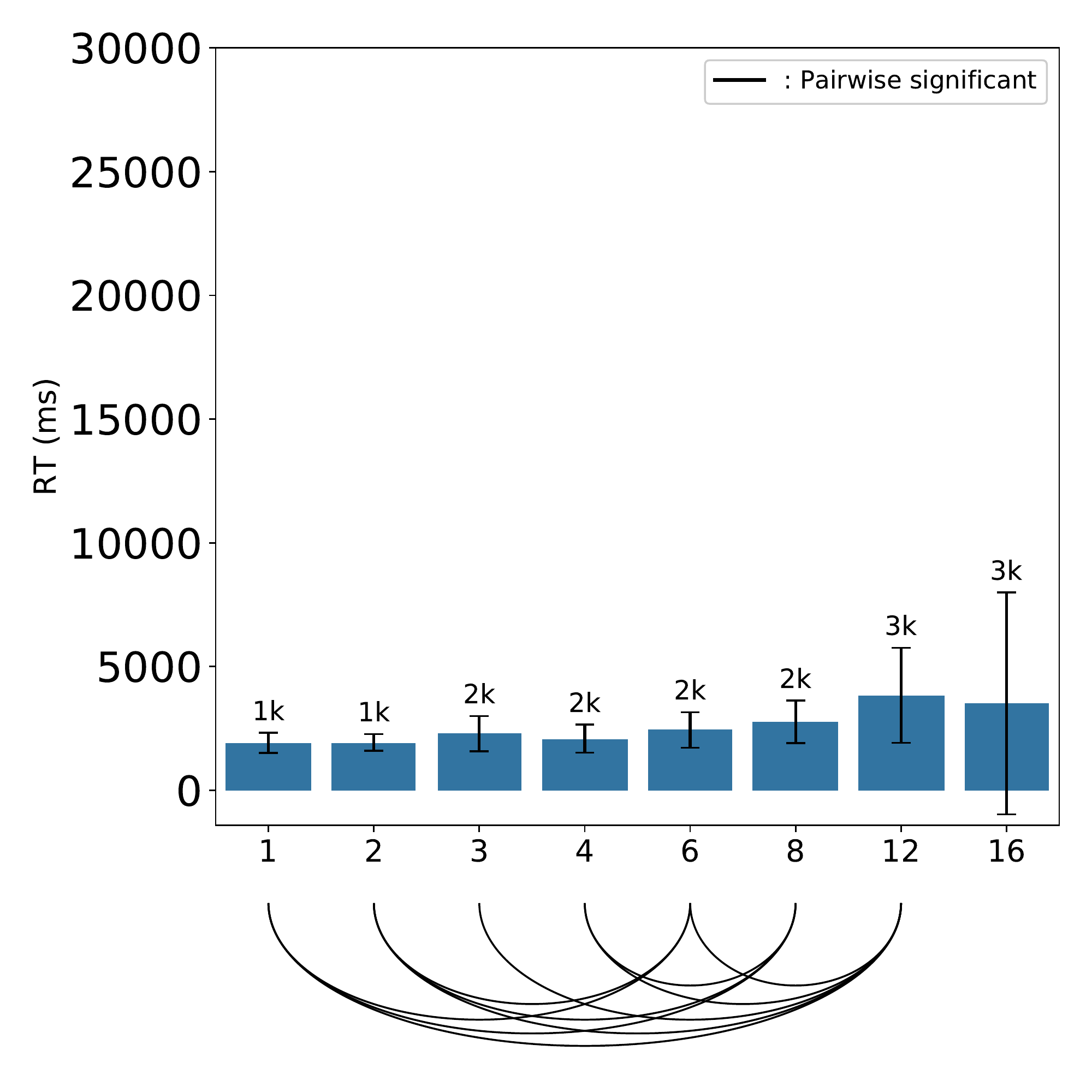}} \\
    \hline
    {\shortstack{\begin{sideways}\hspace{-0.5cm} \emph{Conj.}type\end{sideways}}} &
    {\includegraphics[align=c,width=.40\linewidth]{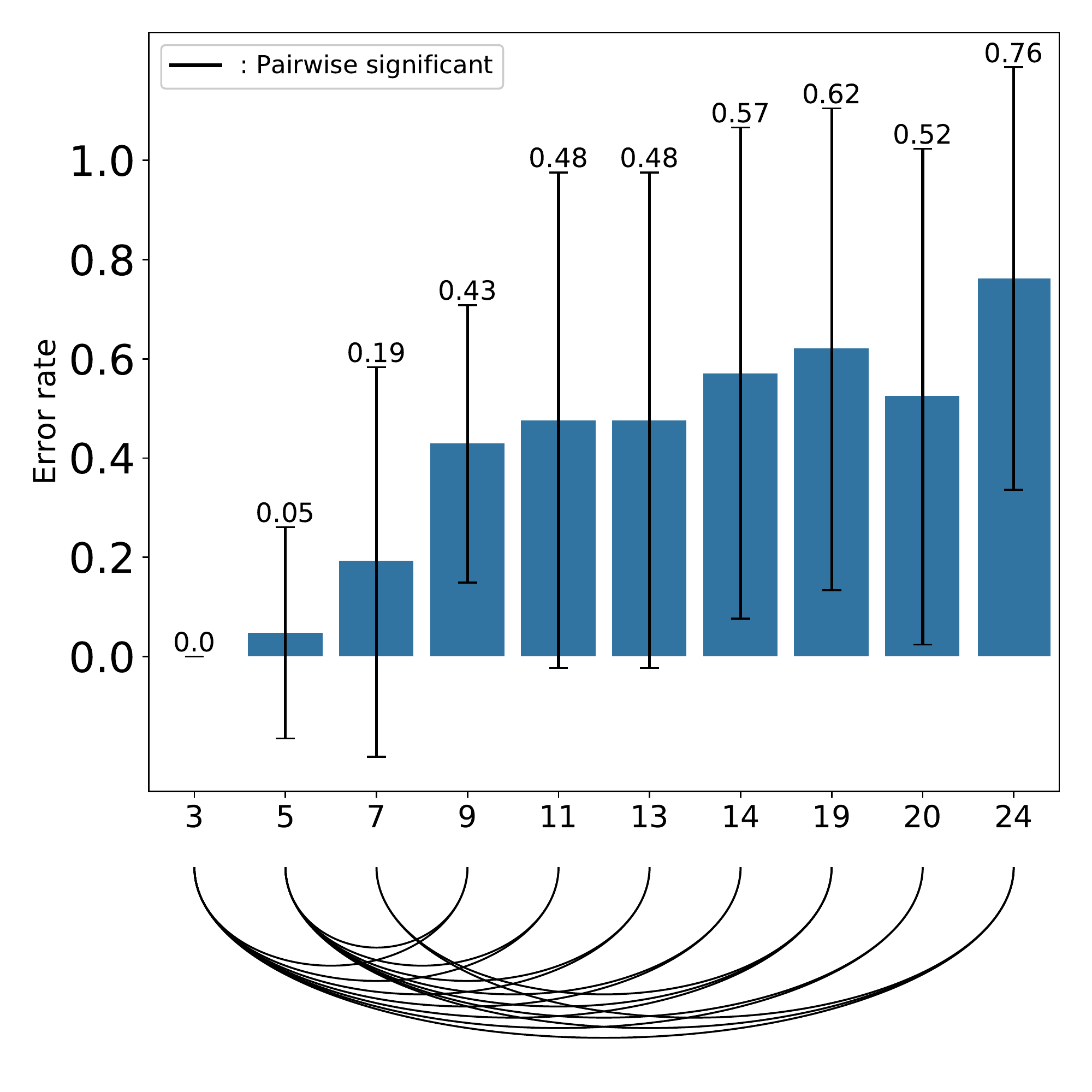}} &
    {\includegraphics[align=c,width=.40\linewidth]{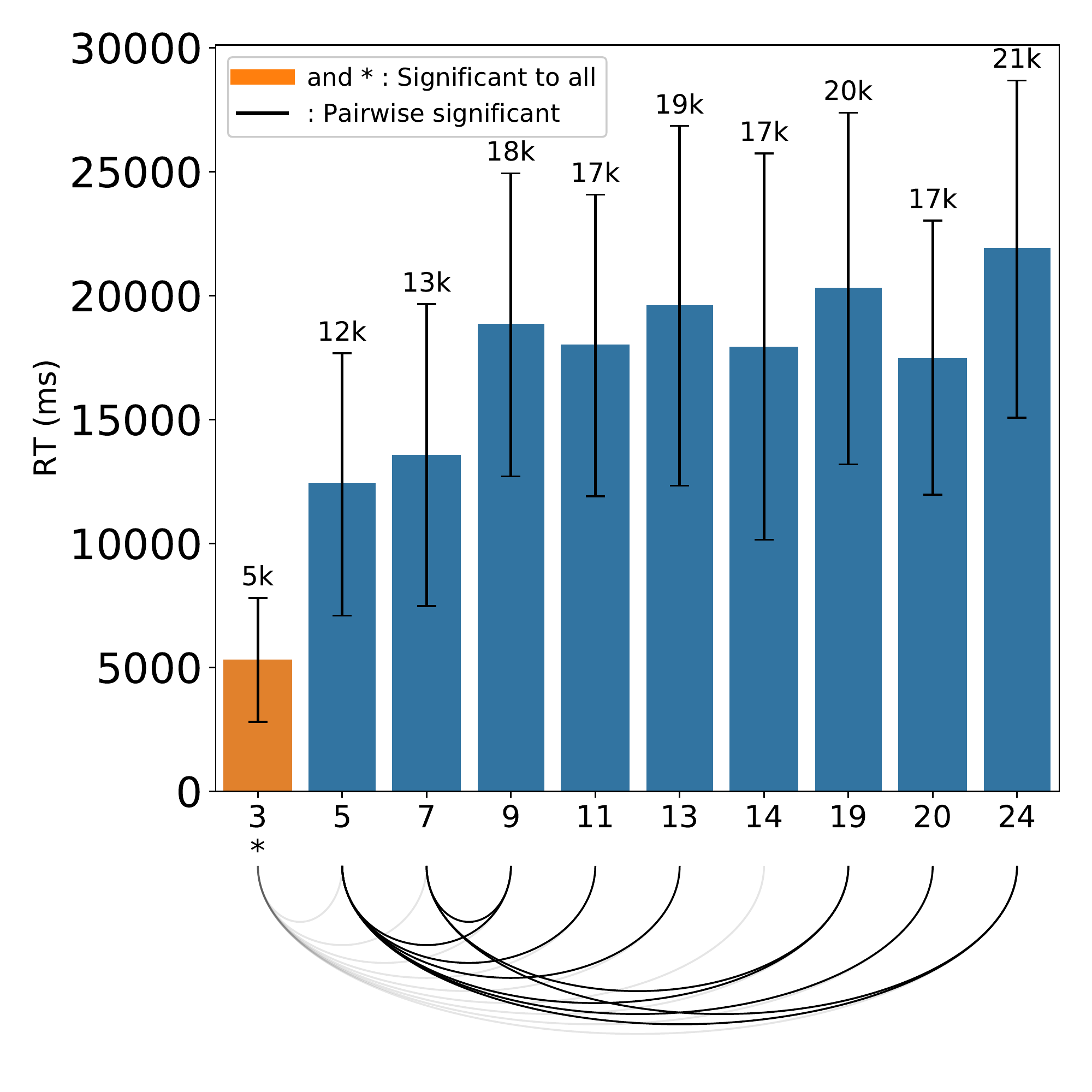}} \\
    \hline
\end{tabular}
\caption{Subjects ER and mean RT performances for \nbDiffStimuli and each \textit{type} value. A reading example is given in \autoref{fig:human_type_perf} caption.}
\label{fig:human_er_rt_nbDiffStimuli}
\end{figure}

In this experiment, we found that \nbCols and \nbShapes did not have significant effects on performance when they were not relevant dimensions for identifying the outlier. This statement extends the findings of the visual search literature and mainly corroborates the work of Pashler~\cite{pashler1988texture}, who specifically studied how a visual search for an unknown target was affected by heterogeneity in irrelevant dimensions. Although this has been verified by both the literature and our experiment, we find it counterintuitive. In a complex display with tens of features, all stimuli cannot be processed at once. In such a case, we expect the visual search for an unknown target to be affected by heterogeneity in irrelevant dimensions. Since subjects are not told which dimension is relevant or not, it is not until the task is solved (\textit{i.e.}, the outlier is found) that they know which dimension was (ir)relevant. Visual searches are processed serially on complex displays, and subjects can be significantly affected by heterogeneity in irrelevant dimensions.

This intuition motivated the study of the subjects' performances through the spectrum of a ``meta-parameter" that merges both \nbCols and \nbShapes. Its definition also depends on the experimental object \textit{type} since it defines which and how many dimensions are reserved for the outlier. We call this meta-parameter \nbDiffStimuli, which stands for the \textit{``number of visually different distractors"}. For example, an experimental object of type \typecol, with 4 \nbCols and 2 \nbShapes, will have $(4-1) * 2 = 6$ \nbDiffStimuli. It is important to note that the experiment was not designed to study this parameter, and its distribution is not uniform. By definition, \nbDiffStimuli quantifies the visual attribute heterogeneity in a trial, whatever the relevant dimension(s) is/are.

Subject performances aggregated by \nbDiffStimuli are presented in \autoref{fig:human_er_rt_nbDiffStimuli}. In type \typecol trials, subjects did not make any errors when there were less than 12 \nbDiffStimuli, and the ER does not significantly increase starting from 12 \nbDiffStimuli. The type \typecol RTs increase starting from 6 or 8 \nbDiffStimuli. The same trend is observed for the type \typeshape performances, where no errors were made in trials with less than 16 \nbDiffStimuli (except for 8), and the RTs substantially increase starting from 8 \nbDiffStimuli. For the \typered type, subjects never made any errors, and the RT performances remain almost constant, although a slight increase in the RT occurs starting from 8 \nbDiffStimuli. Finally, the type \typeconj ER significantly increases starting from 5 and 7 \nbDiffStimuli, and the RTs dramatically increase above 3 and 7 \nbDiffStimuli and does not significantly increase further as the \nbDiffStimuli values increase.

Performances tend to worsen around 8 \nbDiffStimuli. In all plots except those of the type \typecol and type \typered ERs, performances drop when the threshold of 8 \nbDiffStimuli is reached or exceeded. This number is inside the range of Miller~\cite{miller1956magical} magical number. This generic number is defined as ``the capacity of the short term (or working) memory is $7 \pm 2$". This suggests that there are no specific capacity limits for the different visual attributes but rather that representation complexity is driven by the total number of different visual stimuli that are drawn within. The representation is easier to interpret if its number of visually different stimuli does not exceed the short-term memory capacity, regardless of the chosen visual attributes and their features. This hypothesis would certainly be contestable, as we can expect visual attributes to require different kinds of efforts to be remembered, and their perceptual kernels~\cite{demiralp2014kernels} are different. Verifying this hypothesis is beyond the scope of this experiment, but it is an interesting lead for future work.

\subsection{Common Assumption about Color Capacity Limit}

It is commonly assumed that the maximum number of colors in a representation that humans can efficiently handle is $10 \pm 2$. This assumes a capacity limit significantly higher than the Miller~\cite{miller1956magical} magical number, meaning that color should be an efficient visual attribute (\textit{i.e.}, efficient to process and remember). This common assumption seems to be used by many color palette providers, as many palettes are built for up to 12 classes. For example, the ColorBrewer tool \cite{harrower2003colorbrewer} recommends using between 5 and 7 classes for choropleth maps, while isoline maps can safely use more; the online tool provides color palettes with up to 12 classes. To the best of our knowledge, no study has verified this assumption. As our experiment showed the capacity limit of color to be \textbf{strictly less than 7} for a feature search, our results invalidate the commonly assumed limit of $10 \pm 2$ colors. Color remains an efficient visual attribute for encoding data, but its limit is lower than assumed in many representation tools. Some strategies, such as color grouping \cite{haroz2012outlier}, can be used to increase this limit but are not suited for all representation designs.

\subsection{Subsampling in Experiments}

The goal of this paper is to study the capacity limits of the color and shape attributes at a large scale, and we have considered relatively high numbers of parameters and parameter values. Varying these parameters led to the combinatorial explosion of the number of experimental objects (\num{3290}). To address such broad parameter spaces in user evaluations, two methods are commonly used: (i) recruit enough subjects to cover all the parameter configurations or (ii) find a subset of configurations that represent the whole parameter space.

The first method requires many subjects to cover all parameter configurations. The number of subjects can be reduced using \textit{between-subjects} experimental designs and can be found through \textit{crowdsourcing} platforms. In our experiment, we opted for the second method, which requires finding a subset of experimental objects that can statistically represent the whole parameter space. This process, which we call \textit{parameter space reduction}, is more often referred to as \textit{subsampling}.

There already exist well-established strategies for subsampling a design space. Since the data in this experiment are synthetic and generated, we only discuss the sampling methods that work on parameter space distributions (and not datasets). For example, \textit{systematic} and \textit{random} sampling are often used in many evaluations. Sometimes, the design space is also arbitrarily defined according to prior works. Such sampling methods have an important drawback: they assume how task difficulty is correlated with parameter values (\textit{e.g.}, linearly with systematic sampling). In general, there is no evidence that the difficulty of a task is linearly (or not) related to the considered parameter variations, and these sampling methods might result in the loss of some key values for the study.

An answer to this drawback is to base the subsampling process on a metric that assesses the configurations of parameter (\textit{i.e.}, the conditions) efficiency to solve a given task. We computed a task-oriented difficulty metric based on a deep neural network (DNN). The main advantages of this metric design are (i) that it fits any task that can be programmatically expressed and (ii) it does not require any a priori information about the task. The model learns by itself what ``elements" of the display are relevant for solving the task. The main drawback is that the model performances and strategies for solving the task will rely on its architecture. This means that different architectures do not achieve the same performances with the same parameter space conditions, and it becomes difficult to select a model and interpret its results. Some studies have already been confronted with this drawback and proposed mitigation ideas (see \autoref{sec:related_work_dnn}), but this approach would benefit from further research on the correlations that lie between humans and DNNs with respect to representation interpretation tasks.

\subsection{\resnet--Subject Correlations}
Even though it was stated that we did not expect deep neural networks (DNNs) and subjects' results to be strictly correlated (see \autoref{sec:methodology}), it seems necessary to study whether their performances are completely different. To this end, we computed Spearman correlation coefficients between the \resnet ERs and subject performances in terms of \nbCols, \nbShapes, \textit{type}, and \nbDiffStimuli.

For the overall and type \typecol experimental objects, \resnet and subject performances are positively correlated ($\rho\ge 0.68$), although the correlation on \nbShapes is lower ($\rho=0.5$). This lower correlation is probably due to some strategy that we suspect \resnet uses to solve the task (see \autoref{sec:resnet_limitations}). For type \typeshape trials, the results are strongly correlated on \nbShapes only. Finally, both the subject and \resnet results are strongly correlated ($\rho\ge 0.72$) on type \typeconj. Future work to deepen our understanding of DNN-human correlations is a major lead that would benefit DNN-based approaches when assessing visualization efficiency.

\section{Conclusion}
\label{sec:conclusion}
Visual search guidelines are well established cornerstones of the information visualization community for building efficient representations. However, they rely on fine-grained evaluations whose parameter space is limited compared those encountered by modern real-world representations. This paper has presented an experiment to measure the capacity limits of attention with the color and shape visual attributes as a continuation of the information visualization community efforts to quantify visual attribute performances in the information visualization context.

We trained a convolutional neural network to compute a difficulty metric that provided insights into the effects of the considered representation parameter values on the performances of subjects solving the task. This metric helped refine our hypotheses and served as a base for the parameter space reduction needed to design a reasonable user evaluation. This process limited the number of arbitrary choices we had to make to design the user evaluation and eased the experimental reproducibility.
Then, we studied some hypotheses and the capacity limits of attention with respect to color and shape with a user evaluation on 21 valid subjects. The task consisted of identifying an outlier stimulus in a background of randomly laid out distractor stimuli on 44 trials selected from the difficulty metric analysis to represent the whole dataset that was initially considered.
The results of the experiment showed that the capacity limit of attention for the color dimension (strictly less than 7) is higher than that of the shape dimension (strictly less than 5) when each is the relevant dimension for identifying the outlier in an information visualization context. This conclusion implies that colors should be preferred over shapes for representing multiclass data, and this is also what subjects preferred. 
The experiment also provided evidence that redundant encoding is a major simplification factor for representation readability, with performances being constantly good despite the increase in distractor heterogeneity. Finally, we found that mixing these visual attributes to represent data was very harmful to the understanding of the representation. In this scenario, the performances with regard to identifying an outlier dramatically decrease even if there are only a few different features.

This experiment showed that an approach based on a DNN model is a promising means to refine user evaluation designs when parameter value variations lead to a combinatorial explosion in the number of configurations to consider. It also extended theories from the visual search literature, as some results corroborated these theories on the highly complex representations on which the task was evaluated, thereby demonstrating their validity for a specific information visualization-like context.

Many future work ideas have emerged throughout this study. Regarding the experiment itself, a first line of future work could be to search for a set of outlier stimuli instead of a single outlier stimulus. That is, find an outlier cluster in a grid of random stimuli (or random clusters). This problem is also common in information visualization and involves other visual search strategies, such as texture segregation~\cite{callaghan1986textureSegr,pashler1988texture}. Moreover, we studied the color and shape visual attributes in this paper, but many others are also commonly used in representations (\textit{e.g.}, position, size). One could study the different effects of mixing these dimensions to represent data and measure which dimensions are least harmful to representation readability when joined together. Although conjunction search has been shown to make visual search tasks much harder, visualization designers cannot always afford not to mix their visual attributes when representing tens of data classes. However, our experiment showed that mixing color and shapes quickly made representations arduous to read. Hence, more in-depth studies to find conditions that optimize conjunction search in complex displays would be valuable.

Regarding the difficulty metric, it will be mandatory to deepen our knowledge about the correlations between humans and DNNs. Understanding when to trust DNN predictions to infer information about task difficulty would enable us to weight the metric and make it fit closely to what humans are truly capable of. One could also study how bioinspired models behave compared to standard CNNs and whether they are closer to human behaviors on visual search tasks. Finally, considering an ensemble of networks instead of relying on a single network could lead to minimization of the bias induced by specific DNN architectures.

\ifCLASSOPTIONcaptionsoff
  \newpage
\fi

\bibliographystyle{IEEEtran}
\bibliography{outEval}

\vskip 0pt plus -1fil

\begin{IEEEbiography}[{\includegraphics[width=1in,height=1.25in,clip,keepaspectratio]{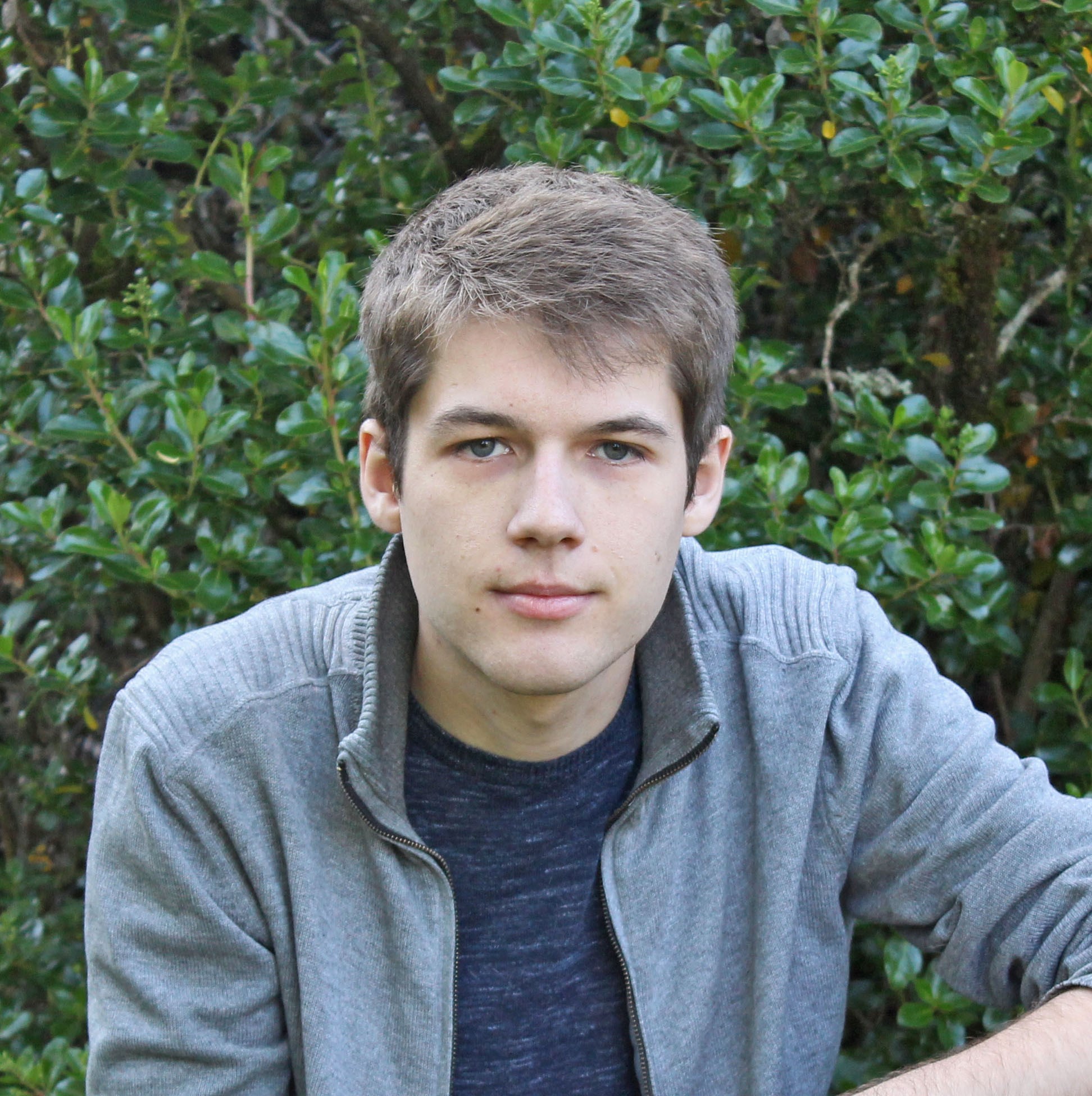}}]{Loann Giovannangeli}
is a PhD. student at the LaBRI, University of Bordeaux, France. He worked one year as a research engineer in the LaBRI. He obtained his Master of Science degree in 2019 from the University of Bordeaux. His research interest include Information Visualization, Machine Learning and especially the evaluation of representations efficiency with computer vision techniques.
\end{IEEEbiography}

\vskip 0pt plus -1fil

\begin{IEEEbiography}[{\includegraphics[width=1in,height=1.25in,clip,keepaspectratio]{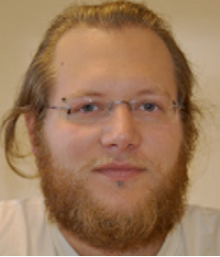}}]{Romain Giot}
is Associate Professor at the LaBRI, University of Bordeaux, France, since 2013.
His research interests include Explainable Machine Learning, Biometric Authentication, Information Visualization.
\end{IEEEbiography}

\vskip 0pt plus -1fil

\begin{IEEEbiography}[{\includegraphics[width=1in,height=1.25in,clip,keepaspectratio]{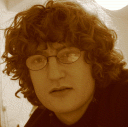}}]{David Auber}
received his PhD degree from the University of Bordeaux I in 2003. He has been an assistant professor in the University of Bordeaux Department of Computer Science since 2004. His current research interests include information visualization, graph drawing, bioinformatics, databases, and software engineering 
\end{IEEEbiography}

\vskip 0pt plus -1fil

\begin{IEEEbiography}[{\includegraphics[width=1in,height=1.25in,clip,keepaspectratio]{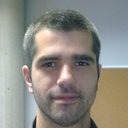}}]{Romain Bourqui}
received his Master and PhD degrees in Computer Science from the University Bordeaux I in 2005 and 2008. He has been an associate professor at the University of Bordeaux since 2009. His research interests include Information Visualization, Large Data Visualization, Explainable Machine Learning.
\end{IEEEbiography}

\end{document}